\documentclass[12pt,oneside]{article}
\usepackage{amsfonts}
\begin{document}
%\frontmatter

%\topmargin=-.35in
%\textheight=8.60in
%\oddsidemargin=0.0in
%\textwidth=6.3in

%\begin{titlepage}
\begin{center}
{\large\bf Cluster Convergence Theorem\\}
\vspace{0.14cm}
\vspace*{.05in}
{Chris Austin\footnote{Email: chrisaustin@ukonline.co.uk}\\
%\vspace*{.3in}
\small 33 Collins Terrace, Maryport, Cumbria CA15 8DL, England\\
}
\end{center}
%\vspace*{0.8in}
\begin{center}
{\bf Abstract}
\end{center}
\noindent A power-counting theorem is presented, that is designed to play an analogous
role, in the proof of a BPHZ convergence theorem, in Euclidean position space,
to the role played by Weinberg's power-counting theorem, in Zimmermann's proof
of the BPHZ convergence theorem, in momentum space.  If $x$ denotes a position
space configuration, of the vertices, of a Feynman diagram, and $\sigma$ is a
real number, such that $0 < \sigma < 1$, a $\sigma$-cluster, of $x$, is a
nonempty subset, $J$, of the vertices of the diagram, such that the maximum
distance, between any two vertices, in $J$, is less than $\sigma$, times the
minimum distance, from any vertex, in $J$, to any vertex, not in $J$.  The set of
all the $\sigma$-clusters, of $x$, has similar combinatoric properties to a
forest, and the configuration space, of the vertices, is cut up into a finite
number of sectors, classified by the set of all their $\sigma$-clusters.  It
is proved that if, for each such sector, the integrand can be bounded by an
expression, that satisfies a certain power-counting requirement, for each
$\sigma$-cluster, then the integral, over the position, of any
one vertex, is absolutely convergent, and the result can be
bounded by the sum of a finite number of expressions, of the same
type, each of which satisfies the corresponding
power-counting requirements.

\vspace{0.05cm}

What follows is the \LaTeX2e transcription of my 1988 paper, with references
added.  Conventions, some definitions, and some combinatoric Lemmas,
are given in Section \ref{Section 2}.  $\sigma$-clusters are introduced, and
some related Lemmas are established, in Section \ref{Section 3}.  The Cluster
Convergence Theorem is stated, and proved, in Section \ref{Section 4}, and the
paper concludes with an \ref{Appendix}\hspace{-0.01cm}ppendix, on $\sigma$-clusters.

\vspace{0.05cm}
\enlargethispage{\baselineskip}

An application of the Cluster Convergence Theorem, to the
proof of a BHPZ convergence theorem, in Euclidean position space,
without exponentiating the propagators, is presently available, at
\cite{BPHZ in EPS}.  A power-counting convergence theorem was
proved, in momentum space, by Weinberg \cite{Weinberg}.  This was
used to prove a BPHZ convergence theorem, in momentum space, by
Zimmermann \cite{Zimmermann}.  BPHZ renormalization was introduced
by Bogoliubov and Parasiuk \cite{Bogoliubov Parasiuk}, and a BPHZ
convergence proof, in the parameter space, of the exponentiated
propagators, was given by Hepp \cite{Hepp}.  The idea, of
$\sigma$-clusters, arose from seeking an analogue, in position
space, of the momentum space clusters, used by 't Hooft
\cite{'t Hooft}.

\section{Introduction}

Many attempts to make QCD into a quantitative theory, for hadrons, involve the
use of Wilson loops.  When a Feynman diagram, contributing to a Wilson loop
expectation value, is calculated, some of the vertices, of the diagram, must be
integrated, along the loop.  For such a calculation, it is desirable to be able
to carry out the calculation, of the diagram, directly in position space, and, in
particular, it is desirable to understand, in as much detail, as possible, how
the cancellation of short-distance divergences, when the diagram is
renormalized, works in position space.  As a preliminary step, in this
direction, I shall present, in this paper, a power-counting theorem, the
``Cluster Convergence Theorem'', for use in Euclidean position space.

\section{Preparations.}
\label{Section 2}

Let $A$, and $B$, be any two sets.  I shall use the convention, that the notation,
$A \subseteq B$, means ``$A$ is a subset of $B$'', and includes the possibility,
that $A = B$.  I shall write $A \subset B$, and say ``$A$ is a strict subset
of $B$'', to indicate that $A$ is a subset of $B$, but not equal to $B$.  The
notation $A \vdash B$, (``$A$ outside $B$''), means the set of all the members,
of $A$, that are {\emph{not}} members, of $B$.

The word ``ifif'' is short for ``if and only if''.

For any finite set, $A$, the notation, $\# \left( A \right)$, indicates the
number of members, of $A$.  If $n$ is an integer, $\geq 0$, an
$n${\emph{-member set}} is a finite set, $A$, such that $\# \left( A \right) = n$.

The symbol, $\emptyset$, denotes the empty set.

A {\emph{map}} is a set, $M$, whose members are all ordered pairs, and which
satisfies the requirement, that if $\left( a, b \right) \in M$, and $\left( e, f
\right) \in M$, then $a = e$ implies $b = f$.

For any map, $M$, I define $\mathcal{D} \left( M \right)$, the {\emph{domain}}
of $M$, to be the set of all the first components, of members, of $M$, and
$\mathcal{R} \left( M \right)$, the {\emph{range}} of $M$, to be the set of
all the second components, of members, of $M$.  Note that $\mathcal{D} \left( M
\right)$ is finite, ifif $M$ is finite, and that if $M$ is finite, then $\#
\left( \mathcal{D} \left( M \right) \right) = \# \left( M \right)$, and $\#
\left( \mathcal{R} \left( M \right) \right) \leq \# \left( M \right)$.

If $M$ is a map, and $i$ is a member of $\mathcal{D} \left( M \right)$, then the
notation, $M_i$, denotes the second component of the unique member, of $M$, whose
first component, is $i$.

Note that if $M$ is a map, then every subset, of $M$, is also a map.

For any ordered pair, $\left( M, A \right)$, of a map, $M$, and a subset, $A$, of
$\mathcal{D} \left( M \right)$, I define $\mathcal{N} \left( M, A \right)$, the
{\emph{restriction, of}} $M$, {\emph{to the domain,}} $A$, to be the unique
subset, of $M$, whose domain, is $A$; that is, $\mathcal{N} \left( M, A \right)$
is the set of all the members, of $M$, whose first components, are members, of
$A$.

For any two sets, $A$, and $B$, I shall denote, by $B^A$, the set, whose members
are all the maps, whose domain is $A$, and whose range is a subset, of $B$.

A {\emph{bijection}} is a map, $M$, such that, if $\left( a, b \right) \in M$, and
$\left( e, f \right) \in M$, then $b = f$ implies $a = e$.

Let $A$, and $B$, be any two distinct sets.  I shall say that the two-member set,
$\left\{ A, B \right\}$, is {\emph{disjoint}}, ifif $A \cap B$ is empty, and
{\emph{compatible}}, ifif at least one, of $A \vdash B$, and $A \cap B$, and $B
\vdash A$, is empty.  I shall say that $A$ {\emph{overlaps}} $B$, ifif the
two-member set, $\left\{ A, B \right\}$, is {\emph{not}} compatible.

If $F$ is any set, all of whose members are, themselves, sets, I shall say that
$F$ is {\emph{disjoint}}, ifif every two-member subset, of $F$, is disjoint, and
{\emph{compatible}}, ifif every two-member subset, of $F$, is compatible.

For any set, $F$, whose members are all, themselves, sets, I define $\mathcal{U}
\left( F \right)$ to be the union, of all the members, of $F$.

For any set, $F$, whose members are all finite sets, I define $\mathbb{B} \left(
F \right)$ to be the set, whose members are all the members, $A$, of $F$, such
that $\# \left( A \right) \geq 2$.

A {\emph{partition}} is a disjoint set, none of whose members is empty.  If $A$
is any set, a {\emph{partition, of}} $A$, is a partition, $F$, such that
$\mathcal{U} \left( F \right) = A$.

Note that if $A$ is any set, and $F$ is a set, such that every member, of $F$, is
a set, no member, of $F$, is empty, $\mathcal{U} \left( F \right) = A$, and for
each member, $i$, of $A$, there is exactly one member, $B$, of $F$, such that $i \in
B$ holds, then $F$ is a partition, of $A$.

And note that, if $F$ is a partition, and $G$ is a subset, of $F$, such that
$\mathcal{U} \left( G \right) =\mathcal{U} \left( F \right)$, then $G = F$.

A {\emph{forest}} is a compatible set, $F$, such that $\mathcal{U} \left( F
\right)$ is finite, and no member, of $F$, is empty.  A {\emph{leaf}} is a
one-member set.

A {\emph{greenwood}} is a compatible set, $F$, such that $\mathcal{U} \left( F
\right)$ is finite, $\# \left( \mathcal{U} \left( F \right) \right) \geq 2$,
the empty set, $\emptyset$, is {\emph{not}} a member of $F$, and every
one-member subset, of $\mathcal{U} \left( F \right)$, {\emph{is}} a member, of
$F$.

Note that if $F$ is a greenwood, then $F$ is finite, and $\# \left( F \right) <
2^{\# \left( \mathcal{U} \left( F \right) \right)}$.

A greenwood, $F$, is {\emph{high}}, ifif $\mathcal{U} \left( F \right)$
{\emph{is}} a member, of $F$, and {\emph{low}}, ifif $\mathcal{U} \left( F
\right)$ is {\emph{not}} a member, of $F$.

If $A$ is any finite set, such that $\# \left( A \right) \geq 2$, a
{\emph{greenwood, of }}$A$, is a greenwood, $F$, such that $\mathcal{U} \left( F
\right) = A$.  I shall denote the set, whose members are all the high
greenwoods, of $A$, by $\mathbb{H} \left( A \right)$.  Note that $\mathbb{H}
\left( A \right)$ is finite, and that $\# \left( \mathbb{H} \left( A \right)
\right) < 2^{\left( 2^{\# \left( A \right)} \right)}$, and that $\# \left(
\mathbb{H} \left( A \right) \right)$ depends, only, on $\# \left( A \right)$.

For any ordered pair, $\left( F, A \right)$, of a greenwood, $F$, and a subset, $A$,
of $\mathcal{U} \left( F \right)$, such that $\# \left( A \right) \geq 2$, I
define $\mathcal{L} \left( F, A \right)$, the {\emph{crown, of}} $A$,
{\emph{in}} $F$, to be the set, whose members are all the members, $B$, of $F$,
such that $B \subset A$.  Note that $\mathcal{L} \left( F, A \right)$ is a low
greenwood, of $A$.

For any ordered pair, $\left( F, A \right)$, of a greenwood, $F$, and a subset, $A$,
of $\mathcal{U} \left( F \right)$, such that $\# \left( A \right) \geq 2$, I
define $\mathcal{S} \left( F, A \right)$, the {\emph{projection, of}} $F$,
{\emph{to}} $A$, to be the set, whose members are all the distinct, nonempty, $B
\cap A$, for $B \in F$.  Note that $\mathcal{S} \left( F, A \right)$ is a
greenwood, of $A$, and that if $F$ is high, then $\mathcal{S} \left( F, A
\right)$ is high, and that $\mathcal{L} \left( F, A \right)$ is a subset, of
$\mathcal{S} \left( F, A \right)$.

Note that if $F$ is a greenwood, $A$ is a subset, of $\mathcal{U} \left( F
\right)$, such that $\# \left( A \right) \geq 2$, and $B$ is any member, of
$\mathcal{S} \left( F, A \right)$, then there is a member, $W$, of $F$, such that
$W \cap A = B$.  In particular if $F$ is a greenwood, such that $\# \left(
\mathcal{U} \left( F \right) \right) \geq 3$, $i$ is any member, of
$\mathcal{U} \left( F \right)$, $J \equiv \left( \mathcal{U} \left( F \right)
\vdash \left\{ i \right\} \right)$, and $A$ is any member, of $\mathcal{S}
\left( F, J \right)$, then at least one, of $A$, and $\left( A \cup \left\{ i
\right\} \right)$, is a member, of $F$.

For any ordered pair, $\left( F, A \right)$, of a greenwood, $F$, and a nonempty
subset, $A$, of $\mathcal{U} \left( F \right)$, I define $\mathcal{E} \left( F, A
\right)$, the {\emph{stem, of }}$A$, {\emph{in}} $F$, to be the set, whose
members are all the members, $B$, of $F$, such that $A \subset B$, and I define
$\mathbb{L} \left( F, A \right)$, the {\emph{level, of}} $A$, {\emph{in}} $F$,
by $\mathbb{L} \left( F, A \right) \equiv \# \left( \mathcal{E} \left( F, A
\right) \right)$.  Note that $\mathbb{L} \left( F, A \right)$ is an integer,
$\geq 0$.

Note that if $F$ is a greenwood, and $A$, and $B$, are members, of $F$, such that
$A \subset B$, then $\mathbb{L} \left( F, B \right) <\mathbb{L} \left( F, A
\right)$.

Now if $F$ is a greenwood, $A$ is a nonempty subset, of $\mathcal{U} \left( F
\right)$, such that $\mathbb{L} \left( F, A \right) \geq 1$, and $B$, and $C$,
are members, of $\mathcal{E} \left( F, A \right)$, then $B \cap C$ is nonempty,
hence exactly one, of $B \subset C$, $B = C$, and $C \subset B$, holds, hence, if
$B \neq C$, then $\mathbb{L} \left( F, B \right) \neq \mathbb{L} \left( F, C
\right)$.  Hence the levels, in $F$, of the $\mathbb{L} \left( F, A \right)$
members, of $\mathcal{E} \left( F, A \right)$, are $\mathbb{L} \left( F, A
\right)$ distinct integers, $\geq 0$, and $\leq \left( \mathbb{L} \left( F, A
\right) - 1 \right)$, hence, for each integer, $n$, such that $0 \leq n \leq
\left( \mathbb{L} \left( F, A \right) - 1 \right)$, there is exactly one
member, $B$, of $\mathcal{E} \left( F, A \right)$, such that $\mathbb{L} \left(
F, B \right) = n$.

For any ordered pair, $\left( F, A \right)$, of a greenwood, $F$, and a subset, $A$,
of $\,\mathcal{U} \left( F \right)$, such that $\mathbb{L} \left( F, A \right)
\geq 1$, I define $\mathcal{C} \left( F, A \right)$, the {\emph{container, of }}$A$,
{\emph{in}} $F$, to be the unique member, $B$, of $\mathcal{E} \left( F, A
\right)$, such that $\mathbb{L} \left( F, B \right) = \left( \mathbb{L}
\left( F, A \right) - 1 \right)$.

Note that this definition implies, that, if $F$ is any greenwood, and $A$ is any
nonempty subset, of $\mathcal{U} \left( F \right)$, such that $\mathbb{L}
\left( F, A \right) \geq 1$, then $A \subset \mathcal{C} \left( F, A \right)$
holds, and that if $K$ is any member, of $F$, such that $A \subset K$ holds,
then $\mathcal{C} \left( F, A \right) \subseteq K$ holds, hence there is no
member, $K$, of $F$, such that $A \subset K$, and $K \subset \mathcal{C} \left( F,
A \right)$, both hold.  Furthermore, if $F$ is any greenwood, $A$ is any
nonempty subset, of $\mathcal{U} \left( F \right)$, such that $\mathbb{L}
\left( F, A \right) \geq 1$ holds, and $B$ is a member, of $F$, such that $A
\subset B$ holds, and there is no member, $K$, of $F$, such that $A \subset K$, and
$K \subset B$, both hold, then $B =\mathcal{C} \left( F, A \right)$.

For any ordered pair, $\left( F, A \right)$, of a greenwood, $F$, and a subset, $A$,
of $\mathcal{U} \left( F \right)$, such that $\# \left( A \right) \geq 2$, I
define $\mathcal{P} \left( F, A \right)$ to be the set, whose members are all
the members, $B$, of $\mathcal{L} \left( F, A \right)$, such that $\mathbb{L}
\left( \mathcal{L} \left( F, A \right), B \right) = 0$.

Now $\mathcal{L} \left( F, A \right)$ is a low greenwood, of $A$.  Let $i$ be
any member, of $A$.  Then $\left\{ i \right\}$ is a member, of $\mathcal{L}
\left( F, A \right)$, and there is exactly one member, $B$, of $\mathcal{E}
\left( \mathcal{L} \left( F, A \right), \left\{ i \right\} \right) \cup
\left\{ \left\{ i \right\} \right\}$, such that $\mathbb{L} \left( \mathcal{L}
\left( F, A \right), B \right) = 0$, hence there is exactly one member, $B$, of
$\mathcal{P} \left( F, A \right)$, such that $i \in B$.  Furthermore, no member,
of $\mathcal{P} \left( F, A \right)$, is empty, and $\mathcal{U} \left(
\mathcal{P} \left( F, A \right) \right) = A$, hence $\mathcal{P} \left( F, A
\right)$ is a partition, of $A$.  I will call $\mathcal{P} \left( F, A \right)$
the {\emph{partition, of}} $A$, {\emph{in}} $F$.

Now if $B$ is any member, of $F$, such that $B \subset A$, and there is no member,
$K$, of $F$, such that $B \subset K$, and $K \subset A$, both hold, then $B$ is a
member, of $\mathcal{L} \left( F, A \right)$, and $\mathbb{L} \left(
\mathcal{L} \left( F, A \right), B \right) = 0$, hence $B$ is a member, of
$\mathcal{P} \left( F, A \right)$.  And if $B$ is any member, of $\mathcal{P}
\left( F, A \right)$, then $B \subset A$ holds, and there is no member, $K$, of
$F$, such that $B \subset K$, and $K \subset A$, both hold.  Hence $\mathcal{P}
\left( F, A \right)$ is the set, whose members are all the members, $B$, of $F$,
such that $B \subset A$ holds, and there is no member, $K$, of $F$, such that $B
\subset K$, and $K \subset A$, both hold.

Note that if $F$ is any greenwood, and $A$ is any member of $\mathbb{B} \left(
F \right)$, then $\mathcal{P} \left( F, A \right)$ is the set, whose members
are all the members, $B$, of $F$, such that $\mathbb{L} \left( F, B \right) \geq
1$, and $\mathcal{C} \left( F, B \right) = A$, both hold, and that if $F$ is any
greenwood, and $A$ is any member, of $F$, such that $\mathbb{L} \left( F, A
\right) \geq 1$ holds, then $A$ is a member, of $\mathcal{P} \left(
F,\mathcal{C} \left( F, A \right) \right)$.

For any ordered triple, $\left( F, A, b \right)$, of a greenwood, $F$, a subset,
$A$, of $\mathcal{U} \left( F \right)$, such that $\# \left( A \right) \geq 2$,
and a member, $b$, of $A$, I define $\mathcal{K} \left( F, A, b \right)$ to be
the unique member, $B$, of $\mathcal{P} \left( F, A \right)$, such that $b \in
B$.

Note that if $F$ is any {\emph{high}} greenwood, and $A$ is any nonempty,
{\emph{strict}} subset, of $\mathcal{U} \left( F \right)$, then $\mathbb{L}
\left( F, A \right) \geq 1$, so $\mathcal{C} \left( F, A \right)$ is defined.

For any ordered pair, $\left( F, i \right)$, of a {\emph{high}} greenwood, $F$,
and a member, $i$, of $\mathcal{U} \left( F \right)$, I define $\mathcal{H}
\left( F, i \right) \equiv \left( \mathcal{C} \left( F, \left\{ i \right\}
\right) \vdash \left\{ i \right\} \right)$.  Note that $\mathcal{H} \left( F,
i \right)$ has at least one member, that $i$ is {\emph{not}} a member of
$\mathcal{H} \left( F, i \right)$, and that $\left( \mathcal{H} \left( F, i
\right) \cup \left\{ i \right\} \right) =\mathcal{C} \left( F, \left\{ i
\right\} \right)$.  Furthermore, if $\# \left( \mathcal{U} \left( F \right)
\right) \geq 3$, then $\mathcal{H} \left( F, i \right)$ is a member of
$\mathcal{S} \left( F, \left( \mathcal{U} \left( F \right) \vdash \left\{ i
\right\} \right) \right)$, since $\mathcal{H} \left( F, i \right) =\mathcal{C}
\left( F, \left\{ i \right\} \right) \cap \left( \mathcal{U} \left( F \right)
\vdash \left\{ i \right\} \right)$, and $\mathcal{H} \left( F, i \right)$ is
nonempty.

For any finite set, $A$, such that $\# \left( A \right) \geq 2$, I define
$\mathcal{Q} \left( A \right)$ to be the set, whose members are all the
two-member subsets, of $A$.  Note that $\# \left( \mathcal{Q} \left( A \right)
\right) = \frac{1}{2} \# \left( A \right) \left( \# \left( A \right) - 1
\right)$.

For each ordered pair, $\left( F, A \right)$, of a greenwood, $F$, and a subset,
$A$, of $\mathcal{U} \left( F \right)$, such that $\# \left( A \right) \geq 2$,
I define $\mathcal{W} \left( F, A \right)$ to be the set, whose members are all
the two-member subsets, of $A$, that are {\emph{not}} subsets, of any member, of
$\mathcal{P} \left( F, A \right)$.  In other words, $\mathcal{W} \left( F, A
\right)$ is the set, whose members are all the two-member subsets, of $A$, that
have nonempty intersection, with two distinct members, of $\mathcal{P} \left( F,
A \right)$.  Note that $\mathcal{W} \left( F, A \right)$ is always nonempty,
since $\mathcal{P} \left( F, A \right)$ always has at least two members.

The symbol, $\mathbb{R}$, denotes the set of all the finite, real numbers.

$\mathbb{E}_d$ denotes $d$-dimensional Euclidean space.  Note that for any
three points $x_1$, $x_2$, and $x_3$ of $\mathbb{E}_d$, the triangle
inequality, $\left| x_1 - x_2 \right| \leq \left| x_1 - x_3 \right| + \left|
x_3 - x_2 \right|$, holds.

I shall assume $d \geq 1$, throughout the paper.

For all $t \in \mathbb{R}$ I define
\[ \mathbb{S} \left( t \right) = \left\{ \begin{array}{ccc}
     1 & \mathrm{ifif} & t \geq 0\\
     0 & \mathrm{ifif} & t < 0
   \end{array} \right\}, \]
and
\[ \mathbb{U} \left( t \right) = \left\{ \begin{array}{ccc}
     t & \mathrm{if} & t \geq 0\\
     0 & \mathrm{if} & t \leq 0
   \end{array} \right\}, \]
and
\[ \mathbb{M} \left( t \right) = \left\{ \begin{array}{ccc}
     t & \mathrm{if} & t \geq 0\\
     - t & \mathrm{if} & t \leq 0
   \end{array} \right\}. \]

\vspace{0.6cm}

\noindent {\bf{Lemma 1.}}  Let $F$ be any greenwood, $B$ be any member, of $F$, such
that $B \neq \mathcal{U} \left( F \right)$, $i$, and $j$, be any members, of $B$,
and $k$ be any member, of $\left( \mathcal{U} \left( F \right) \vdash B
\right)$.  Then for every member, $A$, of $F$, $\left\{ i, k \right\} \subseteq
A$ ifif $\left\{ j, k \right\} \subseteq A$.

For exactly one of the three possibilities, $A \subseteq B$, $A \cap B =
\emptyset$, and $B \subset A$ holds.

First suppose $A \subseteq B$.  Then $k \notin A$, hence neither $\left\{ i, k
\right\}$, nor $\left\{ j, k \right\}$, is a subset, of $A$.

Now suppose $A \cap B = \emptyset$.  Then $i \notin A$, and $j \notin A$, hence
neither $\left\{ i, k \right\}$, nor $\left\{ j, k \right\}$, is a subset, of
$A$.

Finally, suppose $B \subset A$.  Then $i \in A$, and $j \in A$, hence $\left\{
i, k \right\} \subseteq A$, ifif $\left\{ j, k \right\} \subseteq A$.

\vspace{0.6cm}

\noindent {\bf{Lemma 2.}}  Let $F$ be any high greenwood, $J$ be any subset, of
$\mathcal{U} \left( F \right)$, such that $\# \left( J \right) \geq 2$, $E$ be
any greenwood, of $J$, such that $\mathcal{S} \left( F, J \right) \subset E$,
$A$ be any member, of $\left( E \vdash \mathcal{S} \left( F, J \right)
\right)$, $a$, and $b$, be any two members, of $A$, such that $\left\{ a, b
\right\} \in \mathcal{W} \left( E, A \right)$, $e$ be any member, of $A$, and
$f$ be any member, of $\left( \mathcal{C} \left( E, A \right) \vdash A
\right)$.  Then $\left\{ a, b \right\} \in \mathcal{W} \left( F,\mathcal{C}
\left( F, A \right) \right)$, and $\left\{ e, f \right\} \in \mathcal{W} \left(
F,\mathcal{C} \left( F, A \right) \right)$.

\vspace{0.6cm}

\noindent {\bf{Proof.}}  Note that $\mathcal{S} \left( F, J \right) \subset E$
implies that $J \in E$, hence $E$ is high, and note that $A \in \left( E \vdash
\mathcal{S} \left( F, J \right) \right)$ implies that $\# \left( A \right)
\geq 2$, and that $A \subset J$, hence $\mathcal{W} \left( E, A \right)$ is
defined, and nonempty, and $\mathcal{C} \left( E, A \right)$, and $\mathcal{C}
\left( F, A \right)$, are defined.

Note, also, that $A \in \left( E \vdash \mathcal{S} \left( F, J \right) \right)$
implies that A is {\emph{not}} a member, of $\mathcal{S} \left( F, J \right)$,
hence that $A$ is not a member, of $F$.

Now $\left\{ a, b \right\}$ is a member, of $\mathcal{Q} \left( \mathcal{C}
\left( F, A \right) \right)$.  Suppose $K$ is a member, of $\mathcal{P} \left(
F,\mathcal{C} \left( F, A \right) \right)$, such that $\left\{ a, b \right\}
\subseteq K$.  Now $\left\{ a, b \right\} \subseteq J$ holds, hence $\left\{
a, b \right\} \subseteq \left( K \cap J \right)$ holds, hence $\left( K \cap J
\right)$ is nonempty, hence is a member, of $\mathcal{S} \left( F, J \right)$,
hence is a member, of $E$.  Now the definition, of $\mathcal{C} \left( F, A
\right)$, implies that $A$ is not a strict subset, of $K$, and $A \notin F$
implies that $A \neq K$, hence $A$ is not a subset of $K$, hence $A \vdash K$
is nonempty, hence $A \vdash \left( K \cap J \right)$ is nonempty.
Furthermore, $\left\{ a, b \right\} \subseteq \left( A \cap \left( K \cap J
\right) \right)$, hence $A \cap \left( K \cap J \right)$ is nonempty, hence
$\left( K \cap J \right) \subset A$ must hold, since $E$ is a greenwood.  This
contradicts $\left\{ a, b \right\} \in \mathcal{W} \left( E, A \right)$, since
$\left\{ a, b \right\} \subseteq \left( K \cap J \right)$.  Hence there cannot
be any member, $K$, of $\mathcal{P} \left( F, \mathcal{C} \left( F, A \right)
\right)$, such that $\left\{ a, b
\right\} \subseteq K$ holds, hence $\left\{ a, b \right\}$ is a member, of
$\mathcal{W} \left( F,\mathcal{C} \left( F, A \right) \right)$.

Now $A \subseteq \mathcal{C} \left( F, A \right)$, and $A \subseteq J$, both
hold, hence $A \subseteq \left( \mathcal{C} \left( F, A \right) \cap J
\right)$ holds.  Hence $\left( \mathcal{C} \left( F, A \right) \cap J \right)$
is nonempty, hence is a member, of $\mathcal{S} \left( F, J \right)$.  Hence the
assumption, that $A$ is a member, of $\left( E \vdash \mathcal{S} \left( F, J
\right) \right)$, hence that $A$ is {\emph{not}} a member, of $\mathcal{S}
\left( F, J \right)$, implies that $A$ is {\emph{not}} equal, to $\left(
\mathcal{C} \left( F, A \right) \cap J \right)$, hence that $A \subset \left(
\mathcal{C} \left( F, A \right) \cap J \right)$ holds.  Furthermore, the facts
that $\left( \mathcal{C} \left( F, A \right) \cap J \right) \in \mathcal{S}
\left( F, J \right)$, and that $\mathcal{S} \left( F, J \right) \subset E$,
together imply, that $\left( \mathcal{C} \left( F, A \right) \cap J \right) \in
E$.  Hence the definition, of $\mathcal{C} \left( E, A \right)$, implies
$\mathcal{C} \left( E, A \right) \subseteq \left( \mathcal{C} \left( F, A
\right) \cap J \right)$, hence $\mathcal{C} \left( E, A \right)$ is a subset,
of $\mathcal{C} \left( F, A \right)$.

Hence $\left\{ e, f \right\}$ is a member, of $\mathcal{Q} \left( \mathcal{C}\!
\left( F, A \right) \right)$.  Suppose $M$ is a member, of $\mathcal{P}\! \left(
F,\mathcal{C}\! \left( F, A \right) \right)$, such that $\left\{ e, f \right\}
\subseteq M$.  Now $\left. \{ e, f \right\} \subseteq J$ holds, hence $\left\{
e, f \right\} \subseteq \left( M \cap J \right)$ holds, hence $\left( M \cap J
\right)$ is nonempty, hence is a member of $\mathcal{S} \left( F, J \right)$,
hence is a member, of $E$.  Now $e$ is a member, of $A \cap \left( M \cap J
\right)$, and $f$ is a member, of $\left( \left( M \cap J \right) \vdash A
\right)$, hence $A \subset \left( M \cap J \right)$ must hold, since $E$ is a
greenwood.  But this implies $A \subset M$, which is impossible, by the
definition of $\mathcal{C} \left( F, A \right)$.  Hence there cannot be any
member, $M$, of $\mathcal{P} \left( F,\mathcal{C} \left( F, A \right) \right)$,
such that $\left\{ e, f \right\} \subseteq M$ holds, hence $\left\{ e, f
\right\}$ is a member, of $\mathcal{W} \left( F,\mathcal{C} \left( F, A \right)
\right)$.

\vspace{0.6cm}

\noindent {\bf{Lemma 3.}}  Let $F$ be any high greenwood, let $A$ be any member, of
$\mathbb{B} \left( F \vdash \left\{ \mathcal{U} \left( F \right) \right\}
\right)$, let $a$, and $b$, be any two members, of $A$, such that $\mathcal{K}
\left( F, A, a \right) \neq \mathcal{K} \left( F, A, b \right)$, let $e$ be
any member, of $A$, and let $f$ be any member, of $\left( \mathcal{C} \left( F,
A \right) \vdash A \right)$.  Then for every member, $B$, of $F$, such that $B
\neq A$, $\left\{ e, f \right\} \subseteq B$ ifif $\left\{ a, b \right\}
\subseteq B$.

For the facts that $A \in F$, $B \in F$, and $B \neq A$, together imply, that
exactly one of the three possibilities, $B \subset A$, $A \cap B =
\emptyset$, and $A \subset B$, holds.

If $B \subset A$, then $f \notin B$, hence $\left\{ e, f \right\}$ is
{\emph{not}} a subset, of $B$.  And the fact, that $\mathcal{K} \left( F, A, a
\right) \neq \mathcal{K} \left( F, A, b \right)$, implies that $\left\{ a, b
\right\}$ is not a subset, of any member, of $\mathcal{L} \left( F, A \right)$,
and, in particular, $\left\{ a, b \right\}$ is not a subset, of $B$.

If $A \cap B = \emptyset$, then $e \notin B$, hence $\left\{ e, f \right\}$ is
not a subset, of $B$, and $a \notin B$, hence $\left\{ a, b \right\}$ is not a
subset, of $B$.

If $A \subset B$, then $\left\{ a, b \right\} \subseteq B$.  And $A \subset
B$, and the definition, of $\mathcal{C} \left( F, A \right)$, together imply,
that $\mathcal{C} \left( F, A \right) \subseteq B$.  But $\left\{ e, f
\right\} \subseteq \mathcal{C} \left( F, A \right)$, hence $\left\{ e, f
\right\} \subseteq B$.

\vspace{0.6cm}

\noindent {\bf{Lemma 4.}}  Let $F$ be a greenwood, let $A$ be a member, of $F$, such
that $\mathbb{L} \left( F, A \right) \geq 1$, and such that $\left(
\mathcal{C} \left( F, A \right) \vdash A \right)$ has at least two members,
let $J \equiv \left( \mathcal{U} \left( F \right) \vdash A \right)$, and let
$K \equiv \left( \mathcal{C} \left( F, A \right) \vdash A \right)$.  Then
$\mathcal{L} \left( \mathcal{S} \left( F, J \right), K \right) =\mathcal{L}
\left( F, K \right)$.

(Note that $K \subseteq J$, hence $J$ has at least two members, hence
$\mathcal{S} \left( F, J \right)$ is defined.)

\vspace{0.6cm}

\noindent {\bf{Proof.}}  First, let $B$ be any member, of $\mathcal{L} \left( F, K
\right)$.  Then $B$ is a member, of $F$, such that $B \subset K$.  Hence $B$ is
a subset, of $J$, hence $B$ is a member, of $\mathcal{S} \left( F, J \right)$,
hence $B$ is a member, of $\mathcal{L} \left( \mathcal{S} \left( F, J \right),
K \right)$, since $B \subset K$ holds.

Now let $B$ be any member, of $\mathcal{L} \left( \mathcal{S} \left( F, J
\right), K \right)$.  Then $B$ is a member, of $\mathcal{S} \left( F, J
\right)$, such that $B \subset K$.  Now $B \in \mathcal{S} \left( F, J
\right)$ implies that there is a member, $W$, of $F$, such that $B = W \cap J =
W \cap \left( \mathcal{U} \left( F \right) \vdash A \right) = W \vdash A$,
since $W$ is a subset, of $\mathcal{U} \left( F \right)$.

Now $W \in F$, and $A \in F$, together imply, that exactly one of the three
possibilities, $W \subseteq A$, $A \subset W$, and $A \cap W = \emptyset$,
holds.

And $B$ is nonempty, hence $B = W \vdash A$ implies $W \subseteq A$ cannot
hold.

Now the fact, that $B \subset K$, implies that there is a member, say $j$, of
$K$, such that $j \notin B$.  Then $j$ is a member, of $J$, hence $j \notin \left(
W \cap J \right)$ implies $j \notin W$, hence $K \vdash W$ is nonempty.  But $K$
is a subset, of $\mathcal{C} \left( F, A \right)$, hence $\left( \mathcal{C}
\left( F, A \right) \vdash W \right)$ is nonempty, hence $\mathcal{C} \left(
F, A \right)$ is not a subset, of $W$.  Hence $A \subset W$ cannot hold, since
$A \subset W$ implies $\mathcal{C} \left( F, A \right) \subseteq W$.

The only remaining possibility is $A \cap W = \emptyset$, hence $A \cap W =
\emptyset$ holds, hence $B = W \vdash A = W$, hence $B$ is a member, of $F$,
hence $B$ is a member, of $\mathcal{L} \left( F, K \right)$, since $B \subset
K$ holds.

\vspace{0.6cm}

For any ordered pair, $\left( F, i \right)$, of a {\emph{high}} greenwood, $F$,
and a member, $i$, of $\mathcal{U} \left( F \right)$, I defined $\mathcal{H}
\left( F, i \right)$, on page 5, by $\mathcal{H} \left( F, i \right)
\equiv \left( \mathcal{C} \left( F, \left\{ i \right\} \right) \vdash \left\{
i \right\} \right)$, and noted that $\mathcal{H} \left( F, i \right)$ always
has at least one member, that $i$ is {\emph{not}} a member, of $\mathcal{H}
\left( F, i \right)$, that $\left( \mathcal{H} \left( F, i \right) \cup
\left\{ i \right\} \right) =\mathcal{C} \left( F, \left\{ i \right\} \right)$,
and that, if $\# \left( \mathcal{U} \left( F \right) \right) \geq 3$, then
$\mathcal{H} \left( F, i \right)$ is a member, of $\mathcal{S} \left( F, \left(
\mathcal{U} \left( F \right) \vdash \left\{ i \right\} \right) \right)$.  Now
for any ordered pair, $\left( F, A \right)$, of a high greenwood, $F$, and a
member, $A$, of $F$, such that $A \neq \mathcal{U} \left( F \right)$, the
definition, of $\mathcal{C} \left( F, A \right)$, implies that $A$ is a member,
of $\mathcal{P} \left( F,\mathcal{C} \left( F, A \right) \right)$.  Hence
$\left\{ i \right\}$ is always a member, of $\mathcal{P} \left( F,\mathcal{C}
\left( F, \left\{ i \right\} \right) \right)$.

\vspace{0.6cm}

\noindent {\bf{Lemma 5.}}  Let $F$ be any high greenwood, $i$ be any member, of
$\mathcal{U} \left( F \right)$, and $A$ be any member, of $\mathcal{P} \left(
F,\mathcal{C} \left( F, \left\{ i \right\} \right) \right)$, such that $A \neq
\left\{ i \right\}$.  Then $A \subseteq \mathcal{H} \left( F, i \right)$.

For the unique member, $B$, of $\mathcal{P} \left( F,\mathcal{C} \left( F,
\left\{ i \right\} \right) \right)$, such that $i \in B$, is $B = \left\{ i
\right\}$.  Hence if $A$ is any member, of $\mathcal{P} \left( F,\mathcal{C}
\left( F, \left\{ i \right\} \right) \right)$, such that $A \neq \left\{ i
\right\}$, then $i$ is not a member, of $A$, hence $A$ is a subset, of $\left(
\mathcal{C} \left( F, \left\{ i \right\} \right) \vdash \left\{ i \right\}
\right) =\mathcal{H} \left( F, i \right)$.

\vspace{0.6cm}

Note that Lemma 5 implies, that if $F$ is any high greenwood, and $i$ is any
member, of $\mathcal{U} \left( F \right)$, such that $\mathcal{H} \left( F, i
\right)$ {\emph{is}} a member, of $F$, then $\mathcal{H} \left( F, i \right)$
is a member, of $\mathcal{P} \left( F,\mathcal{C} \left( F, \left\{ i \right\}
\right) \right)$, and $\mathcal{P} \left( F,\mathcal{C} \left( F, \left\{ i
\right\} \right) \right) = \left\{ \mathcal{H} \left( F, i \right), \left\{ i
\right\} \right\}$.

\vspace{0.6cm}

\noindent {\bf{Lemma 6.}}  If $F$ is any high greenwood, and $i$ is any member, of
$\mathcal{U} \left( F \right)$, such that $\mathcal{H} \left( F, i \right)$ is
{\emph{not}} a member, of $F$, then there exist two members, $a$, and $b$, of
$\mathcal{H} \left( F, i \right)$, such that $\mathcal{K} \left( F,\mathcal{C}
\left( F, \left\{ i \right\} \right), a \right) \neq \mathcal{K} \left(
F,\mathcal{C} \left( F, \left\{ i \right\} \right), b \right)$.

For let $A$ be any member, of $\mathcal{P} \left( F,\mathcal{C} \left( F,
\left\{ i \right\} \right) \right)$, such that $A \neq \left\{ i \right\}$.
Then $A$ is a subset, of $\mathcal{H} \left( F, i \right)$, by Lemma 5, and $A$
is a member, of $F$, hence $A \neq \mathcal{H} \left( F, i \right)$, hence $A$
is a strict subset, of $\mathcal{H} \left( F, i \right)$, hence $\left(
\mathcal{H} \left( F, i \right) \vdash A \right)$ is nonempty.  Let $a$ be any
member, of $A$, and let $b$ be any member, of $\left( \mathcal{H} \left( F, i
\right) \vdash A \right)$.  Then $\mathcal{K} \left( F,\mathcal{C} \left( F,
\left\{ i \right\} \right), a \right) = A$, and $\mathcal{K} \left(
F,\mathcal{C} \left( F, \left\{ i \right\} \right), b \right) \neq A$.

\vspace{0.6cm}

\noindent {\bf{Lemma 7.}}  Let $F$ be a high greenwood, let $i$ be a member, of
$\mathcal{U} \left( F \right)$, such that $\mathcal{H} \left( F, i \right)$ is
{\emph{not}} a member, of $F$, and let $J \equiv \left( \mathcal{U} \left( F
\right) \vdash \left\{ i \right\} \right)$.  Then $\mathcal{P} \left(
\mathcal{S} \left( F, J \right),\mathcal{H} \left( F, i \right) \right) =
\left( \mathcal{P} \left( F,\mathcal{C} \left( F, \left\{ i \right\} \right)
\right) \vdash \left\{ \left\{ i \right\} \right\} \right)$.

\vspace{0.6cm}

\noindent {\bf{Proof.}}  Note that the assumption, that $\mathcal{H} \left( F, i
\right)$ is {\emph{not}} a member, of $F$, implies that $\# \left( \mathcal{H}
\left( F, i \right) \right) \geq 2$.  Hence Lemma 4, for the case, that $A =
\left\{ i \right\}$, and $K =\mathcal{H} \left( F, i \right)$, implies that
$\left. \mathcal{L} \left( \mathcal{S}( F, J \right),\mathcal{H} \left( F, i
\right) \right) =\mathcal{L} \left( F,\mathcal{H} \left( F, i \right)
\right)$, hence that $ \mathcal{P} \left( \mathcal{S}\left( F, J \right),\mathcal{H} \left( F, i
\right) \right) =\mathcal{P} \left( F,\mathcal{H} \left( F, i \right)
\right)$.  And the assumption, that $\mathcal{H} \left( F, i \right)$ is
{\emph{not}} a member, of $F$, implies that $\left( \mathcal{P} \left(
F,\mathcal{C} \left( F, \left\{ i \right\} \right) \right) \vdash \left\{
\left\{ i \right\} \right\} \right) =\mathcal{P} \left( F,\mathcal{H} \left(
F, i \right) \right)$, hence \\
$\mathcal{P} \left( \mathcal{S} \left( F, J
\right),\mathcal{H} \left( F, i \right) \right) = \left( \mathcal{P} \left(
F,\mathcal{C} \left( F, \left\{ i \right\} \right) \right) \vdash \left\{
\left\{ i \right\} \right\} \right)$ holds.

\section{$\sigma$-Clusters.}
\label{Section 3}

Let $x$ be a map, such that $\mathcal{D} \left( x \right)$ is finite, $\# \left(
\mathcal{D} \left( x \right) \right) \geq 2$, and $\mathcal{R} \left( x
\right)$ is a subset of $\mathbb{E}_d$, and let $\sigma$ be a real number,
such that $0 < \sigma < 1$.  A $\sigma${\emph{-cluster, of}} $x$, is a nonempty
subset, $J$, of $\mathcal{D} \left( x \right)$, such that either $\# \left( J
\right) = 1$, or else $\left| x_i - x_j | < \sigma \left| x_k - x_m \right|
\right.$ holds, for all $i \in J$, $j \in J$, $k \in J$, and $m \in \left(
\mathcal{D} \left( x \right) \vdash J \right)$.

Note that every one-member subset, of $\mathcal{D} \left( x \right)$, is always
a $\sigma$-cluster, of $x$, and that $\mathcal{D} \left( x \right)$ is always a
$\sigma$-cluster, of $x$.

I shall denote the set, whose members are all the $\sigma$-clusters, of $x$, by
$\mathcal{F} \left( x, \sigma \right)$.

\vspace{0.6cm}

\noindent {\bf{Lemma 8.}}  Let $x$ be a map, such that $\mathcal{D} \left( x
\right)$ is finite, $\# \left( \mathcal{D} \left( x \right) \right) \geq 2$,
and $\mathcal{R} \left( x \right)$ is a subset of $\mathbb{E}_d$, and let
$\sigma$ be a real number, such that $0 < \sigma < 1$.  Then no two
$\sigma$-clusters, of $x$, overlap.

For if $A$, and $B$, are two overlapping subsets, of $\mathcal{D} \left( x
\right)$, then $A$ has a member, say $i$, that is not a member, of $B$, $\left(
A \cap B \right)$ has at least one member, say $j$, and $B$ has at least one
member, say $k$, that is not a member, of $A$.  Hence $\# \left( A \right) \geq
2$, and $\# \left( B \right) \geq 2$.  Hence if $A$ is a $\sigma$-cluster, of
$x$, then $\left| x_i - x_j \right| < \sigma \left| x_j - x_k \right|$, and if
$B$ is a $\sigma$-cluster, of $x$, then $\left| x_j - x_k \right| < \sigma
\left| x_i - x_j \right|$, hence $A$, and $B$, cannot both be $\sigma$-clusters,
of $x$.

\vspace{0.6cm}

\noindent {\bf{Corollary.}}  $\mathcal{F} \left( x, \sigma \right)$ is a high
greenwood, of $\mathcal{D} \left( x \right)$.

\vspace{0.6cm}

In the \ref{Appendix}\hspace{-0.01cm}ppendix, I display, for any ordered quintuple, $\left( F, \sigma, d, z,
L \right)$, of a high greenwood, $F,$ a real number, $\sigma$, such that $0 <
\sigma < 1$, an integer, $d \geq 1$, a point, $z$, of $\mathbb{E}_d$, and a real
number, $L > 0$, a subset, $\mathbb{Z} \left( F, \sigma, d, z, L \right)$, of
$\left( \mathbb{E}_d \right)^{\mathcal{U} \left( F \right)}$, such that
$\mathbb{Z} \left( F, \sigma, d, z, L \right)$ has nonzero $\left( \left( \#
\left( \mathcal{U} \left( F \right) \right) \right) d \right)$-volume,
$\mathcal{F} \left( x, \sigma \right) = F$ holds, for all $x \in \mathbb{Z}
\left( F, \sigma, d, z, L \right)$, and $\left| x_i - z \right| \leq L$ holds,
for every member, $i$, of $\mathcal{U} \left( F \right)$, for all $x \in
\mathbb{Z} \left( F, \sigma, d, z, L \right)$.

\vspace{0.6cm}

\noindent {\bf{Lemma 9.}}  Let $x$ be a map, such that $\mathcal{D} \left( x
\right)$ is finite, $\# \left( \mathcal{D} \left( x \right) \right) \geq 2$,
and $\mathcal{R} \left( x \right) \subseteq \mathbb{E}_d$, let $J$ be a
subset, of $\mathcal{D} \left( x \right)$, such that $\# \left( J \right) \geq
2$, let $y \equiv\mathcal{N} \left( x, J \right)$ be the restriction, of $x$, to the
domain, $J$, and let $A$ be a $\sigma$-cluster, of $x$.  Then $A \cap J$ is a
$\sigma$-cluster, of $y$, unless $A \cap J$ is empty.

For if $\# \left( A \cap J \right) = 1$, then $A \cap J$ is automatically a
$\sigma$-cluster, of $y =\mathcal{N} \left( x, J \right)$.  If $\# \left( A
\cap J \right) \geq 2$, then $\# \left( A \right) \geq 2$, hence $\left| x_i -
x_j \right| < \sigma \left| x_k - x_m \right|$ holds, for all $i \in A$, $j
\in A$, $k \in A$, and $m \in \left( \mathcal{D} \left( x \right) \vdash A
\right)$, hence $\left| y_i - y_j \right| < \sigma \left| y_k - y_m \right|$
holds, for all $i \in \left( A \cap J \right)$, $j \in \left( A \cap J
\right)$, $\left. k \in ( A \cap J \right)$, and $m \in \left( J \vdash \left(
A \cap J \right) \right)$, hence $A \cap J$ is a $\sigma$-cluster, of $y$.

(Note that, for each member, $k$, of $J$, $x_k$, and $y_k$, are equally valid,
alternative notations, for the second component, of the unique member, of $y$,
whose first component, is $k$.)

\vspace{0.6cm}

\noindent {\bf{Corollary.}}  Let $x$ be a map, such that $\mathcal{D} \left( x
\right) \!$ is finite, $\# \left( \mathcal{D} \left( x \right) \right) \geq
2$, and $\mathcal{R} \left( x \right) \subseteq \mathbb{E}_d$, let $J$ be a
subset, of $\mathcal{D} \left( x \right)$, such that $\# \left( J \right) \geq
2$, and let $y \equiv \mathcal{N} \left( x, J \right)$ be the restriction, of
$x$, to the domain, $J$.  Then $\mathcal{S} \left( \mathcal{F} \left( x, \sigma
\right), J \right)$ is a subset, of $\mathcal{F} \left( y, \sigma \right)$.

\vspace{0.6cm}

\noindent {\bf{Lemma 10.}}  Let $x$ be a map, such that $\mathcal{D} \left( x
\right)$ is finite, $\# \left( \mathcal{D} \left( x \right) \right) \geq 2$,
and $\mathcal{R} \left( x \right) \subseteq \mathbb{E}_d$, let $J$ be a
subset, of $\mathcal{D} \left( x \right)$, such that $\# \left( J \right) \geq
2$, let $y \equiv \mathcal{N} \left( x, J \right)$ be the restriction, of $x$,
to the domain, $J$, let $A$ be a subset, of $J$, such that $A$ is a
$\sigma$-cluster, of $x$, and let $B$ be any subset, of $A$.  Then $B$ is a
$\sigma$-cluster, of $x$, ifif $B$ is a $\sigma$-cluster, of $y$.

\vspace{0.6cm}

\noindent {\bf{Proof.}}  First assume $B$ is a $\sigma$-cluster, of $x$.  Then, by
Lemma 9, $B \cap J$ is a $\sigma$-cluster, of $y$.  But $B \cap J = B$, hence
$B$ is a $\sigma$-cluster, of $y$.

Now assume $B$ is a $\sigma$-cluster, of $y$.  Then if $\# \left( B \right) =
1$, $B$ is automatically a $\sigma$-cluster, of $x$.  Assume, now, that $B$ is
a $\sigma$-cluster, of $y$, and that $\# \left( B \right) \geq 2$.  Then
$\left| x_i - x_j \right| < \sigma \left| x_k - x_m \right|$ holds, for all $i
\in B$, $j \in B$, $k \in B$, and $m \in \left( J \vdash B \right)$, and this
is true, in particular, for all $i \in B$, $j \in B$, $k \in B$, and $m \in
\left( A \right. \left. \vdash B \right)$.  Now $\# \left( B \right) \geq 2$,
and $B \subseteq A$, hence $\# \left( A \right) \geq 2$, hence the fact, that
$A$ is a $\sigma$-cluster, of $x$, implies that $\left| x_i - x_j \right| <
\sigma \left| x_k - x_m \right|$ holds, for all $i \in A$, $j \in A$, $k \in
A$, and $m \in \left( \mathcal{D} \left( x \right) \vdash A \right)$, and this
is true, in particular, for all $i \in B$, $j \in B$, $k \in B$, and $m \in
\left( \mathcal{D} \left( x \right) \vdash A \right)$.  Hence $\left| x_i -
x_j \right| < \sigma \left| x_k - x_m \right|$ holds, for all $i \in B$, $j
\in B$, $k \in B$, and $m \in \left( \mathcal{D} \left( x \right) \vdash B
\right)$, hence $B$ is a $\sigma$-cluster, of $x$.

\vspace{0.6cm}

\noindent {\bf{Corollary 1.}}  Let $x$ be a map, such that $\mathcal{D} \left( x
\right)$ is finite, $\# \left( \mathcal{D} \left( x \right) \right) \geq 2$,
and $\mathcal{R} \left( x \right) \subseteq \mathbb{E}_d$, let $J$ be a
subset, of $\mathcal{D} \left( x \right)$, such that $\# \left( J \right) \geq
2$, let $y \equiv\mathcal{N} \left( x, J \right)$ be the restriction, of $x$, to the
domain, $J$, and let $A$ be a subset, of $J$, such that $\# \left( A \right)
\geq 2$, and $A$ is a $\sigma$-cluster, of $x$.  Then $\mathcal{L} \left(
\mathcal{F} \left( y, \sigma \right), A \right) =\mathcal{L} \left(
\mathcal{F} \left( x, \sigma \right), A \right)$.

In particular, this is true, for all $A \in \mathbb{B} \left( \mathcal{P}
\left( \mathcal{F} \left( x, \sigma \right), J \right) \right)$.

\vspace{0.6cm}

\noindent {\bf{Corollary 2.}}  Let $x$ be a map, such that $\mathcal{D} \left( x
\right)$ is finite, $\# \left( \mathcal{D} \left( x \right) \right) \geq 2$,
and $\mathcal{R} \left( x \right) \subseteq \mathbb{E}_d$, let $A$ be a
$\sigma$-cluster, of $x$, such that $\# \left( A \right) \geq 2$, let $y \equiv
\mathcal{N} \left( x, A \right)$ be the restriction, of $x$, to the domain, $A$,
and let $B$ be any subset, of $A$.  Then $B$ is a $\sigma$-cluster, of $x$, ifif
$B$ is a $\sigma$-cluster, of $y$.

\vspace{0.6cm}

\noindent {\bf{Lemma 11.}}  Let $x$ be a map, such that $\mathcal{D} \left( x
\right)$ is finite, $\# \left( \mathcal{D} \left( x \right) \right) \geq 2$,
and $\mathcal{R} \left( x \right) \subseteq \mathbb{E}_d$, let $\sigma$ be a
real number, such that $0 < \sigma < 1$, let $A$ be a $\sigma$-cluster, of $x$,
such that $\# \left( A \right) \geq 2$, let $a$, and $b$, be any two members, of
$A$, such that $\mathcal{K} \left( \mathcal{F} \left( x, \sigma \right), A, a
\right) \neq \mathcal{K} \left( \mathcal{F} \left( x, \sigma \right), A, b
\right)$, let $k$, and $m$, be any members, of $A$, and let $N \equiv \# \left(
\mathcal{P} \left( \mathcal{F} \left( x, \sigma \right), A \right) \right)$.
Note that $2 \leq N \leq \# \left( A \right)$.

Then $\left| x_k - x_m \right| \leq \sigma \left( \frac{1 + 2 \sigma}{\sigma}
\right)^{N - 1} \left| x_a - x_b \right|$.

\vspace{0.6cm}

\noindent {\bf{Proof.}}  For each integer, $p$, such that $1 \leq p \leq N$, let
$\mathcal{I} \left( p \right)$ be the set, whose members are the $p$ integers
$\geq 1$ and $\leq p$, and let $R_p \equiv \sigma \left( \frac{1 + 2
\sigma}{\sigma} \right)^{p - 1} \left| x_a - x_b \right|$.

I will construct a bijection, $B$, whose domain is $\mathcal{I} \left( N
\right)$, and whose range is \\
$\mathcal{P} \left( \mathcal{F} \left( x, \sigma
\right), A \right)$, such that, for each integer, $p \in \mathcal{I} \left( N
\right)$, $\left| x_k - x_m \right| \leq R_p$ holds, for all members, $k$, and
$m$, of $\mathcal{U} \left( \mathcal{R} \left( \mathcal{N} \left(
B,\mathcal{I} \left( p \right) \right) \right) \right)$.  This will prove the
Lemma, since $\mathcal{U} \left( \mathcal{R} \left( \mathcal{N} \left(
B,\mathcal{I} \left( N \right) \right) \right) \right) =\mathcal{U} \left(
\mathcal{R} \left( B \right) \right) =\mathcal{U} \left( \mathcal{P} \left(
\mathcal{F} \left( x, \sigma \right), A \right) \right) = A$, and \\
$R_N =
\sigma \left( \frac{1 + 2 \sigma}{\sigma} \right)^{N - 1} \left| x_a - x_b
\right|$.

For all $p \in \mathcal{I} \left( N \right)$, I define $T_p \equiv \mathcal{U}
\left( \mathcal{R} \left( \mathcal{N} \left( B,\mathcal{I} \left( p \right)
\right) \right) \right)$.  Hence $T_1 = B_1$, and $T_p = T_{p - 1} \cup B_p$,
for $2 \leq p \leq N$.

Let $B_1 \equiv \mathcal{K} \left( \mathcal{F} \left( x, \sigma \right), A, a
\right)$, and $B_2 \equiv \mathcal{K} \left( \mathcal{F} \left( x, \sigma
\right), A, b \right)$.  Now the fact, that $B_1$ is a $\sigma$-cluster, of $x$,
implies that $\left| x_k - x_m \right| \leq \sigma \left| x_a - x_b \right|$
holds, for all members, $k$, and $m$, of $B_1$, hence $\left| x_k - x_m \right|
\leq R_1$ holds, for all members, $k$, and $m$, of $T_1 = B_1$.  And the fact,
that $B_2$ is a $\sigma$-cluster, of $x$, implies that $\left| x_k - x_m
\right| \leq \sigma \left| x_a - x_b \right|$ holds, for all members, $k$, and
$m$, of $B_2$, hence $\left| x_k - x_m \right| \leq R_1$ holds, for all
members, $k$, and $m$, of $B_2$.  Now let $k$ be any member, of $T_1 = B_1$, and
$m$ be any member, of $B_2$.  Then
\[ \left| x_k - x_m \right| \leq \left| x_k - x_a \right| + \left| x_a - x_b
   \right| + \left| x_b - x_m \right| \leq R_1 + \left( \frac{1}{\sigma}
   \right) R_1 + R_1 = \left( \frac{1 + 2 \sigma}{\sigma} \right) R_1 = R_2 .
\]
Now $R_1 \leq R_2$, hence $\left| x_k - x_m \right| \leq R_2$ holds, for all
members $k$, and $m$, of $T_2 = B_1 \cup B_2$.

Now if $N = 2$, the proof is now complete.

Assume, now, that $N \geq 3$.  I will proceed by induction, on $p$.

For $3 \leq p \leq N$, assume that $\left( p - 1 \right)$ distinct members,
$B_1, \ldots, B_{p - 1}$, of \\
$\mathcal{P} \left( \mathcal{F} \left( x, \sigma
\right), A \right)$, have been identified, such that $\left| x_k - x_m \right|
\leq R_{p - 1}$ holds, for all members, $k$, and $m$, of $T_{p - 1} = T_{p - 2}
\cup B_{p - 1}$.

Now the facts, that $A$ {\emph{is}} a $\sigma$-cluster, of $x$, and that $T_{p -
1}$ is {\emph{not}} a $\sigma$-cluster, of $x$, together imply, that there is a
member, $e$, of $T_{p - 1}$, and a member, $f$, of $\left( A \vdash T_{p - 1}
\right)$, such that $\left| x_e - x_f \right| \leq \left( \frac{1}{\sigma}
\right) R_{p - 1}$.

For if there is a member, $g$, of $T_{p - 1}$, and a member, $h$, of $\left(
\mathcal{D} \left( x \right) \vdash A \right)$, such that $\left| x_g - x_h
\right| \leq \left( \frac{1}{\sigma} \right) R_{p - 1}$, then the facts, that
$T_{p - 1} \subseteq A$, and that $A$ is a $\sigma$-cluster, of $x$, together
imply, that for all $u \in A$, and $v \in A$, $\left| x_u - x_v \right| <
\sigma \left| x_g - x_h \right| \leq R_{p - 1} < \left( \frac{1}{\sigma}
\right) R_{p - 1}$ holds, hence, since $T_{p - 1} \subseteq A$ holds, and
neither $T_{p - 1}$, nor $\left( A \vdash T_{p - 1} \right)$, is empty, there
is a member, $e$, of $T_{p - 1}$, and a member, $f$, of $\left( A \vdash T_{p
- 1} \right)$, such that $\left| x_e - x_f \right| \leq \left(
\frac{1}{\sigma} \right) R_{p - 1}$.

And if $\left| x_g - x_h \right| > \left( \frac{1}{\sigma} \right) R_{p - 1}$
holds, for every member, $g$, of $T_{p - 1}$, and every member, $h$, of
$\left( \mathcal{D} \left( x \right) \vdash A \right)$, then the induction
assumption, that $\left| x_k - x_m \right| \leq R_{p - 1}$ holds, for all
members, $k$, and $m$, of $T_{p - 1}$, and the fact, that $T_{p - 1}$ is
{\emph{not}} a $\sigma$-cluster of, $x$, together imply, that there must be a
member, $e$, of $T_{p - 1}$, and a member, $f$, of $\left( A \vdash T_{p - 1}
\right)$, such that $\left| x_e - x_f \right| \leq \left( \frac{1}{\sigma}
\right) R_{p - 1}$.

Now choose a member, $e$, of $T_{p - 1}$, and a member, $f$, of $\left( A
\vdash T_{p - 1} \right)$, such that $\left| x_e - x_f \right| \leq \left(
\frac{1}{\sigma} \right) R_{p - 1}$, and let $B_p \equiv \mathcal{K} \left(
\mathcal{F} \left( x, \sigma \right), A, f \right)$.  Now the fact, that $B_p$
is a $\sigma$-cluster, of $x$, implies that $\left| x_k - x_m \right| \leq
\sigma \left| x_e - x_f \right| \leq R_{p - 1}$ holds, for all members, $k$,
and $m$, of $B_p$.  And by the induction assumption, $\left| x_k - x_m \right|
\leq R_{p - 1}$ holds, for all members, $k$, and $m$, of $T_{p - 1}$.  Hence, if
$k$ is any member, of $T_{p - 1}$, and $m$ is any member, of $B_p$,
\[ \left| x_k - x_m \right| \leq \left| x_k - x_e \right| + \left| x_e - x_f
   \right| + \left| x_f - x_m \right| \qquad\qquad\qquad\qquad \]
\[ \qquad\qquad\leq R_{p - 1} + \left( \frac{1}{\sigma} \right) R_{p - 1} + R_{p - 1} =
   \left( \frac{1 + 2 \sigma}{\sigma} \right) R_{p - 1} = R_p \]
hence, since $R_{p - 1} \leq R_p$, $\left| x_k - x_m \right| \leq R_p$ holds,
for all members, $k$, and $m$, of $T_p = T_{p - 1} \bigcup B_p$, and the
induction step is complete.

\vspace{0.6cm}

If $I$ is a finite set, such that $\# \left( I \right) \geq 2$, and $\sigma$
is a real number, such that $0 < \sigma < 1$, then the $\left( \# \left( I
\right) d \right)$-dimensional space, $\left( \mathbb{E}_d \right)^I$, of all
maps, $x$, such that $\mathcal{D} \left( x \right) = I$, and $\mathcal{R}
\left( x \right)$ is a subset, of $\mathbb{E}_d$, may be divided into a finite
number, $\# \left( \mathbb{H} \left( I \right) \right)$, of sectors, one
sector being associated with each high greenwood, of $I$, by assigning each
point, $x$, of $\left( \mathbb{E}_d \right)^I$, to the sector associated with the
high greenwood, $\mathcal{F} \left( x, \sigma \right)$, of $I$.

For any ordered triple, $\left( x, \sigma, F \right)$, whose first component is
a map, $x$, such that $\mathcal{D} \left( x \right)$ is finite, $\# \left(
\mathcal{D} \left( x \right) \right) \geq 2$, and $\mathcal{R} \left( x
\right) \subseteq \mathbb{E}_d$, and whose second component is a real number,
$\sigma$, such that $0 < \sigma < 1$, and whose third component is a high
greenwood, $F$, of $\mathcal{D} \left( x \right)$, I define the number,
$\mathcal{A} \left( x, \sigma, F \right)$, by
\[ \mathcal{A} \left( x, \sigma, F \right) \equiv \left\{ \begin{array}{ccc}
     1 & \mathrm{ifif} & \mathcal{F} \left( x, \sigma \right) = F\\
     0 & \mathrm{ifif} & \mathcal{F} \left( x, \sigma \right) \neq F
   \end{array} \right\} . \]

Note that for any finite set, $I$, such that $\# \left( I \right) \geq 2$, and
any real number, $\sigma$, such that $0 < \sigma < 1$, the following identity
holds, for all $x \in \left( \mathbb{E}_d \right)^I$.
\[ \sum_{F \in \mathbb{H}\, \left( I \right)} \mathcal{A} \left( x, \sigma, F
   \right) = 1. \]

For any ordered pair, $\left( x, L \right)$, whose first component is a map,
$x$, such that $\mathcal{D} \left( x \right)$ is finite, $\# \left(
\mathcal{D} \left( x \right) \right) \geq 2$, and $\mathcal{R} \left( x
\right) \subseteq \mathbb{E}_d$, and whose second component is a real number,
$L \geq 0$, the number, $\mathcal{B} \left( x, L \right)$, is defined, by
\[ \mathcal{B} \left( x, L \right) \equiv \left\{ \begin{array}{cc}
     1 & \mathrm{ifif}\, \left| x_i - x_j \right| \leq L,\, \mathrm{for}
     \,\, \mathrm{all}\,\, i
     \in \mathcal{D} \left( x \right),\, \mathrm{and}\,\, j \in \mathcal{D} \left( x
     \right)\\
     0 & \mathrm{otherwise}
   \end{array} \right\} . \]

Note that
\[ \mathcal{B} \left( x, L \right) = \prod_{\Delta \equiv \left\{ i, j
   \right\} \in \mathcal{Q} \left( \mathcal{D} \left( x \right) \right)}
   \mathbb{S} \left( L - \left| x_i - x_j \right| \right) . \]

\vspace{0.25cm}

A {\emph{set of powers}} is a map, $\alpha$, such that $\mathcal{D} \left(
\alpha \right)$ is nonempty, every member, of $\mathcal{D} \left( \alpha
\right)$, is a two-member set, $\mathcal{U} \left( \mathcal{D} \left( \alpha
\right) \right)$ is a finite set, every two-member subset, of $\mathcal{U}
\left( \mathcal{D} \left( \alpha \right) \right)$, is a member, of $\mathcal{D}
\left( \alpha \right)$, and $\mathcal{R} \left( \alpha \right)$ is a subset, of
$\mathbb{R}$.

Note that if $\alpha$ is a set of powers, then $\# \left( \mathcal{U} \left(
\mathcal{D} \left( \alpha \right) \right) \right) \geq 2$, and $\mathcal{D}
\left( \alpha \right) =\mathcal{Q} \left( \mathcal{U} \left( \mathcal{D}
\left( \alpha \right) \right) \right)$.

If $A$ is any finite set, such that $\# \left( A \right) \geq 2$, a {\emph{set
of powers, for}} $A$, is a set of powers, $\alpha$, such that $\mathcal{U}
\left( \mathcal{D} \left( \alpha \right) \right) = A$.  Note that if $\alpha$
is a set of powers, for $A$, then $\mathcal{D} \left( \alpha \right)
=\mathcal{Q} \left( A \right)$.

For any ordered pair, $\left( x, \alpha \right)$, whose first component is a
map, $x$, such that $\mathcal{D} \left( x \right)$ is finite, $\# \left(
\mathcal{D} \left( x \right) \right) \geq 2$, and $\mathcal{R} \left( x
\right) \subseteq \mathbb{E}_d$, and whose second component is a set of
powers, $\alpha$, for $\mathcal{D} \left( x \right)$, I define
\[ \Psi \left( x, \alpha \right) \equiv \prod_{\Delta \equiv \left\{ i, j \right\}
   \in \mathcal{Q} \left( \mathcal{D} \left( x \right) \right)} \left| x_i -
   x_j \right|^{- \alpha_{\Delta}} . \]
For any ordered pair, $\left( \alpha, A \right)$, of a set of powers,
$\alpha$, and a subset, $A$, of $\mathcal{U} \left( \mathcal{D} \left( \alpha
\right) \right)$, such that $\# \left( A \right) \geq 2$, I define
\[ \Gamma \left( \alpha, A \right) \equiv \sum_{\Delta \in \mathcal{Q} \left(
   A \right)} \alpha_{\Delta} . \]
For any greenwood, $F$, a {\emph{good set of powers, for}} $F$, is a set of
powers, $\alpha$, for $\mathcal{U} \left( F \right)$, such that $\Gamma \left(
\alpha, A \right) < d \left( \# \left( A \right) - 1 \right)$ holds, for every
member, $A$, of $\mathbb{B} \left( F \right)$.  (Note that the definition, of
a good set of powers, for $F$, depends, implicitly, on $d$.)

For any ordered triple, $\left( \alpha, i, A \right)$, of a set of powers,
$\alpha$, a member, $i$, of $\mathcal{U} \left( \mathcal{D} \left( \alpha
\right) \right)$, and a subset, $A$, of $\left( \mathcal{U} \left( \mathcal{D}
\left( \alpha \right) \right) \vdash \left\{ i \right\} \right)$, I define
\[ \xi \left( \alpha, i, A \right) \equiv \sum_{j \in A} \alpha_{\left\{ i, j
   \right\}} . \]

\section{Cluster Convergence Theorem.}
\label{Section 4}

{\bf{Proposition.}}  Let $I$ be any finite set, such that $\# \left( I
\right) \geq 3$, $i$ be any member, of $I$, $F$ be any high greenwood, of $I$,
$\alpha$ be any good set of powers, for $F$, $\sigma$ be any real number, such
that $0 < \sigma < 1$, and $L$ be any real number, $\geq 0$.

Let $J \equiv \left( I \vdash \left\{ i \right\} \right)$.

Let $\mathbb{K} \left( F, J \right)$ be the set, whose members are all the
high greenwoods, $E$, of $J$, such that $\mathcal{S} \left( F, J \right)
\subseteq E$, and for each member, $A$, of $\mathbb{B} \left( \mathcal{P}
\left( F, J \right) \right)$, $\mathcal{L} \left( E, A \right) =\mathcal{L}
\left( F, A \right)$ holds.  (Note that $\mathbb{K} \left( F, J \right)$ is a
finite set.)

Then there exists a map, $\eta$, whose domain is $\mathbb{K} \left( F, J
\right)$, and a map, $C$, whose domain is $\mathbb{K} \left( F, J \right)$,
such that the following three statements, $\mathcal{X}_1 \left( \eta \right)$,
$\mathcal{X}_2 \left( C \right)$, and $\mathcal{X}_3 \left( \eta, C \right)$,
all hold.

\vspace{0.3cm}

\noindent $\mathcal{X}_1 \left( \eta \right)$.  For each member, $E$, of $\mathbb{K}
\left( F, J \right)$, $\eta_E$ is a good set of powers, for $E$.

\vspace{0.3cm}

\noindent $\mathcal{X}_2 \left( C \right)$.  For each member, $E$, of $\mathbb{K}
\left( F, J \right)$, $C_E$ is a finite real number, $\geq 0$.

\vspace{0.3cm}

\noindent $\mathcal{X}_3 \left( \eta, C \right)$.  For all maps, $y$, whose domain is
$J$, and whose range is a subset, of $\mathbb{E}_d$, the following inequality,
in which $x \equiv y \cup \left\{ \left( i, x_i \right) \right\}$, holds.
\[ \int d^d x_i \mathcal{A} \left( x, \sigma, F \right) \mathcal{B} \left( x,
   L \right) \Psi \left( x, \alpha \right) \leq \sum_{E \in\, \mathbb{K} \left(
   F, J \right)} C_E \mathcal{A} \left( y, \sigma, E \right) \mathcal{B}
   \left( y, L \right) \Psi \left( y, \eta_E \right) \]

\vspace{0.6cm}

\noindent {\bf{Corollary.}}  The integral, of $\mathcal{A} \left( x, \sigma, F
\right) \mathcal{B} \left( x, L \right) \Psi \left( x, \alpha \right)$, with
respect to any $\left( \# \left( I \right) - 1 \right)$ members, of $x$, is
finite.

\vspace{0.6cm}

\noindent {\bf{Proof.}}  It is enough to prove the main Proposition, since the
Corollary is a simple consequence, of the main Proposition.

I shall proceed to bound $\int d^d x_i \mathcal{A} \left( x, \sigma, F \right)
\mathcal{B} \left( x, L \right) \Psi \left( x, \alpha \right)$, where $x
\equiv y \cup \left\{ \left( i, x_i \right) \right\}$, and $y$ is any map,
whose domain, is $J$, and whose range is a subset, of $\mathbb{E}_d$, in three
main steps.

A central role will be played by $\mathcal{H} \left( F, i \right)$, which, for
any ordered pair, $\left( F, i \right)$, of a high greenwood, $F$, and a
member, $i$, of $\mathcal{U} \left( F \right)$, was defined, on page 5, by
$\mathcal{H} \left( F, i \right) \equiv \left( \mathcal{C} \left( F, \left\{ i
\right\} \right) \vdash \left\{ i \right\} \right) =\mathcal{C} \left( F,
\left\{ i \right\} \right) \cap \left( \mathcal{U} \left( F \right) \vdash
\left\{ i \right\} \right)$.

In the present case, $\mathcal{U} \left( F \right) = I$, $\left( \mathcal{U}
\left( F \right) \vdash \left\{ i \right\} \right) = J$, and $\mathcal{H}
\left( F, i \right)$ is a subset, of $J$, and a member, of $\mathcal{S} \left(
F, J \right)$.

The first step, is to find a real number, $S \geq 0$, and a good set of powers,
$\beta$, for $F$, such that
\[ \mathcal{A} \left( x, \sigma, F \right) \mathcal{B} \left( x, L \right)
   \Psi \left( x, \alpha \right) \leq S\mathcal{A} \left( x, \sigma, F \right)
   \mathcal{B} \left( x, L \right) \Psi \left( x, \beta \right) \]
holds, for all $x \in \left( \mathbb{E}_d \right)^I$, and, in addition,
$\beta_{\left\{ i, k \right\}} = 0$ holds, for all $k \in \left( J \vdash
\mathcal{H} \left( F, i \right) \right)$.

If $\left( J \vdash \mathcal{H} \left( F, i \right) \right)$ is nonempty, let
$k$ be any member, of $\left( J \vdash \mathcal{H} \left( F, i \right)
\right)$, and let $j$ be any member, of $\mathcal{H} \left( F, i \right)$.  Now
$\left( \mathcal{H} \left( F, i \right) \cup \left\{ i \right\} \right)
=\mathcal{C} \left( F, \left\{ i \right\} \right)$ is a member, of $F$, hence,
for all $x \in \left( \mathbb{E}_d \right)^I$, such that $\mathcal{A} \left(
x, \sigma, F \right)$ is nonzero, $\left( \mathcal{H} \left( F, i \right) \cup
\left\{ i \right\} \right)$ is a $\sigma$-cluster, of $x$, hence
\[ \left| y_k - y_j \right| \leq \left| y_k - x_i \right| + \left| x_i - y_j
   \right| \leq \left| y_k - x_i \right| + \sigma \left| y_k - x_i \right| =
   \left( 1 + \sigma \right) \left| y_k - x_i \right| \]
and
\[ \left| y_k - x_i \right| \leq \left| y_k - y_j \right| + \left| y_j - x_i
   \right| \leq \left| y_k - y_j \right| + \sigma \left| y_k - y_j \right| =
   \left( 1 + \sigma \right) \left| y_k - y_j \right| \]
both hold, for all $x \in \left( \mathbb{E}_d \right)^I$, such that
$\mathcal{A} \left( x, \sigma, F \right)$ is nonzero, hence
\[ \left( 1 + \sigma \right)^{\mathbb{M} \left( \alpha_{\left\{ i, k
   \right\}} \right)} \left( \frac{\left| x_i - y_k \right|}{\left| y_j - y_k
   \right|} \right)^{\alpha_{\left\{ i, k \right\}}} \geq 1 \]
holds, for all $x \in \left( \mathbb{E}_d \right)^I$, such that $\mathcal{A}
\left( x, \sigma, F \right)$ is nonzero, where $\mathbb{M} \left( t \right)$
was defined, for all $t \in \mathbb{R}$, on page 6, as the absolute value of
$t$.

(Note that, for each member, $m$, of $J$, $x_m$, and $y_m$, are equally valid
notations, for the second component, of the unique member, of $x$, whose first
component, is $m$, since $y$ is the restriction, of $x$, to the domain, $J$.)

Choose a member, $g$, of $\mathcal{H} \left( F, i \right)$, and let $\beta$ be
the set of powers, for $I$, such that $\beta_{\left\{ g, k \right\}} = \left(
\alpha_{\left\{ g, k \right\}} + \alpha_{\left\{ i, k \right\}} \right)$
holds, for all $k \in \left( J \vdash \mathcal{H} \left( F, i \right)
\right)$, $\beta_{\left\{ i, k \right\}} = 0$ holds, for all $k \in \left( J
\vdash \mathcal{H} \left( F, i \right) \right)$, and $\beta_{\Delta} =
\alpha_{\Delta}$ holds, for all other members, $\Delta$, of $\mathcal{Q}
\left( I \right)$.  Then
\[ \Psi \left( x, \beta \right) = \Psi \left( x, \alpha \right) \prod_{k \in
   \left( J \,\vdash \mathcal{H} \left( F, i \right) \right)} \left(
   \frac{\left| x_i - y_k \right|}{\left| y_g - y_k \right|}
   \right)^{\alpha_{\left\{ i, k \right\}}} \]
holds, for all $x \in \left( \mathbb{E}_d \right)^I$, and
\[ \mathcal{A} \left( x, \sigma, F \right) \mathcal{B} \left( x, L \right)
   \Psi \left( x, \alpha \right) \leq \qquad\qquad\qquad\qquad\qquad\qquad\qquad\qquad\qquad\qquad \]
\[ \qquad\qquad\leq \left( \prod_{k \in \left( J \,\vdash \mathcal{H} \left( F, i \right)
   \right)} \left( 1 + \sigma \right)^{\mathbb{M} \left( \alpha_{\left\{ i, k
   \right\}} \right)} \right) \mathcal{A} \left( x, \sigma, F \right)
   \mathcal{B} \left( x, L \right) \Psi \left( x, \beta \right) \]
holds, for all $x \in \left( \mathbb{E}_d \right)^I$.

Now, if $\left( J \vdash \mathcal{H} \left( F, i \right) \right) = \left( I
\vdash \left( \mathcal{H} \left( F, i \right) \cup \left\{ i \right\} \right)
\right)$ is nonempty, let $k$ be any member, of $\left( J \vdash \mathcal{H}
\left( F, i \right) \right)$, and let $j$ be any member, of $\mathcal{H} \left(
F, i \right)$.  Note that $\left( \mathcal{H} \left( F, i \right) \cup \left\{
i \right\} \right) =\mathcal{C} \left( F, \left\{ i \right\} \right)$ is a
member, of $F$.  Then Lemma 1, for the case that $B =\mathcal{C} \left( F,
\left\{ i \right\} \right)$, implies that for every member, $A$, of $F$,
$\left\{ i, k \right\} \subseteq A$, ifif $\left\{ j, k \right\} \subseteq A$,
and this is true, in particular, for the case $j = g$.  Hence, for every
member, $A$, of $F$, $\Gamma \left( \beta, A \right) = \Gamma \left( \alpha, A
\right)$, hence $\beta$ is a good set of powers, for $F$.

Hence, it is now sufficient, to bound $\int d^d x_i \mathcal{A} \left( x,
\sigma, F \right) \mathcal{B} \left( x, L \right) \Psi \left( x, \beta
\right)$, where $\beta$ is a good set of powers, for $F$, such that
$\beta_{\left\{ i, k \right\}} = 0$ holds, for all $k \in \left( J \vdash
\mathcal{H} \left( F, i \right) \right)$.

Now, the following identity holds, for all $y \in \left( \mathbb{E}_d
\right)^J$.
\[ 1 \equiv \sum_{E \in \mathbb{H}\, \left( J \right)} \mathcal{A} \left( y,
   \sigma, E \right) . \]
Hence, the following identity holds, for all $x \in \left( \mathbb{E}_d
\right)^I$.
\[ \mathcal{A} \left( x, \sigma, F \right) \equiv \sum_{E \in \,\mathbb{H}\,
   \left( J \right)} \mathcal{A} \left( y, \sigma, E \right) \mathcal{A}
   \left( x, \sigma, F \right) . \]
Now the Corollary, to Lemma 9, implies that the product, $\mathcal{A} \left(
y, \sigma, E \right) \mathcal{A} \left( x, \sigma, F \right)$, is zero, for
all $x \in \left( \mathbb{E}_d \right)^I$, unless $\mathcal{S} \left( F, J
\right) \subseteq E$, for $\mathcal{A} \left( y, \sigma, E \right)$ is zero,
unless $\mathcal{F} \left( y, \sigma \right) = E$, and $\mathcal{A} \left( x,
\sigma, F \right)$ is zero, unless $\mathcal{F} \left( x, \sigma \right) = F$,
and the Corollary, to Lemma 9, implies that $\mathcal{S} \left( \mathcal{F}
\left( x, \sigma \right), J \right) \subseteq \mathcal{F} \left( y, \sigma
\right)$.

And Corollary 1, of Lemma 10, implies that the product, $\mathcal{A} \left( y,
\sigma, E \right) \mathcal{A} \left( x, \sigma, F \right)$, is zero, for all $x
\in \left( \mathbb{E}_d \right)^I$, unless $\mathcal{L} \left( E, A \right)
=\mathcal{L} \left( F, A \right)$ holds, for all $A \in \mathbb{B} \left(
\mathcal{P} \left( F, J \right) \right)$, for $\mathcal{A} \left( y, \sigma, E
\right)$ is zero, unless $\mathcal{F} \left( y, \sigma \right) = E$, and
$\mathcal{A} \left( x, \sigma, F \right)$ is zero, unless $\mathcal{F} \left(
x, \sigma \right) = F$, and Corollary 1, of Lemma 10, implies that
$\mathcal{L} \left( \mathcal{F} \left( y, \sigma \right), A \right)
=\mathcal{L} \left( \mathcal{F} \left( x, \sigma \right), A \right)$ holds,
for every member, $A$, of $\mathbb{B} \left( \mathcal{P} \left( \mathcal{F}
\left( x, \sigma \right), J \right) \right)$.

Now let $\mathbb{K} \left( F, J \right)$ be the set, defined on page 15,
whose members are all the high greenwoods, $E$, of $J$, such that $\mathcal{S}
\left( F, J \right) \subseteq E$ holds, and for every member, $A$, of
$\mathbb{B} \left( \mathcal{P} \left( F, J \right) \right)$, $\mathcal{L}
\left( E, A \right) =\mathcal{L} \left( F, A \right)$ holds.

Then the following identity holds, for all $x \in \left( \mathbb{E}_d
\right)^I$.
\[ \mathcal{A} \left( x, \sigma, F \right) = \sum_{E \in \mathbb{K} \left( F,
   J \right)} \mathcal{A} \left( y, \sigma, E \right) \mathcal{A} \left( x,
   \sigma, F \right) . \]
Now let $E$ be any member, of $\mathbb{K} \left( F, J \right)$, such that
$\left( E \vdash \mathcal{S} \left( F, J \right) \right)$ is nonempty, $A$ be
any member, of $\left( E \vdash \mathcal{S} \left( F, J \right) \right)$, (so
$\# \left( A \right) \geq 2$, and $A \subset J$ both hold), $a$, and $b$, be
any two members, of $A$, such that $\left\{ a, b \right\} \in \mathcal{W}
\left( E, A \right)$, $e$ be any member, of $A$, and $f$ be any member, of
$\left( \mathcal{C} \left( E, A \right) \vdash A \right)$.

Then Lemma 2 implies that $\left\{ a, b \right\}$ is a member of $\mathcal{W}
\left( F,\mathcal{C} \left( F, A \right) \right)$, hence \\
$\mathcal{K} \left(
F,\mathcal{C} \left( F, A \right), a \right) \neq \mathcal{K} \left(
F,\mathcal{C} \left( F, A \right), b \right)$, and that $\left\{ e, f
\right\}$ is a member, of $\mathcal{Q} \left( \mathcal{C} \left( F, A \right)
\right)$.

Hence Lemma 11 implies that
\[ \left| y_a - y_b \right| \geq \left( \frac{1}{\sigma} \right) \left(
   \frac{\sigma}{1 + 2 \sigma} \right)^{\left( \# \left( \mathcal{P} \left(
   F,\mathcal{C} \left( F, A \right) \right) \right) - 1 \right)} \left| y_e -
   y_f \right| \]
holds, for all $x \in \left( \mathbb{E}_d \right)^I$, such that the product,
$\mathcal{A} \left( y, \sigma, E \right) \mathcal{A} \left( x, \sigma, F
\right)$, is nonzero.

Hence
\[ \mathbb{S} \left( \left| y_a - y_b \right| - \left( \frac{1}{\sigma}
   \right) \left( \frac{\sigma}{1 + 2 \sigma} \right)^{\left( \# \left(
   \mathcal{P} \left( F,\mathcal{C} \left( F, A \right) \right) \right) - 1
   \right)} \left| y_e - y_f \right| \right), \]
where $\mathbb{S} \left( t \right)$ is the step function, defined on page 6,
for all $t \in \mathbb{R}$, is equal to $1$, for all $x \in \left(
\mathbb{E}_d \right)^I$, such that the product, $\mathcal{A} \left( y,
\sigma, E \right) \mathcal{A} \left( x, \sigma, F \right)$, is nonzero.

For each nonempty, {\emph{strict}} subset, $A$, of $I$, let
\[ n \left( F, A \right) \equiv \left( \# \left( \mathcal{P} \left(
   F,\mathcal{C} \left( F, A \right) \right) \right) - 1 \right) . \]
And for each member, $E$, of $\mathbb{K} \left( F, J \right)$, let
\[ \Theta \left( F, y, \sigma, E \right) \equiv \qquad\qquad\qquad\qquad\qquad\qquad\qquad\qquad
\qquad\qquad\qquad\qquad\qquad\qquad\qquad \]
\[ \qquad\!\!\!\qquad\equiv \prod_{A \in \left( E\, \vdash \mathcal{S} \left( F, J \right)
   \right)} \prod_{\begin{array}{c}\\[-25pt]\scriptstyle{
     \Delta \equiv \left\{ a, b \right\} \in \mathcal{W} \left( E, A \right)}\\[-8pt]
     \scriptstyle{e \in A}\\[-8pt]
     \scriptstyle{f \in \left( \mathcal{C} \left( E, A \right)\, \vdash A \right)
   }\end{array}} \mathbb{S} \left( \left| y_a - y_b \right| - \left(
   \frac{1}{\sigma} \right) \left( \frac{\sigma}{1 + 2 \sigma} \right)^{n \left( F,
   A \right)} \left| y_e - y_f \right| \right), \]
so that $\Theta \left( F, y, \sigma, E \right)$ is equal to $1$, for all $x
\in \left( \mathbb{E}_d \right)^I$, such that the product, \\
$\mathcal{A}
\left( y, \sigma, E \right) \mathcal{A} \left( x, \sigma, F \right)$, is
nonzero.

Then the following identity holds, for all $x \in \left( \mathbb{E}_d
\right)^I$.
\[ \mathcal{A} \left( x, \sigma, F \right) = \sum_{E \in \mathbb{K} \left( F,
   J \right)} \Theta \left( F, y, \sigma, E \right) \mathcal{A} \left( y,
   \sigma, E \right) \mathcal{A} \left( x, \sigma, F \right) . \]
Hence
\[ \int d^d x_i \mathcal{A} \left( x, \sigma, F \right) \mathcal{B} \left( x,
   L \right) \Psi \left( x, \beta \right) = \qquad\qquad\qquad\qquad\qquad\qquad
   \qquad\qquad\qquad\qquad\]
\[ \qquad\qquad= \sum_{E \in\, \mathbb{K} \left( F, J \right)} \int d^d x_i \Theta \left(
   F, y, \sigma, E \right) \mathcal{A} \left( y, \sigma, E \right) \mathcal{A}
   \left( x, \sigma, F \right) \mathcal{B} \left( x, L \right) \Psi \left( x,
   \beta \right) \]
\[ \qquad\qquad= \sum_{E \in\, \mathbb{K} \left( F, J \right)} \Theta \left( F, y, \sigma,
   E \right) \mathcal{A} \left( y, \sigma, E \right) \int d^d x_i \mathcal{A}
   \left( x, \sigma, F \right) \mathcal{B} \left( x, L \right) \Psi \left( x,
   \beta \right) \]
holds, for all $y \in \left( \mathbb{E}_d \right)^J$.

Now
\[ \mathcal{B} \left( x, L \right) = \prod_{\Delta \equiv \left\{ j, k
   \right\} \in \mathcal{Q} \left( I \right)} \mathbb{S} \left( L - \left|
   x_j - x_k \right| \right) =\mathcal{B} \left( y, L \right) \prod_{j \in J}
   \mathbb{S} \left( L - \left| x_i - y_j \right| \right) \]
holds, for all $x \in \left( \mathbb{E}_d \right)^I$, hence
\[ \int d^d x_i \mathcal{A} \left( x, \sigma, F \right) \mathcal{B} \left( x,
   L \right) \Psi \left( x, \beta \right) = \qquad\qquad\qquad\qquad\qquad\qquad
   \qquad\qquad\qquad\qquad \]
\[ = \sum_{E \in\, \mathbb{K} \left( F, J \right)} \Theta \left( F, y, \sigma,
   E \right) \mathcal{A} \left( y, \sigma, E \right) \mathcal{B} \left( y, L
   \right) \times \qquad\qquad\quad\]
\[ \qquad\qquad\qquad\qquad\qquad\qquad
\times \int d^d x_i \mathcal{A} \left( x, \sigma, F \right) \left( \prod_{j
   \in J} \mathbb{S} \left( L - \left| x_i - y_j \right| \right) \right) \Psi
   \left( x, \beta \right) \]
holds, for all $y \in \left( \mathbb{E}_d \right)^J$.

Let $\mu$ be the restriction, of $\beta$, to the domain, $\mathcal{Q} \left( J
\right)$.

Then
\[ \Psi \left( x, \beta \right) = \Psi \left( y, \mu \right) \prod_{j \in
   \mathcal{H} \left( F, i \right)} \left| x_i - y_j \right|^{- \beta_{\left\{
   i, j \right\}}} \]
holds, for all $x \in \left( \mathbb{E}_d \right)^I$, since $\beta_{\left\{
i, j \right\}} = 0$, for all $j \in \left( J \vdash \mathcal{H} \left( F, i
\right) \right)$.  Hence
\[ \int d^d x_i \mathcal{A} \left( x, \sigma, F \right) \mathcal{B} \left( x,
   L \right) \Psi \left( x, \beta \right) = \qquad\qquad\qquad\qquad\qquad\qquad
   \qquad\qquad\qquad\qquad\!\!\!\!\qquad \]
\[ = \sum_{E \in\, \mathbb{K} \left( F, J \right)} \Theta \left( F, y, \sigma,
   E \right) \mathcal{A} \left( y, \sigma, E \right) \mathcal{B} \left( y, L
   \right) \Psi \left( y, \mu \right) \,\,\times \]
\[ \qquad\qquad\qquad
\times \int d^d x_i \mathcal{A} \left( x, \sigma, F \right) \left( \prod_{j
   \in J} \mathbb{S} \left( L - \left| x_i - y_j \right| \right) \right)
   \left( \prod_{j \in \mathcal{H} \left( F, i \right)} \left| x_i - y_j
   \right|^{- \beta_{\left\{ i, j \right\}}} \right) \]
holds, for all $y \in \left( \mathbb{E}_d \right)^J$.

Now, $\mathcal{H} \left( F, i \right) \cup \left\{ i \right\} =\mathcal{C}
\left( F, \left\{ i \right\} \right)$ is a member, of $F$, $\mu$ is the
restriction, of $\beta$, to the domain, $\mathcal{Q} \left( J \right)$, and
$\beta$ is a good set of powers, for $F$, hence the following statement holds,
provided that $\# \left( \mathcal{H} \left( F, i \right) \right) \geq 2$.
\[ \Gamma \left( \mu,\mathcal{H} \left( F, i \right) \right) + \xi \left(
   \beta, i,\mathcal{H} \left( F, i \right) \right) < d \# \left( \mathcal{H}
   \left( F, i \right) \right) . \]
And, if $\# \left( \mathcal{H} \left( F, i \right) \right) = 1$, then the
following statement holds.
\[ \xi \left( \beta, i,\mathcal{H} \left( F, i \right) \right) < d \]
Now, if there is any member, $A$, of $\mathcal{S} \left( F, J \right)$, such
that $\mathcal{H} \left( F, i \right) \subset A$, let $A$ be a member, of
$\mathcal{S} \left( F, J \right)$, such that $\mathcal{H} \left( F, i \right)
\subset A$.  Now $A \in \mathcal{S} \left( F, J \right)$ implies that at least
one, of $A$, and $A \cup \left\{ i \right\}$, is a member, of $F$.  But $A$
overlaps $\mathcal{H} \left( F, i \right) \cup \left\{ i \right\}$, and
$\mathcal{H} \left( F, i \right) \cup \left\{ i \right\}$ is a member, of $F$,
hence $A$ cannot be a member, of $F$, hence $A \cup \left\{ i \right\}$ is a
member, of $F$.  Now $\mathcal{H} \left( F, i \right) \subset A$ means that $\#
\left( A \right) \geq 2$, since $\left. \mathcal{H} \left( F, i \right) \equiv
(\mathcal{C} \left( F, \left\{ i \right\} \right) \vdash \left\{ i \right\}
\right)$, is nonempty.  Hence the facts, that $\mu$ is the restriction, of
$\beta$, to the domain, $\mathcal{Q} \left( J \right)$, and that $\beta$ is a
good set of powers, for $F$, imply that the following statement holds.
\[ \Gamma \left( \mu, A \right) + \xi \left( \beta, i, A \right) < d \# \left(
   A \right) . \]
Now $\beta_{\left\{ i, j \right\}} = 0$, for all $j \in \left( J \vdash
\mathcal{H} \left( F, i \right) \right)$.  Hence, if $\mathcal{H} \left( F, i
\right) \subset A$, then
\[ \xi \left( \beta, i, A \right) = \sum_{j \in A} \beta_{\left\{ i, j
   \right\}} = \sum_{j \in \mathcal{H} \left( F, i \right)} \beta_{\left\{ i,
   j \right\}} = \xi \left( \beta, i,\mathcal{H} \left( F, i \right) \right) .
\]
Hence, for all members, $A$, of $\mathbb{B} \left( \mathcal{S} \left( F, J
\right) \right)$, such that $\mathcal{H} \left( F, i, \right) \subseteq A$, the
following statement, $\mathcal{T}_1 \left( A \right)$, holds.

\vspace{0.5cm}

\noindent $\mathcal{T}_1 \left( A \right)$.
$\qquad\qquad\qquad\quad$
$\Gamma \left( \mu, A \right) + \xi \left(
\beta, i,\mathcal{H} \left( F, i \right) \right) < d \# \left( A \right)$.

\vspace{0.5cm}

And if $\# \left( \mathcal{H} \left( F, i \right) \right) = 1$, so
$\mathcal{H} \left( F, i \right) \notin \mathbb{B} \left( \mathcal{S} \left( F,
J \right) \right)$, then $\xi \left( \beta, i,\mathcal{H} \left( F, i \right)
\right) < d$ holds.

Now let $A$ be any member, of $\mathbb{B} \left( \mathcal{S} \left( F, J
\right) \right)$, such that $A \subset \mathcal{H} \left( F, i \right)$.  Then
$A \in \mathcal{S} \left( F, J \right)$ again implies, that at least one, of
$A$, and $A \cup \left\{ i \right\}$, is a member, of $F$.  Now $\left(
\mathcal{H} \left( F, i \right) \cup \left\{ i \right\} \right) =\mathcal{C}
\left( F, \left\{ i \right\} \right)$, hence $\left\{ i \right\} \subset
\left( A \cup \left\{ i \right\} \right)$, and $\left( A \cup \left\{ i
\right\} \right) \subset \mathcal{C} \left( F, \left\{ i \right\} \right)$,
both hold, hence the definition, of $\mathcal{C} \left( F, \left\{ i \right\}
\right)$, implies that $\left( A \cup \left\{ i \right\} \right)$ is not a
member, of $F$, hence $A$ is a member, of $F$, hence the facts, that $\mu$ is
the restriction, of $\beta$, to the domain, $\mathcal{Q} \left( J \right)$, and
that $\beta$ is a good set of powers, for $F$, imply that $\Gamma \left( \mu,
A \right) < d \left( \# \left( A \right) - 1 \right)$ holds.

Now let $A$ be any member, of $\mathbb{B} \left( \mathcal{S} \left( F, J
\right) \right)$, such that $A \cap \mathcal{H} \left( F, i \right) =
\emptyset$.  Then $A \in \mathcal{S} \left( F, J \right)$ again implies, that
at least one, of $A$, and $A \cup \left\{ i \right\}$, is a member, of $F$.  But
$A \cup \left\{ i \right\}$ overlaps $\mathcal{H} \left( F, i \right) \cup
\left\{ i \right\}$, and $\mathcal{H} \left( F, i \right) \cup \left\{ i
\right\}$ is a member, of $F$, hence $A \cup \left\{ i \right\}$ cannot be a
member of, $F$, hence $A$ is a member of $F$, hence, again, the facts, that
$\mu$ is the restriction, of $\beta$, to the domain, $\mathcal{Q} \left( J
\right)$, and that $\beta$ is a good set of powers, for $F$, imply that
$\Gamma \left( \mu, A \right) < d \left( \# \left( A \right) - 1 \right)$
holds.

Hence, for each member, $A$, of $\mathbb{B} \left( \mathcal{S} \left( F, J
\right) \right)$, such that either $A \subset \mathcal{H} \left( F, i
\right)$, or $A \cap \mathcal{H} \left( F, i \right) = \emptyset$, the
following statement, $\mathcal{T}_2 \left( A \right)$, holds.

\vspace{0.5cm}

\noindent $\mathcal{T}_2 \left( A \right)$.
$\qquad\qquad\qquad\qquad\quad\quad$
$\Gamma \left( \mu, A \right) < d \left( \#
\left( A \right) - 1 \right)$.

\vspace{0.5cm}

For all integers, $n \geq 0$, let $\mathbb{A}_n$ denote the area, of the
$n$-sphere, of unit radius, so
\[ \mathbb{A}_n = \left\{ \begin{array}{cc}
     \displaystyle{\frac{2 \left( 2 \pi \right)^{\left( \frac{n}{2} \right)}}{\left( n - 1
     \right) !!}} & \qquad\quad\quad( n \textrm{ even and } \geq 0 )\\[0.38cm]
     \displaystyle{\frac{2 \pi^{\left( \frac{n + 1}{2} \right)}}{\left( \frac{n - 1}{2}
     \right) !}} & \qquad\quad\quad( n \textrm{ odd and } \geq 1 )
   \end{array} \right\}, \]
where $\left( - 1 \right) !! \equiv 1$, and, for $n$ even, $n \geq 2$, $\left(
n - 1 \right) !! \equiv \left( n - 1 \right) \left( n - 3 \right) !!$.

In the next step, I shall show how to find a real number, $T \geq 0$, and a set
of powers, $\nu$, for $J$, such that
\[ \int d^d x_i \mathcal{A} \left( x, \sigma, F \right) \left( \prod_{j \in J}
   \mathbb{S} \left( L - \left| x_i - y_j \right| \right) \right) \left(
   \prod_{j \in \mathcal{H} \left( F, i \right)} \left| x_i - y_j \right|^{-
   \beta_{\left\{ i, j \right\}}} \right) \leq T \Psi \left( y, \nu \right) \]
holds, for all $y \in \left( \mathbb{E}_d \right)^J$, and such that, if a set
of powers, $\rho$, for $J$, is defined, by $\rho_{\Delta} \equiv \left(
\mu_{\Delta} + \nu_{\Delta} \right)$, for all $\Delta \in \mathcal{Q} \left( J
\right)$, then $\rho$ is a good set of powers, for $\mathcal{S} \left( F, J
\right)$.  ($\nu_{\Delta}$ will, in fact, be nonzero, for no more than two
members, $\Delta$, of $\mathcal{Q} \left( J \right)$.)

I shall treat, separately, the case, where $\mathcal{H} \left( F, i \right) \in
F$, and the case, where $\mathcal{H} \left( F, i \right) \notin F$.

\subsection*{Case A.  $\mathcal{H} \left( F, i \right)$ {\emph{is}}
a member, of $F$.}

The assumption, that $\mathcal{H} \left( F, i \right)$ is a member, of $F$, and
the facts, that $\mu$ is the restriction, of $\beta$, to the domain,
$\mathcal{Q} \left( J \right)$, and that $\beta$ is a good set of powers, for
$F$, together imply, that the following statement, $\mathcal{T}_3$, holds,
provided that $\# \left( \mathcal{H} \left( F, i \right) \right) \geq 2$.

\vspace{0.5cm}

\noindent $\mathcal{T}_3$.
$\qquad\qquad\qquad\qquad\quad$
$\Gamma \left( \mu,\mathcal{H} \left( F, i \right) \right) <
d \left( \# \left( \mathcal{H} \left( F, i \right) \right) - 1 \right)$.

\vspace{0.5cm}

Now the assumption, that $\mathcal{H} \left( F, i \right)$ is a member, of $F$,
implies, that, for all $x \in \left( \mathbb{E}_d \right)^I$, such that
$\mathcal{A} \left( x, \sigma, F \right)$ is nonzero, $\mathcal{H} \left( F, i
\right)$ is a $\sigma$-cluster, of $x$, hence, that, for all $j \in \mathcal{H}
\left( F, i \right)$, and $k \in \mathcal{H} \left( F, i \right)$,
\[ \left| x_i - y_k \right| \leq \left| x_i - y_j \right| + \left| y_j - y_k
   \right| \leq \left| x_i - y_j \right| + \sigma \left| x_i - y_j \right| =
   \left( 1 + \sigma \right) \left| x_i - y_j \right| \]
holds, for all $x \in \left( \mathbb{E}_d \right)^I$, such that $\mathcal{A}
\left( x, \sigma, F \right)$ is nonzero.

Hence, for all $j \in \mathcal{H} \left( F, i \right)$, and $k \in \mathcal{H}
\left( F, i \right)$,
\[ \left( 1 + \sigma \right)^{\mathbb{M} \left( \beta_{\left\{ i, j \right\}}
   \right)} \left( \frac{\left| x_i - y_j \right|}{\left| x_i - y_k \right|}
   \right)^{\beta_{\left\{ i, j \right\}}} \geq 1 \]
holds, for all $x \in \left( \mathbb{E}_d \right)^I$, such that $\mathcal{A}
\left( x, \sigma, F \right)$ is nonzero.

Hence, choosing a member, $h$, of $\mathcal{H} \left( F, i \right)$,
\[ \int d^d x_i \mathcal{A} \left( x, \sigma, F \right) \left( \prod_{j \in J}
   \mathbb{S} \left( L - \left| x_i - y_j \right| \right) \right) \left(
   \prod_{j \in \mathcal{H} \left( F, i \right)} \left| x_i - y_j \right|^{-
   \beta_{\left\{ i, j \right\}}} \right) \leq \qquad\qquad\quad\quad \]
\[ \qquad\qquad \leq \int d^d x_i \mathcal{A} \left( x, \sigma, F \right) \left( \prod_{j
   \in J} \mathbb{S} \left( L - \left| x_i - y_j \right| \right) \right)
   \left( \prod_{j \in \mathcal{H} \left( F, i \right)} \left( \frac{\left( 1
   + \sigma \right)^{\mathbb{M} \left( \beta_{\left\{ i, j \right\}}
   \right)}}{\left| x_i - y_h \right|^{\beta_{\left\{ i, j \right\}}}} \right)
   \right) \]
holds, for all $y \in \left( \mathbb{E}_d \right)^J$.

Let
\[ \Xi \equiv \prod_{j \in \mathcal{H} \left( F, i \right)} \left( 1 + \sigma
   \right)^{\mathbb{M} \left( \beta_{\left\{ i, j \right\}} \right)} . \]
Now
\[ \prod_{j \in \mathcal{H} \left( F, i \right)} \left| x_i - y_h \right|^{-
   \beta_{\left\{ i, j \right\}}} = \left| x_i - y_h \right|^{- \xi \left(
   \beta, i,\mathcal{H} \left( F, i \right) \right)}, \]
where, for any ordered triple, $\left( \alpha, i, A \right)$, of a set of
powers, $\alpha$, a member, $i$, of $\mathcal{U} \left( \mathcal{D} \left(
\alpha \right) \right)$, and a subset, $A$, of $\left( \mathcal{U} \left(
\mathcal{D} \left( \alpha \right) \right) \vdash \left\{ i \right\} \right)$,
I defined $\xi \left( \alpha, i, A \right)$, on page 15, by
\[ \xi \left( \alpha, i, A \right) \equiv \sum_{j \in A} \alpha_{\left\{ i, j
   \right\}} . \]
Hence
\[ \int d^d x_i \mathcal{A} \left( x, \sigma, F \right) \left( \prod_{j \in J}
   \mathbb{S} \left( L - \left| x_i - y_j \right| \right) \right) \left(
   \prod_{j \in \mathcal{H} \left( F, i \right)} \left| x_i - y_j \right|^{-
   \beta_{\left\{ i, j \right\}}} \right) \leq \qquad\qquad\quad\quad \]
\[ \qquad\qquad \leq \Xi \int d^d x_i \mathcal{A} \left( x, \sigma, F \right) \left(
   \prod_{j \in J} \mathbb{S} \left( L - \left| x_i - y_j \right| \right)
   \right) \left| x_i - y_h \right|^{- \xi \left( \beta, i,\mathcal{H} \left(
   F, i \right) \right)} \]
holds, for all $y \in \left( \mathbb{E}_d \right)^J$.

I shall now treat, separately, the cases $\left( d - \xi \left( \beta,
i,\mathcal{H} \left( F, i \right) \right) \right) < 0$, \\
$\left( d - \xi \left(
\beta, i,\mathcal{H} \left( F, i \right) \right) \right) > 0$, and $\left( d -
\xi \left( \beta, i,\mathcal{H} \left( F, i \right) \right) \right) = 0$.

\subsubsection*{Case A1.  $\mathcal{H} \left( F, i \right)$
{\emph{is}} a member, of $F$, and $\left( d - \xi
\left( \beta, i,\mathcal{H} \left( F, i \right) \right) \right) < 0$.}

Now, if $\# \left( \mathcal{H} \left( F, i \right) \right)$ is equal to $1$,
then the facts, that $\beta$ is a good set of powers, for $F$, and that
$\mathcal{H} \left( F, i \right) \cup \left\{ i \right\}$ is a member, of $F$,
imply that $\xi \left( \beta, i,\mathcal{H} \left( F, i \right) \right) < d$,
hence, since $\left( d - \xi \left( \beta, i,\mathcal{H} \left( F, i \right)
\right) \right) < 0$ holds, by assumption, in the present case, $\mathcal{H}
\left( F, i \right)$ must have at least two members, in the present case.

Choose two members, $a$, and $b$, of $\mathcal{H} \left( F, i \right)$, such
that $\mathcal{K} \left( \mathcal{S} \left( F, J \right),\mathcal{H} \left( F,
i \right), a \right) \neq \mathcal{K} \left( \mathcal{S} \left( F, J
\right),\mathcal{H} \left( F, i \right), b \right)$.  Now, by assumption,
$\mathcal{H} \left( F, i \right)$ is a member, of $F$, hence \\
$\left| y_a - y_b
\right| < \sigma \left| y_h - x_i \right|$ holds, for all $x \in \left(
\mathbb{E}_d \right)^I$, such that $\mathcal{A} \left( x, \sigma, F \right)$
is nonzero, hence $\mathcal{A} \left( x, \sigma, F \right) \leq \mathbb{S}
\left( \left| x_i - y_h \right| - \left( \frac{1}{\sigma} \right) \left| y_a -
y_b \right| \right)$ holds, for all $x \in \left( \mathbb{E}_d \right)^I$,
hence
\[ \int d^d x_i \mathcal{A} \left( x, \sigma, F \right) \left( \prod_{j \in J}
   \mathbb{S} \left( L - \left| x_i - y_j \right| \right) \right) \left| x_i
   - y_h \right|^{- \xi \left( \beta, i,\mathcal{H} \left( F, i \right)
   \right)} \qquad\qquad\qquad\qquad \]
\[ \qquad\qquad\qquad
 \leq \int d^d x_i \mathbb{S} \left( \left| x_i - y_h \right| - \left(
   \frac{1}{\sigma} \right) \left| y_a - y_b \right| \right) \left| x_i - y_h
   \right|^{- \xi \left( \beta, i,\mathcal{H} \left( F, i \right) \right)} =
\]
\[ \qquad\qquad\qquad\qquad\qquad\qquad
 = \frac{\mathbb{A}_{d - 1} \left( \left( \frac{1}{\sigma} \right) \left|
   y_a - y_b \right| \right)^{\left( d - \xi \left( \beta, i,\mathcal{H}
   \left( F, i \right) \right) \right)}}{\left( \xi \left( \beta,
   i,\mathcal{H} \left( F, i \right) \right) - d \right)} \]
holds, for all $y \in \left( \mathbb{E}_d \right)^J$.

Hence, in the present case, $\nu_{\left\{ a, b \right\}} = \left( \xi \left(
\beta, i,\mathcal{H} \left( F, i \right) \right) - d \right)$, and
$\nu_{\Delta} = 0$, for all other members, $\Delta$, of $\mathcal{Q} \left( J
\right)$, hence $\rho_{\left\{ a, b \right\}} = \mu_{\left\{ a, b \right\}} +
\left( \xi \left( \beta, i,\mathcal{H} \left( F, i \right) \right) - d
\right)$, and $\rho_{\Delta} = \mu_{\Delta}$, for all other members, $\Delta$,
of $\mathcal{Q} \left( J \right)$.

Now let $A$ be any member, of $\mathbb{B} \left( \mathcal{S} \left( F, J
\right) \right)$.  Then exactly one of the three possibilities, $\mathcal{H}
\left( F, i \right) \subseteq A$, $A \subset \mathcal{H} \left( F, i \right)$,
and $A \cap \mathcal{H} \left( F, i \right) = \emptyset$, holds.

If $\mathcal{H} \left( F, i \right) \subseteq A$, then $\left\{ a, b \right\}
\subseteq A$, hence $\Gamma \left( \rho, A \right) = \Gamma \left( \mu, A
\right) + \xi \left( \beta, i,\mathcal{H} \left( F, i \right) \right) - d$.
Hence $\mathcal{T}_1 \left( A \right)$ implies that $\Gamma \left( \rho, A
\right) < d \left( \# \left( A \right) - 1 \right)$.

If $A \subset \mathcal{H} \left( F, i \right)$, then $\left\{ a, b \right\}$
is {\emph{not}} a subset, of $A$, hence $\Gamma \left( \rho, A \right) = \Gamma
\left( \mu, A \right)$, hence $\mathcal{T}_2 \left( A \right)$ implies that
$\Gamma \left( \rho, A \right) < d \left( \# \left( A \right) - 1 \right)$.

And if $A \cap \mathcal{H} \left( F, i \right) = \emptyset$, then $\left\{ a,
b \right\}$ is {\emph{not}} a subset, of $A$, hence $\Gamma \left( \rho, A
\right) = \Gamma \left( \mu, A \right)$, hence, again, $\mathcal{T}_2 \left( A
\right)$ implies that $\Gamma \left( \rho, A \right) < d \left( \# \left( A
\right) - 1 \right)$.

Hence, in this case, $\rho$ is a good set of powers, for $\mathcal{S} \left(
F, J \right)$.

\subsubsection*{Case A2.  $\mathcal{H} \left( F, i \right)$
{\emph{is}} a member, of $F$, and $\left( d - \xi
\left( \beta, i,\mathcal{H} \left( F, i \right) \right) \right) > 0$.}

This case includes, in particular, the case where $\# \left( \mathcal{H}
\left( F, i \right) \right) = 1$.

I shall treat separately, the two cases, $\mathcal{H} \left( F, i \right)
\subset J$, and $\mathcal{H} \left( F, i \right) = J$.

\subsubsection*{Case A2a.  $\mathcal{H} \left( F, i \right)$
{\emph{is}} a member, of $F$, $\left( d - \xi \left( \beta,
i,\mathcal{H} \left( F, i \right) \right) \right) > 0$, and
$\mathcal{H} \left( F, i \right) \subset J$.}

Choose a member, $e$, of $\mathcal{H} \left( F, i \right)$, and a member, $f$,
of $\left( \mathcal{C} \left( \mathcal{S} \left( F, J \right),\mathcal{H}
\left( F, i \right) \right)\vdash\right. $ \\
$\left.\mathcal{H} \left( F, i \right) \right)$.
Now $\mathcal{H} \left( F, i \right) \cup \left\{ i \right\}$ is a member, of
$F$, hence $\left| x_i - y_h \right| < \sigma\! \left| y_e - y_f \right|$ holds,
for all $x \in \left( \mathbb{E}_d \right)^I$, such that $\mathcal{A} \left(
x, \sigma, F \right)$ is nonzero, hence $\mathcal{A} \left( x, \sigma, F
\right) \leq \mathbb{S}\! \left( \sigma\! \left| y_e - y_f \right| - \left| x_i -
y_h \right| \right)$ holds, for all $x \in \left( \mathbb{E}_d \right)^I$,
hence
\[ \int d^d x_i \mathcal{A} \left( x, \sigma, F \right) \left( \prod_{j \in J}
   \mathbb{S} \left( L - \left| x_i - y_j \right| \right) \right) \left| x_i
   - y_h \right|^{- \xi \left( \beta, i,\mathcal{H} \left( F, i \right)
   \right)} \leq \qquad\qquad\qquad\qquad \]
\[ \qquad\qquad
 \leq \int d^d x_i \mathbb{S} \left( \sigma \left| y_e - y_f \right| -
   \left| x_i - y_h \right| \right) \left| x_i - y_h \right|^{- \xi \left(
   \beta, i,\mathcal{H} \left( F, i \right) \right)} \]
\[ \qquad\qquad\qquad\qquad\qquad\qquad\qquad
 = \frac{\mathbb{A}_{d - 1} \left( \sigma \left| y_e - y_f \right|
   \right)^{\left( d - \xi \left( \beta, i,\mathcal{H} \left( F, i \right)
   \right) \right)}}{\left( d - \xi \left( \beta, i,\mathcal{H} \left( F, i
   \right) \right) \right)} \]
holds, for all $y \in \left( \mathbb{E}_d \right)^J$.

Hence, in the present case, $\nu_{\left\{ e, f \right\}} = \left( \xi \left(
\beta, i,\mathcal{H} \left( F, i \right) \right) - d \right)$, and
$\nu_{\Delta} = 0$, for all other members, $\Delta$, of $\mathcal{Q} \left( J
\right)$, hence $\rho_{\left\{ e, f \right\}} = \mu_{\left\{ e, f \right\}} +
\left( \xi \left( \beta, i,\mathcal{H} \left( F, i \right) \right) - d
\right)$, and $\rho_{\Delta} = \mu_{\Delta}$, for all other members, $\Delta$,
of $\mathcal{Q} \left( J \right)$.

Now let $A$ be any member, of $\mathbb{B} \left( \mathcal{S} \left( F, J
\right) \right)$.  Then exactly one of the four possibilities, $A =\mathcal{H}
\left( F, i \right)$, $\mathcal{H} \left( F, i \right) \subset A$, $A \subset
\mathcal{H} \left( F, i, \right)$, and $A \cap \mathcal{H} \left( F, i \right)
= 0$, holds.

If $A =\mathcal{H} \left( F, i \right)$, then $\left\{ e, f \right\}$ is
{\emph{not}} a subset, of $A$, hence $\Gamma \left( \rho, A \right) = \Gamma
\left( \mu, A \right) = \Gamma \left( \mu,\mathcal{H} \left( F, i \right)
\right)$, hence, since $\# \left( \mathcal{H} \left( F, i \right) \right) \geq
2$, if $\mathcal{H} \left( F, i \right) \in \mathbb{B} \left( \mathcal{S}
\left( F, J \right) \right)$, $\mathcal{T}_3$ implies that $\Gamma \left(
\rho,\mathcal{H} \left( F, i \right) \right) < d \left( \# \left( \mathcal{H}
\left( F, i \right) \right) - 1 \right)$.

If $\mathcal{H} \left( F, i \right) \subset A$, then the definition, of
$\mathcal{C} \left( \mathcal{S} \left( F, J \right),\mathcal{H} \left( F, i
\right) \right)$, implies that \\
$\left( \mathcal{C} \left( \mathcal{S} \left(
F, J \right),\mathcal{H} \left( F, i \right) \right) \subseteq A \right.$,
hence, since $\left\{ e, f \right\}$ is a subset of $\left( \mathcal{C} \left(
\mathcal{S} \left( F, J \right),\mathcal{H} \left( F, i \right) \right)
\right.$, $\left\{ e, f \right\}$ {\emph{is}} a subset, of $A$, hence $\Gamma
\left( \rho, A \right) = \Gamma \left( \mu, A \right) + \xi \left( \beta,
i,\mathcal{H} \left( F, i \right) \right) - d$, hence
$\mathcal{T}_1 \left( A \right)$ implies that $\Gamma \left( \rho, A \right) <
d \left( \# \left( A \right) - 1 \right)$.

And if $A \subset \mathcal{H} \left( F, i \right)$, or $A \cap \mathcal{H}
\left( F, i \right) = \emptyset$, then $\left\{ e, f \right\}$ is {\emph{not}}
a subset of $A$, hence $\Gamma \left( \rho, A \right) = \Gamma \left( \mu, A
\right)$, hence $\mathcal{T}_2 \left( A \right)$ implies that $\Gamma \left(
\rho, A \right) < d \left( \# \left( A \right) - 1 \right)$.

Hence, in this case, $\rho$ is a good set of powers, for $\mathcal{S} \left(
F, J \right)$.

\subsubsection*{Case A2b.  $\mathcal{H} \left( F, i \right)$
{\emph{is}} a member, of $F$, $\left( d - \xi \left(
\beta, i,\mathcal{H}\! \left( F, i \right) \right) \right) > 0$,
and $\mathcal{H} \left( F, i \right) = J$.}

Now
\[ \int d^d x_i \mathcal{A} \left( x, \sigma, F \right) \left( \prod_{j \in J}
   \mathbb{S} \left( L - \left| x_i - y_j \right| \right) \right) \left| x_i
   - y_h \right|^{- \xi \left( \beta, i,\mathcal{H} \left( F, i \right)
   \right)} \qquad\qquad\qquad\qquad \]
\[ \qquad\qquad\qquad
 \leq \int d^d x_i \mathbb{S} \left( L - \left| x_i - y_h \right| \right)
   \left| x_i - y_h \right|^{- \xi \left( \beta, i,\mathcal{H} \left( F, i
   \right) \right)} = \]
\[ \qquad\qquad\qquad\qquad\qquad\qquad\qquad\qquad\qquad
 = \frac{\mathbb{A}_{d - 1} L^{\left( d - \xi \left( \beta, i,\mathcal{H}
   \left( F, i \right) \right) \right)}}{\left( d - \xi \left( \beta,
   i,\mathcal{H} \left( F, i \right) \right) \right)} \]
holds, for all $y \in \left( \mathbb{E}_d \right)^J$, so, in the present
case, $\nu_{\Delta} = 0$, for all $\Delta \in \mathcal{Q} \left( J \right)$,
hence $\rho_{\Delta} = \mu_{\Delta}$, for all members, $\Delta$, of
$\mathcal{Q} \left( J \right)$, hence $\Gamma \left( \rho, A \right) = \Gamma
\left( \mu, A \right)$, for all members, $A$, of $\mathbb{B} \left(
\mathcal{S} \left( F, J \right) \right)$.

Now let $A$ be any member, of $\mathbb{B} \left( \mathcal{S} \left( F, J
\right) \right)$.  Then exactly one of the two possibilities, $A =\mathcal{H}
\left( F, i \right)$, and $A \subset \mathcal{H} \left( F, i \right)$, holds.
If $A =\mathcal{H} \left( F, i \right)$, then $\mathcal{T}_3$ implies, that
$\Gamma \left( \rho, A \right) < d \left( \# \left( A \right) - 1 \right)$,
and if $A \subset \mathcal{H} \left( F, i \right)$, then $\mathcal{T}_2 \left(
A \right)$ implies, that $\Gamma \left( \rho, A \right) < d \left( \# \left( A
\right) - 1 \right)$.

Hence $\rho$ is a good set of powers, for $\mathcal{S} \left( F, J \right)$,
in this case.

\subsubsection*{Case A3.  $\mathcal{H} \left( F, i \right)$
{\emph{is}} a member, of $F$, and $\left( d - \xi
\left( \beta, i,\mathcal{H} \left( F, i \right) \right) \right) = 0$.}

Now if $\# \left( \mathcal{H} \left( F, i \right) \right)$ is equal to $1$,
then the facts, that $\beta$ is a good set of powers, for $F$, and that
$\mathcal{H} \left( F, i \right) \cup \left\{ i \right\}$ is a member, of $F$,
imply that $\xi \left( \beta, i,\mathcal{H} \left( F, i \right) \right) < d$,
hence $\# \left( \mathcal{H} \left( F, i \right) \right) \geq 2$ must hold, in
the present case.

Let
\[ \lambda \equiv \frac{1}{2} \left( d \left( \# \left( \mathcal{H} \left( F,
   i \right) \right) - 1 \right) - \Gamma \left( \mu,\mathcal{H} \left( F, i
   \right) \right) \right), \]
which is $> 0$, by $\mathcal{T}_3$.

Choose two members, $a$, and $b$, of $\mathcal{H} \left( F, i \right)$, such
that $\mathcal{K} \left( \mathcal{S} \left( F, J \right),\mathcal{H} \left( F,
i \right), a \right) \neq \mathcal{K} \left( \mathcal{S} \left( F, J
\right),\mathcal{H} \left( F, i \right), b \right)$, which implies that
$\left\{ a, b \right\}$ is not a subset, of any member, of $\mathcal{L} \left(
\mathcal{S} \left( F, J \right),\mathcal{H} \left( F, i \right) \right)$.

I shall, again, treat separately, the two cases, $\mathcal{H} \left( F, i
\right) \subset J$, and $\mathcal{H} \left( F, i \right) = J$.

\subsubsection*{Case A3a.  $\mathcal{H} \left( F, i \right)$
{\emph{is}} a member, of $F$, $\left( d - \xi \left(
\beta, i,\mathcal{H} \left( F, i \right) \right) \right) = 0$,
and $\mathcal{H} \left( F, i \right) \subset J$.}

Choose a member, $e$, of $\mathcal{H} \left( F, i \right)$, and a member, $f$,
of $\left( \mathcal{C} \left( \mathcal{S} \left( F, J \right),\mathcal{H}
\left( F, i \right) \right) \vdash\right.$ \\
$\left.\mathcal{H} \left( F, i \right) \right)$.

Then the following two alternative approaches, (i), and (ii), give the same
bound, on
\[ \int d^d x_i \mathcal{A} \left( x, \sigma, F \right) \left( \prod_{j \in J}
   \mathbb{S} \left( L - \left| x_i - y_j \right| \right) \right) \left| x_i
   - y_h \right|^{- \xi \left( \beta, i,\mathcal{H} \left( F, i \right)
   \right)} = \qquad\qquad\qquad \]
\[ \qquad\qquad\qquad\qquad
 = \int d^d x_i \mathcal{A} \left( x, \sigma, F \right) \left( \prod_{j \in
   J} \mathbb{S} \left( L - \left| x_i - y_j \right| \right) \right) \left|
   x_i - y_h \right|^{- d}. \]
(i)  The fact, that $\mathcal{H} \left( F, i \right) \in F$, in the present
case, implies that $\left( \frac{\sigma \left| x_i - y_h \right|}{\left| y_a -
y_b \right|} \right)^{\lambda} \geq 1$ holds, for all $x \in \left(
\mathbb{E}_d \right)^I$, such that $\mathcal{A} \left( x, \sigma, F \right)$
is nonzero, hence
\[ \int d^d x_i \mathcal{A} \left( x, \sigma, F \right) \left( \prod_{j \in J}
   \mathbb{S} \left( L - \left| x_i - y_j \right| \right) \right) \left| x_i
   - y_h \right|^{- d} \qquad\qquad\qquad\qquad\qquad\qquad\qquad \]
\[\qquad\qquad
 \leq \int d^d x_i \mathcal{A} \left( x, \sigma, F \right) \left( \prod_{j
   \in J} \mathbb{S} \left( L - \left| x_i - y_j \right| \right) \right)
   \left( \frac{\sigma \left| x_i - y_h \right|}{\left| y_a - y_b \right|}
   \right)^{\lambda} \left| x_i - y_h \right|^{- d} \]
holds, for all $y \in \left( \mathbb{E}_d \right)^J$.

Furthermore, the fact, that $\left( \mathcal{H} \left( F, i \right) \cup
\left\{ i \right\} \right) \in F$, implies that $\left| x_i - y_h \right| <
\sigma \left| y_e - y_f \right|$ holds, for all $x \in \left( \mathbb{E}_d
\right)^I$, such that $\mathcal{A} \left( x, \sigma, F \right)$ is nonzero,
hence that $\mathcal{A} \left( x, \sigma, F \right) \leq \mathbb{S} \left(
\sigma \left| y_e - y_f \right| - \left| x_i - y_h \right| \right)$ holds, for
all $x \in \left( \mathbb{E}_d \right)^I$, hence that
\[ \int d^d x_i \mathcal{A} \left( x, \sigma, F \right) \left( \prod_{j \in J}
   \mathbb{S} \left( L - \left| x_i - y_j \right| \right) \right) \left(
   \frac{\sigma \left| x_i - y_h \right|}{\left| y_a - y_b \right|}
   \right)^{\lambda} \left| x_i - y_h \right|^{- d} \qquad\qquad \]
\[ \qquad\qquad\qquad\qquad
 \leq \int d^d x_i \mathbb{S} \left( \sigma \left| y_e - y_f \right| -
   \left| x_i - y_h \right| \right) \left( \frac{\sigma \left| x_i - y_h
   \right|}{\left| y_a - y_b \right|} \right)^{\lambda} \left| x_i - y_h
   \right|^{- d} \]
\[ \qquad\qquad\qquad\qquad\quad\quad
 =\mathbb{A}_{d - 1} \left( \frac{1}{\lambda} \right) \left( \frac{\sigma^2
   \left| y_e - y_f \right|}{\left| y_a - y_b \right|} \right)^{\lambda} \]
holds, for all $y \in \left( \mathbb{E}_d \right)^J$.

(ii)  Alternatively, note that the fact, that $\left( \mathcal{H} \left( F, i
\right) \cup \left\{ i \right\} \right) \in F$, implies that \\
$\left(
\frac{\sigma \left| y_e - y_f \right|}{\left| x_i - y_h \right|}
\right)^{\lambda} \geq 1$ holds, for all $x \in \left( \mathbb{E}_d
\right)^I$, such that $\mathcal{A} \left( x, \sigma, F \right)$ is nonzero,
hence that
\[ \int d^d x_i \mathcal{A} \left( x, \sigma, F \right) \left( \prod_{j \in J}
   \mathbb{S} \left( L - \left| x_i - y_j \right| \right) \right) \left| x_i
   - y_h \right|^{- d} \qquad\qquad\qquad\qquad\qquad\qquad \]
\[ \qquad\qquad
 \leq \int d^d x_i \mathcal{A} \left( x, \sigma, F \right) \left( \prod_{j
   \in J} \mathbb{S} \left( L - \left| x_i - y_j \right| \right) \right)
   \left( \frac{\sigma \left| y_e - y_f \right|}{\left| x_i - y_h \right|}
   \right)^{\lambda} \left| x_i - y_h \right|^{- d} \]
holds, for all $y \in \left( \mathbb{E}_d \right)^J$.

Furthermore, the fact, that $\mathcal{H} \left( F, i \right) \in F$, in the
present case, implies that $\left| y_a - y_b \right| < \sigma \left| x_i - y_h
\right|$ holds, for all $x \in \left( \mathbb{E}_d \right)^I$, such that
$\mathcal{A} \left( x, \sigma, F \right)$ is nonzero, hence, that $\mathcal{A}
\left( x, \sigma, F \right) \leq \mathbb{S} \left( \left| x_i - y_h \right| -
\left( \frac{1}{\sigma} \right) \left| y_a - y_b \right| \right)$ holds, for
all $x \in \left( \mathbb{E}_d \right)^I$, hence that
\[ \int d^d x_i \mathcal{A} \left( x, \sigma, F \right) \left( \prod_{j \in J}
   \mathbb{S} \left( L - \left| x_i - y_j \right| \right) \right) \left(
   \frac{\sigma \left| y_e - y_f \right|}{\left| x_i - y_h \right|}
   \right)^{\lambda} \left| x_i - y_h \right|^{- d} \qquad\qquad\qquad \]
\[ \qquad\qquad\qquad\qquad
 \leq \int d^d x_i \mathbb{S} \left( \left| x_i - y_h \right| - \left(
   \frac{1}{\sigma} \right) \left| y_a - y_b \right| \right) \left(
   \frac{\sigma \left| y_e - y_f \right|}{\left| x_i - y_h \right|}
   \right)^{\lambda} \left| x_i - y_h \right|^{- d} \]
\[ \qquad\qquad\qquad\qquad\qquad
 =\mathbb{A}_{d - 1} \left( \frac{1}{\lambda} \right) \left( \frac{\sigma^2
   \left| y_e - y_f \right|}{\left| y_a - y_b \right|} \right)^{\lambda} \]
holds, for all $y \in \left( \mathbb{E}_d \right)^J$, in agreement, with the
bound, given by the first approach.

Hence, in the present case, $\nu_{\left\{ a, b \right\}} = \lambda$,
$\nu_{\left\{ e, f \right\}} = - \lambda$, and $\nu_{\Delta} = 0$, for all
other members, $\Delta$, of $\mathcal{Q} \left( J \right)$, hence
$\rho_{\left\{ a, b \right\}} = \mu_{\left\{ a, b \right\}} + \lambda$,
$\rho_{\left\{ e, f \right\}} = \mu_{\left\{ e, f \right\}} - \lambda$, and
$\rho_{\Delta} = \mu_{\Delta}$, for all other members, $\Delta$, of
$\mathcal{Q} \left( J \right)$.

Now, in the present case, $\# \left( \mathcal{H} \left( F, i \right) \right)
\geq 2$, and $\mathcal{H} \left( F, i \right) \subset J$, hence Lemma 4
implies, that, for every member, $B$, of $\mathcal{S} \left( F, J \right)$,
such that $B \neq \mathcal{H} \left( F, i \right)$, $\left\{ e, f \right\}$ is
a subset, of $B$, ifif $\left\{ a, b \right\}$ is a subset, of $B$.

Hence, in the present case, $\Gamma \left( \rho, B \right) = \Gamma \left(
\mu, B \right)$ holds, for every member, $B$, of $\mathbb{B} \left(
\mathcal{S} \left( F, J \right) \right)$, such that $B \neq \mathcal{H} \left(
F, i \right)$.

Now let $A$ be any member, of $\mathbb{B} \left( \mathcal{S} \left( F, J
\right) \right)$.  Then exactly one, of the four possibilities, $A
=\mathcal{H} \left( F, i \right)$, $\mathcal{H} \left( F, i \right) \subset
A$, $A \subset \mathcal{H} \left( F, i \right)$, and $A \cap \mathcal{H}
\left( F, i \right) = \emptyset$, holds.

If $A =\mathcal{H} \left( F, i \right)$, then $\left\{ a, b \right\}$
{\emph{is}} a subset, of $A$, and $\left\{ e, f \right\}$ is {\emph{not}} a
subset, of $A$, hence
\[ \Gamma \left( \rho,\mathcal{H} \left( F, i \right) \right) = \Gamma \left(
   \mu,\mathcal{H} \left( F, i \right) \right) + \lambda
   \qquad\qquad\qquad\qquad\qquad\qquad\qquad\qquad\quad \]
\[ \qquad\qquad\!\!\quad\quad
 = \Gamma \left( \mu,\mathcal{H} \left( F, i \right) \right) + \frac{1}{2}
   \left( d \left( \# \left( \mathcal{H} \left( F, i \right) \right) - 1
   \right) - \Gamma \left( \mu,\mathcal{H} \left( F, i \right) \right) \right)
\]
\[ \qquad = \left( \frac{1}{2} \right) d \left( \# \left( \mathcal{H} \left( F, i
   \right) \right) - 1 \right) + \left( \frac{1}{2} \right) \Gamma \left(
   \mu,\mathcal{H} \left( F, i \right) \right), \]
and this is $< d \left( \# \left( \mathcal{H} \left( F, i \right) \right) - 1
\right)$, by $\mathcal{T}_3$.

If $\mathcal{H} \left( F, i \right) \subset A$, then $\Gamma \left( \rho, A
\right) = \Gamma \left( \mu, A \right)$, hence $\mathcal{T}_1 \left( A
\right)$ implies, for the present case, that $\xi \left( \beta, i,\mathcal{H}
\left( F, i \right) \right) = d$, that $\Gamma \left( \rho, A \right) < d \left(
\# \left( A \right) - 1 \right)$.

And if $A \subset \mathcal{H} \left( F, i \right)$, or $A \cap \mathcal{H}
\left( F, i \right) = \emptyset$, then, again, $\Gamma \left( \rho, A \right)
= \Gamma \left( \mu, A \right)$, hence, $\mathcal{T}_2 \left( A \right)$
implies, that $\Gamma \left( \rho, A \right) < d \left( \# \left( A \right) -
1 \right)$, in the present case.

Hence $\rho$ is a good set of powers, for $\mathcal{S} \left( F, J \right)$,
in the present case.

\subsubsection*{Case A3b.  $\mathcal{H} \left( F, i \right)$
{\emph{is}} a member, of $F$, $\left( d - \xi \left(
\beta, i,\mathcal{H}\! \left( F, i \right) \right) \right) = 0$,
and $\mathcal{H} \left( F, i \right) = J$.}

The following two alternative approaches, (i), and (ii), give the same bound
on
\[ \int d^d x_i \mathcal{A} \left( x, \sigma, F \right) \left( \prod_{j \in J}
   \mathbb{S} \left( L - \left| x_i - y_j \right| \right) \right) \left| x_i
   - y_h \right|^{- \xi \left( \beta, i,\mathcal{H} \left( F, i \right)
   \right)} \qquad\qquad\qquad \]
\[ \qquad\qquad\qquad
 = \int d^d x_i \mathcal{A} \left( x, \sigma, F \right) \left( \prod_{j \in
   J} \mathbb{S} \left( L - \left| x_i - y_j \right| \right) \right) \left|
   x_i - y_h \right|^{- d} . \]
(i)  The fact, that $\mathcal{H} \left( F, i \right) \in F$, in the present
case, implies that $\left( \frac{\sigma \left| x_i - y_h \right|}{\left| y_a -
y_b \right|} \right)^{\lambda} \geq 1$ holds, for all $x \in \left(
\mathbb{E}_d \right)^I$, such that $\mathcal{A} \left( x, \sigma, F \right)$
is nonzero, hence
\[ \int d^d x_i \mathcal{A} \left( x, \sigma, F \right) \left( \prod_{j \in J}
   \mathbb{S} \left( L - \left| x_i - y_j \right| \right) \right) \left| x_i
   - y_h \right|^{- d} \qquad\qquad\qquad\qquad\qquad\qquad\quad \]
\[ \qquad\quad\quad\quad
 \leq \int d^d x_i \mathcal{A} \left( x, \sigma, F \right) \left( \prod_{j
   \in J} \mathbb{S} \left( L - \left| x_i - y_j \right| \right) \right)
   \left( \frac{\sigma \left| x_i - y_h \right|}{\left| y_a - y_b \right|}
   \right)^{\lambda} \left| x_i - y_h \right|^{- d} \]
\[ \qquad\qquad\qquad\qquad\quad\quad
 \leq \int d^d x_i \mathbb{S} \left( L - \left| x_i - y_h \right| \right)
   \left( \frac{\sigma \left| x_i - y_h \right|}{\left| y_a - y_b \right|}
   \right)^{\lambda} \left| x_i - y_h \right|^{- d} \]
\[\qquad\qquad\qquad\qquad\qquad\quad
 =\mathbb{A}_{d - 1} \left( \frac{1}{\lambda} \right) \left( \frac{\sigma
   L}{\left| y_a - y_b \right|} \right)^{\lambda} \]
holds, for all $y \in \left( \mathbb{E}_d \right)^J$.

(ii)  Alternatively, note that $\left( \frac{L}{\left| x_i - y_h \right|}
\right)^{\lambda} \geq 1$ holds, for all $x \in \left( \mathbb{E}_d
\right)^I$, such that
\[ \left( \prod_{j \in J} \mathbb{S} \left( L - \left| x_i - y_j \right|
   \right) \right) \]
is nonzero, hence
\[ \int d^d x_i \mathcal{A} \left( x, \sigma, F \right) \left( \prod_{j \in J}
   \mathbb{S} \left( L - \left| x_i - y_j \right| \right) \right) \left| x_i
   - y_h \right|^{- d} \qquad\qquad\qquad\qquad\qquad\quad\quad \]
\[ \qquad\quad\quad
 \leq \int d^d x_i \mathcal{A} \left( x, \sigma, F \right) \left( \prod_{j
   \in J} \mathbb{S} \left( L - \left| x_i - y_j \right| \right) \right)
   \left( \frac{L}{\left| x_i - y_h \right|} \right)^{\lambda} \left| x_i -
   y_h \right|^{- d} \]
holds, for all $y \in \left( \mathbb{E}_d \right)^J$.

Furthermore, the fact, that $\mathcal{H} \left( F, i \right) \in F$, in the
present case, implies that $\left| y_a - y_b \right| < \sigma \left| x_i - y_h
\right|$ holds, for all $x \in \left( \mathbb{E}_d \right)^I$, such that
$\mathcal{A} \left( x, \sigma, F \right)$ is nonzero, hence, that \\
$\mathcal{A}
\left( x, \sigma, F \right) \leq \mathbb{S} \left( \left| x_i - y_h \right| -
\left( \frac{1}{\sigma} \right) \left| y_a - y_b \right| \right)$ holds, for
all $x \in \left( \mathbb{E}_d \right)^I$, hence, that
\[ \int d^d x_i \mathcal{A} \left( x, \sigma, F \right) \left( \prod_{j \in J}
   \mathbb{S} \left( L - \left| x_i - y_j \right| \right) \right) \left(
   \frac{L}{\left| x_i - y_h \right|} \right)^{\lambda} \left| x_i - y_h
   \right|^{- d} \qquad\qquad\qquad\qquad \]
\[ \leq \int d^d x_i \mathcal{A} \left( x, \sigma, F \right) \left(
   \frac{L}{\left| x_i - y_h \right|} \right)^{\lambda} \left| x_i - y_h
   \right|^{- d} \qquad\qquad\qquad\qquad\!\qquad \]
\[ \qquad\qquad\qquad\quad\quad
 \leq \int d^d x_i \mathbb{S} \left( \left| x_i - y_h \right| - \left(
   \frac{1}{\sigma} \right) \left| y_a - y_b \right| \right) \left(
   \frac{L}{\left| x_i - y_h \right|} \right)^{\lambda} \left| x_i - y_h
   \right|^{- d} \]
\[ =\mathbb{A}_{d - 1} \left( \frac{1}{\lambda} \right) \left( \frac{\sigma
   L}{\left| y_a - y_b \right|} \right)^{\lambda} \]
holds, for all $y \in \left( \mathbb{E}_d \right)^J$, in agreement, with the
bound, given by the first approach.

Hence, in the present case, $\nu_{\left\{ a, b \right\}} = \lambda$, and
$\nu_{\Delta} = 0$, for all other members, $\Delta$, of $\mathcal{Q} \left( J
\right)$, hence $\rho_{\left\{ a, b \right\}} = \mu_{\left\{ a, b \right\}} +
\lambda$, and $\rho_{\Delta} = \mu_{\Delta}$, for all other members, $\Delta$,
of $\mathcal{Q} \left( J \right)$.

Now, in the present case, $\mathcal{H} \left( F, i \right) = J$, hence, if $A$
is any member, of $\mathcal{S} \left( F, J \right)$, such that $A \neq
\mathcal{H} \left( F, i \right)$, then $A \subset \mathcal{H} \left( F, i
\right)$ holds.

Now $\left\{ a, b \right\}$ is not a subset, of any member, of $\mathcal{L}
\left( \mathcal{S} \left( F, J \right),\mathcal{H} \left( F, i \right)
\right)$, hence, in the present case, if $A$ is any member, of $\mathcal{S}
\left( F, J \right)$, such that $A \neq \mathcal{H} \left( F, i \right)$, then
$\left\{ a, b \right\}$ is {\emph{not}} a subset, of $A$, hence $\Gamma \left(
\rho, A \right) = \Gamma \left( \mu, A \right)$.

Hence, in the present case, if $A$ is any member, of $\mathbb{B} \left(
\mathcal{S} \left( F, J \right) \right)$, such that $A \neq \mathcal{H} \left(
F, i \right)$, then $\Gamma \left( \rho, A \right) = \Gamma \left( \mu, A
\right)$, and $A \subset \mathcal{H} \left( F, i \right)$ both hold, hence
$\mathcal{T}_2 \left( A \right)$ implies, that $\Gamma \left( \rho, A \right)
< d \left( \# \left( A \right) - 1 \right)$.

Furthermore,
\[ \Gamma \left( \rho,\mathcal{H} \left( F, i \right) \right) = \Gamma \left(
   \mu,\mathcal{H} \left( F, i \right) \right) + \lambda
   \qquad\qquad\qquad\qquad\qquad\qquad\qquad\qquad\quad \]
\[ \qquad\qquad\!\qquad
 = \Gamma \left( \mu,\mathcal{H} \left( F, i \right) \right) + \frac{1}{2}
   \left( d \left( \# \left( \mathcal{H} \left( F, i \right) \right) - 1
   \right) - \Gamma \left( \mu,\mathcal{H} \left( F, i \right) \right) \right)
\]
\[ \qquad\,
 = \left( \frac{1}{2} \right) d \left( \# \left( \mathcal{H} \left( F, i
   \right) \right) - 1 \right) + \left( \frac{1}{2} \right) \Gamma \left(
   \mu,\mathcal{H} \left( F, i \right) \right), \]
and this is $< d \left( \# \left( \mathcal{H} \left( F, i \right) \right) - 1
\right)$, by $\mathcal{T}_3$.
\enlargethispage{0.1cm}

Hence $\rho$ is a good set of powers, for $\mathcal{S} \left( F, J \right)$,
in the present case, and all cases, where $\mathcal{H} \left( F, i \right) \in
F$, have now been dealt with.  Thus, in every case, where $\mathcal{H} \left(
F, i \right) \in F$ holds, I have found a finite real number, $T \geq 0$, and a
set of powers, $\nu$, for $J$, such that
\[ \int d^d x_i \mathcal{A} \left( x, \sigma, F \right) \left( \prod_{j \in J}
   \mathbb{S} \left( L - \left| x_i - y_j \right| \right) \right) \left(
   \prod_{j \in \mathcal{H} \left( F, i \right)} \left| x_i - y_j \right|^{-
   \beta_{\left\{ i, j \right\}}} \right) \leq T \Psi \left( y, \nu \right) \]
holds, for all $y \in \left( \mathbb{E}_d \right)^J$, and such that, if a set
of powers, $\rho$, for $J$, is defined, by $\rho_{\Delta} \equiv \left(
\mu_{\Delta} + \nu_{\Delta} \right)$, for all $\Delta \in \mathcal{Q} \left( J
\right)$, then $\rho$ is a good set of powers, for $\mathcal{S} \left( F, J
\right)$.

\subsection*{Case B.  $\mathcal{H} \left( F, i \right)$ is
{\emph{not}} a member, of $F$.}

Note that the assumption, that $\mathcal{H} \left( F, i \right)$ is
{\emph{not}} a member, of $F$, implies that \\
$\# \left( \mathcal{H} \left( F, i
\right) \right) \geq 2$.

Now, by Lemma 6, the assumption, that $\mathcal{H} \left( F, i \right)$ is
{\emph{not}} a member, of $F$, implies that there exist two members, $a$, and
$b$, of $\mathcal{H} \left( F, i \right)$, such that $\mathcal{K} \left(
F,\mathcal{C} \left( F, \left\{ i \right\} \right), a \right) \neq$ \\
$\mathcal{K} \left(
F,\mathcal{C} \left( F, \left\{ i \right\} \right), b \right)$.

Let $j$ be any member, of $\mathcal{H} \left( F, i \right)$, and let $a$, and
$b$, be any two members, of $\mathcal{H} \left( F, i \right)$, such that
$\mathcal{K} \left( F,\mathcal{C} \left( F, \left\{ i \right\} \right), a
\right) \neq \mathcal{K} \left( F,\mathcal{C} \left( F, \left\{ i \right\}
\right), b \right)$.  Then $\left\{ i, j \right\}$, and $\left\{ a, b \right\}$,
are both members, of $\mathcal{W} \left( F,\mathcal{C} \left( F, \left\{ i
\right\} \right) \right)$, hence, by Lemma 11,
\[ \sigma \left( \frac{1 + 2 \sigma}{\sigma} \right)^{\left( \# \left(
   \mathcal{P} \left( F,\mathcal{C} \left( F, \left\{ i \right\} \right)
   \right) \right) - 1 \right)} \left| x_i - y_j \right| \geq \left| y_a - y_b
   \right| \]
and
\[ \sigma \left( \frac{1 + 2 \sigma}{\sigma} \right)^{\left( \# \left(
   \mathcal{P} \left( F,\mathcal{C} \left( F, \left\{ i \right\} \right)
   \right) \right) - 1 \right)} \left| y_a - y_b \right| \geq \left| x_i - y_j
   \right| \]
both hold, for all $x \in \left( \mathbb{E}_d \right)^I$, such that
$\mathcal{A} \left( x, \sigma, F \right) \neq 0$.

Hence
\[ \mathcal{A} \left( x, \sigma, F \right) \left( \sigma \left( \frac{1 + 2
   \sigma}{\sigma} \right)^{\left( \# \left( \mathcal{P} \left( F,\mathcal{C}
   \left( F, \left\{ i \right\} \right) \right) \right) - 1 \right)}
   \right)^{\mathbb{M} \left( \beta_{\left\{ i, j \right\}} \right)} \left(
   \frac{\left| x_i - y_j \right|}{\left| y_a - y_b \right|}
   \right)^{\beta_{\left\{ i, j \right\}}} \geq \mathcal{A} \left( x, \sigma,
   F \right) \]
holds, for all $x \in \left( \mathbb{E}_d \right)^I$.

Now choose two members, $a$, and $b$, of $\mathcal{H} \left( F, i \right)$,
such that $\mathcal{K} \left( F,\mathcal{C} \left( F, \left\{ i \right\}
\right), a \right) \neq \mathcal{K} \left( F,\mathcal{C} \left( F, \left\{ i
\right\} \right), b \right)$.  Let
\[ \Omega \equiv \prod_{j \in \mathcal{H} \left( F, i \right)}
   \left( \left( \sigma \left( \frac{1 + 2 \sigma}{\sigma} \right)^{\left( \#
   \left( \mathcal{P} \left( F,\mathcal{C} \left( F, \left\{ i \right\}
   \right) \right) \right) - 1 \right)} \right)^{\mathbb{M} \left(
   \beta_{\left\{ i, j \right\}} \right)} \right) . \]
Then
\[ \int d^d x_i \mathcal{A} \left( x, \sigma, F \right) \left( \prod_{j \in J}
   \mathbb{S} \left( L - \left| x_i - y_j \right| \right) \right) \left(
   \prod_{j \in \mathcal{H} \left( F, i \right)} \left| x_i - y_j \right|^{-
   \beta_{\left\{ i, j \right\}}} \right) \qquad\qquad\qquad\qquad \]
\[ \leq \int d^d x_i \mathcal{A} \left( x, \sigma, F \right) \left( \prod_{j
   \in \mathcal{H} \left( F, i \right)} \left| x_i - y_j \right|^{-
   \beta_{\left\{ i, j \right\}}} \right) \qquad\qquad\qquad\qquad\qquad \]
\[ \leq \Omega \int d^d x_i \mathcal{A} \left( x, \sigma, F \right) \left(
   \prod_{j \in \mathcal{H} \left( F, i \right)} \left| y_a - y_b \right|^{-
   \beta_{\left\{ i, j \right\}}} \right) \quad \]
\[ \qquad\qquad
 = \Omega \int d^d x_i \mathcal{A} \left( x, \sigma, F \right) \left| y_a -
   y_b \right|^{- \xi \left( \beta, i,\mathcal{H} \left( F, i \right) \right)}
\]
\[ \qquad\qquad\qquad\qquad\qquad\qquad
 = \Omega \left| y_a - y_b \right|^{- \xi \left( \beta, i,\mathcal{H} \left(
   F, i \right) \right)} \int d^d x_i \mathcal{A} \left( x, \sigma, F \right)
\]
holds, for all $y \in \left( \mathbb{E}_d \right)^J$.

Now, if $j$ is any member, of $\mathcal{H} \left( F, i \right)$,
\[ \sigma \left( \frac{1 + 2 \sigma}{\sigma} \right)^{\left( \# \left(
   \mathcal{P} \left( F,\mathcal{C} \left( F, \left\{ i \right\} \right)
   \right) \right) - 1 \right)} \left| y_a - y_b \right| \geq \left| x_i - y_j
   \right| \]
holds, in the present case, for all $x \in \left( \mathbb{E}_d \right)^I$,
such that $\mathcal{A} \left( x, \sigma, F \right)$ is nonzero, hence
\[ \mathcal{A} \left( x, \sigma, F \right) \leq \mathbb{S} \left( \sigma
   \left( \frac{1 + 2 \sigma}{\sigma} \right)^{\left( \# \left( \mathcal{P}
   \left( F,\mathcal{C} \left( F, \left\{ i \right\} \right) \right) \right) -
   1 \right)} \left| y_a - y_b \right| - \left| x_i - y_j \right| \right) \]
holds, for all $x \in \left( \mathbb{E}_d \right)^I$.

Let $\mathbb{N} \left( F, i \right) \equiv \left( \# \left( \mathcal{P}
\left( F,\mathcal{C} \left( F, \left\{ i \right\} \right) \right) \right) - 1
\right)$.  (Note that the assumption, that $\mathcal{H} \left( F, i \right)$
is {\emph{not}} a member, of $F$, in the present case, implies that
$\mathbb{N} \left( F, i \right)$ is equal to \\
$\# \left( \mathcal{P} \left(
\mathcal{S} \left( F, J \right),\mathcal{H} \left( F, i \right) \right)
\right)$, in the present case.)

Choose a member, $k$, of $\mathcal{H} \left( F, i \right)$.  Then
\[ \int d^d x_i \mathcal{A} \left( x, \sigma, F \right) \left( \prod_{j \in J}
   \mathbb{S} \left( L - \left| x_i - y_j \right| \right) \right) \left(
   \prod_{j \in \mathcal{H} \left( F, i \right)} \left| x_i - y_j \right|^{-
   \beta_{\left\{ i, j \right\}}} \right) \qquad\qquad\qquad\qquad \]
\[ \leq \Omega \left| y_a - y_b \right|^{- \xi \left( \beta, i,\mathcal{H}
   \left( F, i \right) \right)} \int d^d x_i \mathcal{A} \left( x, \sigma, F
   \right) \leq \qquad\qquad\qquad\qquad\qquad\qquad\,\, \]
\[ \qquad\qquad\quad
 \leq \Omega \left| y_a - y_b \right|^{- \xi \left( \beta, i,\mathcal{H}
   \left( F, i \right) \right)} \int d^d x_i \mathbb{S} \left( \sigma \left(
   \frac{1 + 2 \sigma}{\sigma} \right)^{\mathbb{N} \left( F, i \right)}
   \left| y_a - y_b \right| - \left| x_i - y_k \right| \right) \]
\[ \quad\,\,\,\,
 = \Omega \frac{\mathbb{A}_{d - 1}}{d} \left( \sigma \left( \frac{1 + 2
   \sigma}{\sigma} \right)^{\mathbb{N} \left( F, i \right)} \right)^d \left|
   y_a - y_b \right|^{\left( d - \xi \left( \beta, i,\mathcal{H} \left( F, i
   \right) \right) \right)} . \]
Hence, in the present case, $\nu_{\left\{ a, b \right\}} = \left( \xi \left(
\beta, i,\mathcal{H} \left( F, i \right) \right) - d \right)$, and
$\nu_{\Delta} = 0$, for all other members, $\Delta$, of $\mathcal{Q} \left( J
\right)$, hence $\rho_{\left\{ a, b \right\}} = \mu_{\left\{ a, b \right\}} +
\left( \xi \left( \beta, i,\mathcal{H} \left( F, i \right) \right) - d
\right)$, and $\rho_{\Delta} = \mu_{\Delta}$, for all other members, $\Delta$,
of $\mathcal{Q} \left( J \right)$.

Now let $A$ be any member, of $\mathbb{B} \left( \mathcal{S} \left( F, J
\right) \right)$.  Now $\mathcal{H} \left( F, i \right)$ is a member, of
$\mathcal{S} \left( F, J \right)$, hence exactly one of the three
possibilities, $\mathcal{H} \left( F, i \right) \subseteq A$, $A \subset
\mathcal{H} \left( F, i \right)$, and $A \cap \mathcal{H} \left( F, i \right)
= \emptyset$, holds.

If $\mathcal{H} \left( F, i \right) \subseteq A$, then $\left\{ a, b \right\}$
{\emph{is}} a subset, of $A$, hence $\Gamma \left( \rho, A \right) = \Gamma
\left( \mu, A \right) +$
\\$\xi \left( \beta, i,\mathcal{H} \left( F, i \right)
\right) - d$, hence $\mathcal{T}_1 \left( A \right)$ implies, that $\Gamma
\left( \rho, A \right) < d \left( \# \left( A \right) - 1 \right)$.

Now suppose $A \subset \mathcal{H} \left( F, i \right)$.  Now, by Lemma 7, the
assumption, that $\mathcal{H} \left( F, i \right)$ is {\emph{not}} a member, of
$F$, in the present case, implies, that $\mathcal{P} \left( \mathcal{S} \left(
F, J \right),\mathcal{H} \left( F, i \right) \right) =$ \\
$\left( \mathcal{P}
\left( F,\mathcal{C} \left( F, \left\{ i \right\} \right) \right) \vdash
\left\{ \left\{ i \right\} \right\} \right)$ holds, in the present case.
Hence, the facts, that $a$, and $b$, are members, of $\mathcal{H} \left( F, i
\right)$, and the fact, that $\mathcal{K} \left( F,\mathcal{C} \left( F,
\left\{ i \right\} \right), a \right) \neq \mathcal{K} \left( F,\mathcal{C}
\left( F, \left\{ i \right\} \right), b \right)$, together imply, that
$\mathcal{K} \left( \mathcal{S} \left( F, J \right),\mathcal{H} \left( F, i
\right), a \right) \neq \mathcal{K} \left( \mathcal{S} \left( F, J
\right),\mathcal{H} \left( F, i \right), b \right)$, hence, since $A \subset
\mathcal{H} \left( F, i \right)$, that $\left\{ a, b \right\}$ is {\emph{not}}
a subset, of $A$.  Hence $\Gamma \left( \rho, A \right) = \Gamma \left( \mu, A
\right)$, hence $\mathcal{T}_2 \left( A \right)$ implies, that $\Gamma \left(
\rho, A \right) < d \left( \# \left( A \right) - 1 \right)$.

Now suppose $A \cap \mathcal{H} \left( F, i \right) = \emptyset$.  Then
$\left\{ a, b \right\}$ is {\emph{not}} a subset, of $A$, hence $\Gamma \left(
\rho, A \right) = \Gamma \left( \mu, A \right)$, and $\mathcal{T}_2 \left( A
\right)$, again, implies, that $\Gamma \left( \rho, A \right) < d \left( \#
\left( A \right) - 1 \right)$.

Hence $\rho$ is a good set of powers, for $\mathcal{S} \left( F, J \right)$,
in the present case.

Thus, in the case where $\mathcal{H} \left( F, i \right) \notin F$ holds, I have
now found a finite real number, $T \geq 0$, and a set of powers, $\nu$, for
$J$, such that
\[ \int d^d x_i \mathcal{A} \left( x, \sigma, F \right) \left( \prod_{j \in J}
   \mathbb{S} \left( L - \left| x_i - y_j \right| \right) \right) \left(
   \prod_{j \in \mathcal{H} \left( F, i \right)} \left| x_i - y_j \right|^{-
   \beta_{\left\{ i, j \right\}}} \right) \leq T \Psi \left( y, \nu \right) \]
holds, for all $y \in \left( \mathbb{E}_d \right)^J$, and such that, if a set
of powers, $\rho$, for $J$, is defined, by $\rho_{\Delta} \equiv \left(
\mu_{\Delta} + \nu_{\Delta} \right)$, for all $\Delta \in \mathcal{Q} \left( J
\right)$, then $\rho$ is a good set of powers, for $\mathcal{S} \left( F, J
\right)$.

Every case has now been covered.  Hence I have shown, in this step, that if
$\beta$ is any good set of powers, for $F$, such that $\beta_{\left\{ i, j
\right\}} = 0$ holds, for all $j \in \left( J \vdash \mathcal{H} \left( F, i
\right) \right)$, and $\mu$ is the restriction, of $\beta$, to the domain,
$\mathcal{Q} \left( J \right)$, then there exists a finite real number, $T \geq
0$, and a good set of powers, $\rho$, for $\mathcal{S} \left( F, J \right)$,
such that
\[ \int d^d x_i \mathcal{A} \left( x, \sigma, F \right) \mathcal{B} \left( x,
   L \right) \Psi \left( x, \beta \right) =
   \qquad\qquad\qquad\qquad\qquad\qquad\qquad\qquad\qquad\qquad\!\qquad \]
\[ = \sum_{E \in \mathbb{K} \left( F, J \right)} \Theta \left( F, y, \sigma,
   E \right) \mathcal{A} \left( y, \sigma, E \right) \mathcal{B} \left( y, L
   \right) \Psi \left( y, \mu \right)\, \times
    \qquad\qquad\qquad\qquad\quad\quad \]
\[ \qquad\qquad\qquad\quad
 \times \int d^d x_i \mathcal{A} \left( x, \sigma, F \right) \left( \prod_{j
   \in J} \mathbb{S} \left( L - \left| x_i - y_j \right| \right) \right)
   \left( \prod_{j \in \mathcal{H} \left( F, i \right)} \left| x_i - y_j
   \right|^{- \beta_{\left\{ i, j \right\}}} \right) \]
\[ \qquad\qquad\qquad\quad\,\,
 \leq T \sum_{E \in \mathbb{K} \left( F, J \right)} \Theta \left( F, y,
   \sigma, E \right) \mathcal{A} \left( y, \sigma, E \right) \mathcal{B}
   \left( y, L \right) \Psi \left( y, \rho \right) \]
holds, for all $y \in \left( \mathbb{E}_d \right)^J$, where $\mathbb{K}
\left( F, J \right)$ was defined, on page 15, as the set, whose members are
all the high greenwoods, $E$, of $J$, such that $\mathcal{S} \left( F, J
\right) \subseteq E$ holds, and such that, for each member, $A$, of
$\mathbb{B} \left( \mathcal{P} \left( F, J \right) \right)$, $\mathcal{L}
\left( E, A \right) =\mathcal{L} \left( F, A \right)$ holds, and for each
member, $E$, of $\mathbb{K} \left( F, J \right)$, $\Theta \left( F, y,
\sigma, E \right)$ was defined, on page 18, as
\[ \Theta \left( F, y, \sigma, E \right) \equiv \qquad\qquad\qquad\qquad\qquad\qquad\qquad\qquad
\qquad\qquad\qquad\qquad\qquad\qquad\qquad \]
\[ \qquad\!\!\!\qquad\equiv \prod_{B \in \left( E\, \vdash \mathcal{S} \left( F, J \right)
   \right)} \prod_{\begin{array}{c}\\[-25pt]\scriptstyle{
     \Delta \equiv \left\{ a, b \right\} \in \mathcal{W} \left( E, B \right)}\\[-8pt]
     \scriptstyle{e \in B}\\[-8pt]
     \scriptstyle{f \in \left( \mathcal{C} \left( E, B \right)\, \vdash B \right)
   }\end{array}} \mathbb{S} \left( \left| y_a - y_b \right| - \left(
   \frac{1}{\sigma} \right) \left( \frac{\sigma}{1 + 2 \sigma} \right)^{n \left( F,
   B \right)} \left| y_e - y_f \right| \right), \]
where, for each nonempty, {\emph{strict}} subset, $B$, of $I$, $n \left( F, B
\right)$ was defined, on page 18, as
\[ n \left( F, B \right) \equiv \left( \# \left( \mathcal{P} \left(
   F,\mathcal{C} \left( F, B \right) \right) \right) - 1 \right) . \]
In the final step, I shall find, for each member, $E$, of $\mathbb{K} \left(
F, J \right)$, a finite real number, $U_E \geq 0$, and a good set of powers,
$\eta_E$, for $E$, such that
\[ \Theta \left( F, y, \sigma, E \right) \mathcal{A} \left( y, \sigma, E
   \right) \mathcal{B} \left( y, L \right) \Psi \left( y, \rho \right) \leq
   U_E \mathcal{A} \left( y, \sigma, E \right) \mathcal{B} \left( y, L \right)
   \Psi \left( y, \eta_E \right) \]
holds, for all $y \in \left( \mathbb{E}_d \right)^J$.

Choose a finite real number, $\tau$, such that $\tau > 0$, and for each
ordered pair, $\left( E, B \right)$, of a member, $E$, of $\mathbb{K} \left(
F, J \right)$, such that $\left( E \vdash \mathcal{S} \left( F, J \right)
\right)$ is nonempty, and a member, $B$, of $\left( E \vdash \mathcal{S}
\left( F, J \right) \right)$, choose two members, $a \left( E, B \right)$, and
$b \left( E, B \right)$, of $B$, such that $\mathcal{K} \left( E, B, a \left(
E, B \right) \right) \neq \mathcal{K} \left( E, B, b \left( E, B \right)
\right)$, and a member, $e \left( E, B \right)$, of $B$, and a member, $f
\left( E, B \right)$, of $\left( \mathcal{C} \left( E, B \right) \vdash B
\right)$.

Note that, for all $s \in \mathbb{R}$, $\mathbb{U} \left( s \right)$ was
defined, on page 6, as
\[ \mathbb{U} \left( s \right) \equiv \left\{ \begin{array}{ccc}
     s & \mathrm{if} & s \geq 0\\
     0 & \mathrm{if} & s \leq 0
   \end{array} \right\} . \]
Then, for each ordered pair, $\left( E, B \right)$, of a member, $E$, of
$\mathbb{K} \left( F, J \right)$, such that \\
$\left( E \vdash \mathcal{S}
\left( F, J \right) \right)$ is nonempty, and a member, $B$, of $\left( E
\vdash \mathcal{S} \left( F, J \right) \right)$,
\[ \left( \sigma \left( \frac{1 + 2 \sigma}{\sigma} \right)^{n \left( F, B
   \right)} \left( \frac{\left| y_{a \left( E, B \right)} - y_{b \left( E, B
   \right)} \right|}{\left| y_{e \left( E, B \right)} - y_{f \left( E, B
   \right)} \right|} \right) \right)^{\mathbb{U} \left( \tau + \Gamma \left(
   \rho, B \right) - d \left( \# \left( B \right) - 1 \right) \right)} \geq 1
\]
holds, for all $y \in \left( \mathbb{E}_d \right)^J$, such that $\Theta
\left( F, y, \sigma, E \right) \neq 0$, since $\mathbb{U} \left( s \right)
\geq 0$, for all $s \in \mathbb{R}$.

For each member, $E$, of $\mathbb{K} \left( F, J \right)$, let $\eta_E$ be
the set of powers, for $E$, such that
\[ \Psi \left( y, \eta_E \right) = \Psi \left( y, \rho \right) \left( \prod_{B
   \in \left( E\, \vdash \mathcal{S} \left( F, J \right) \right)} \left(
   \frac{\left| y_{a \left( E, B \right)} - y_{b \left( E, B \right)}
   \right|}{\left| y_{e \left( E, B \right)} - y_{f \left( E, B \right)}
   \right|} \right)^{\mathbb{U} \left( \tau + \Gamma \left( \rho, B \right) -
   d \left( \# \left( B \right) - 1 \right) \right)} \right) \]
holds, for all $y \in \left( \mathbb{E}_d \right)^J$, and let
\[ U_E \equiv \left( \prod_{B \in \left( E\, \vdash \mathcal{S} \left( F, J
   \right) \right)} \left( \sigma \left( \frac{1 + 2 \sigma}{\sigma}
   \right)^{n \left( F, B \right)} \right)^{\mathbb{U} \left( \tau + \Gamma
   \left( \rho, B \right) - d \left( \# \left( B \right) - 1 \right) \right)}
   \right) . \]
Then
\[ \Theta \left( F, y, \sigma, E \right) \mathcal{A} \left( y, \sigma, E
   \right) \mathcal{B} \left( y, L \right) \Psi \left( y, \rho \right)
   \qquad\qquad\qquad\qquad\qquad\qquad\qquad\qquad\qquad \]
\[ \qquad\qquad
 \leq U_E \Theta \left( F, y, \sigma, E \right) \mathcal{A} \left( y,
   \sigma, E \right) \mathcal{B} \left( y, L \right) \Psi \left( y, \eta_E
   \right) \]
\[ \qquad\qquad\qquad\qquad\qquad\qquad\qquad\qquad\qquad\qquad
 \leq U_E \mathcal{A} \left( y, \sigma, E \right) \mathcal{B} \left( y, L
   \right) \Psi \left( y, \eta_E \right) \]
holds, for all $y \in \left( \mathbb{E}_d \right)^J$.

Now let $E$ be any member, of $\mathbb{K} \left( F, J \right)$, such that
$\left( E \vdash \mathcal{S} \left( F, J \right) \right)$ is nonempty, let $A$
be any member, of $E$, and let $B$ be any member, of $\left( E \vdash
\mathcal{S} \left( F, J \right) \right)$, such that $B \neq A$.  Then, by
Lemma 3, $\left\{ e \left( E, B \right), f \left( E, B \right) \right\}$ is a
subset, of $A$, ifif $\left\{ a \left( E, B \right), b \left( E, B \right)
\right\}$ is a subset, of $A$.

Hence, if $E$ is any member, of $\mathbb{K} \left( F, J \right)$, and $A$ is
any member, of $\mathcal{S} \left( F, J \right)$, then $\Gamma \left( \eta_E, A
\right) = \Gamma \left( \rho, A \right)$, hence, the fact, that $\rho$ is a
good set of powers, for $\mathcal{S} \left( F, J \right)$, implies that $\Gamma
\left( \eta_E, A \right) < d \left( \# \left( A \right) - 1 \right)$.
\enlargethispage{0.1cm}

And if $E$ is any member, of $\mathbb{K} \left( F, J \right)$, such that
$\left( E \vdash \mathcal{S} \left( F, J \right) \right)$ is nonempty, and $A$
is any member, of $\left( E \vdash \mathcal{S} \left( F, J \right) \right)$,
then
\[ \Gamma \left( \eta_E, A \right) = \Gamma \left( \rho, A \right)
   -\mathbb{U} \left( \tau + \Gamma \left( \rho, A \right) - d \left( \#
   \left( A \right) - 1 \right) \right) \qquad\qquad\qquad\qquad\qquad\qquad \]
\[ \qquad = \left\{ \begin{array}{cc}
     d \left( \# \left( A \right) - 1 \right) - \tau & \qquad
     \mathrm{if}\,\,\, \Gamma \left(
     \rho, A \right) \geq \left( d \left( \# \left( A \right) - 1 \right) -
     \tau \right)\\
     \Gamma \left( \rho, A \right) & \qquad
     \mathrm{if}\,\,\, \Gamma \left( \rho, A \right)
     \leq \left( d \left( \# \left( A \right) - 1 \right) - \tau \right)
   \end{array} \right\} \]
\[ \qquad\qquad
 = \min \left( \left( d \left( \# \left( A \right) - 1 \right) - \tau
   \right), \Gamma \left( \rho, A \right) \right) \]
\[ \qquad\qquad\qquad\qquad\qquad\qquad
 \leq \left( d \left( \# \left( A \right) - 1 \right) - \tau \right) \]
\[ \qquad\qquad\qquad\qquad\qquad\qquad\qquad\qquad\qquad\qquad\qquad\qquad
 < d \left( \# \left( A \right) - 1 \right) . \]
Hence $\eta_E$ is a good set of powers, for $E$.

For each member, $E$, of $\mathbb{K} \left( F, J \right)$, let $C_E \equiv
STU_E$.

Then, for any good set of powers, $\alpha$, for $F$, I have found a map,
$\eta$, whose domain is $\mathbb{K} \left( F, J \right)$, and a map, $C$,
whose domain is $\mathbb{K} \left( F, J \right)$, such that, for each member,
$E$, of $\mathbb{K} \left( F, J \right)$, $\eta_E$ is a good set of powers,
for $E$, and for each member, $E$, of $\mathbb{K} \left( F, J \right)$, $C_E$
is a finite real number, $\geq 0$, and for all maps, $y$, whose domain is $J$,
and whose range is a subset, of $\mathbb{E}_d$, the following inequality, in
which $x \equiv y \cup \left\{ \left( i, x_i \right) \right\}$, holds.
\[ \int d^d x_i \mathcal{A} \left( x, \sigma, F \right) \mathcal{B} \left( x,
   L \right) \Psi \left( x, \alpha \right) \leq \sum_{E \in\, \mathbb{K} \left(
   F, J \right)} C_E \mathcal{A} \left( y, \sigma, E \right) \mathcal{B}
   \left( y, L \right) \Psi \left( y, \eta_E \right) \]
Hence the proof, of the Cluster Convergence Theorem, is now complete.

\vspace{0.4cm}

There is a second case, of the main Proposition, of the Cluster Convergence
Theorem, which differs, from that, stated on page 15, only in that $\alpha$ is
assumed to be a good set of powers, only for $\left( F \vdash \left\{
\mathcal{U} \left( F \right) \right\} \right)$, rather than for $F$, and in
that, for each member, $E$, of $\mathbb{K} \left( F, J \right)$, $\eta_E$ is
proved to be a good set of powers, only for $\left( E \vdash \left\{
\mathcal{U} \left( E \right) \right\} \right)$, rather than for $E$.  The
proof, of this second case, may be obtained from the proof I have given, of
the Proposition, stated on page 15, by checking, that, in the proof I have
given, no use is made, of the assumed bound, $\Gamma \left( \alpha, I \right)
< d \left( \# \left( I \right) - 1 \right)$, other than to obtain the bounds,
$\Gamma \left( \eta_E, J \right) < d \left( \# \left( J \right) - 1 \right)$,
for each member, $E$, of $\mathbb{K} \left( F, J \right)$.

\vspace{0.4cm}

In a paper, to follow this one, I shall use the Cluster Convergence Theorem,
to prove part of a BPHZ convergence theorem, directly in Euclidean position
space.

\appendix
\section{\hspace{-0.69cm}ppendix.}
\label{Appendix}

In this appendix, I will display, for any ordered quintuple, $\left( F, \sigma,
d, z, L \right)$, of a high greenwood, $F$, a real number, $\sigma$, such that
$0 < \sigma < 1$, an integer, $d \geq\! 1$,
a point, $z$, of $\mathbb{E}_d$, and
a real number, $L > 0$,
a subset, $\mathbb{Z}\! \left( F, \sigma, d, z, L
\right)$, of $\left( \mathbb{E}_d \right)^{\mathcal{U} \left( F \right)}$,
such that $\mathbb{Z}\! \left( F, \sigma, d, z, L \right)$ has nonzero $\left(
\left( \# \left( \mathcal{U} \left( F \right) \right) \right) d
\right)$-volume, $\mathcal{F} \left( x, \sigma \right) = F$ holds, for all $x
\in \mathbb{Z} \left( F, \sigma, d, z, L \right)$, and $\left| x_i - z
\right| \leq L$ holds, for every member, $i$, of $\mathcal{U} \left( F
\right)$, for all $x \in \mathbb{Z} \left( F, \sigma, d, z, L \right)$.

For any ordered triple, $\left( d, z, L \right)$, of an integer, $d \geq 1$, a
point, $z$, of $\mathbb{E}_d$, and a real number, $L > 0$, I define
$\mathbb{V} \left( d, z, L \right)$, the
$d${\emph{-ball, of centre, }}$z${\emph{, and radius, }}$L$,
to be the set, whose members are all the points,
$s$, of $\mathbb{E}_d$, such that $\left| s - z \right| \leq L$.  A
$d${\emph{-ball}} is a subset, $A$, of $\mathbb{E}_d$, such that there is a
point, $z$, of $\mathbb{E}_d$, and a real number, $L > 0$, such that $A
=\mathbb{V} \left( d, z, L \right)$.  Note that every $d$-ball has nonzero
$d$-volume.

For each integer, $N \geq 1$, let $\mathcal{I} \left( N \right)$ be the set,
whose members are the $N$ integers, $\geq 1$, and $\leq N$.

Note that a {\emph{bijection}} was defined, on page 3, to be a map, $M$, such
that if $\left( a, b \right) \in M$, and $\left( e, f \right) \in M$, then $b
= f$ implies $a = e$.

I will first construct a map, $w$, and a map, $R$, with the following five
properties, $\mathcal{Y}_1$, $\mathcal{Y}_2$, $\mathcal{Y}_3$,
$\mathcal{Y}_4$, and $\mathcal{Y}_5$.

\vspace{0.18cm}

\noindent $\mathcal{Y}_1$.  $\mathcal{D} \left( w \right) =\mathcal{D} \left( R \right)
= F$.

\vspace{0.18cm}

\noindent $\mathcal{Y}_2$.  $\mathcal{R} \left( w \right)$ is a subset, of
$\mathbb{E}_d$, and every member, of $\mathcal{R} \left( R \right)$, is a
real number, $> 0$.

\vspace{0.18cm}

\noindent $\mathcal{Y}_3$.  $w_{\begin{array}{c}\\[-25pt]\scriptstyle{
{\!\!\!\mathcal{U} \left( F \right)\!\!\!\!}
   }\end{array}} = z$, and $R_{\mathcal{U}
\left( F \right)} = L$.

\vspace{0.18cm}

\noindent $\mathcal{Y}_4$.  If $A$, and $B$, are members, of $F$, such that $A\! \subseteq
\!B$, then $\mathbb{V}\! \left( d, w_A, R_A \right)\! \subseteq\! \mathbb{V}\! \left(
d, w_B, R_B \right)$.

\vspace{0.18cm}

\noindent $\mathcal{Y}_5$.  If $A$ is any member, of $\mathbb{B} \left( F \right)$, and
$x$ is any map, such that $\mathcal{D} \left( x \right) =\mathcal{U} \left( F
\right)$, $\mathcal{R} \left( x \right) \subseteq \mathbb{E}_d$, and for
every member, $B$, of $\mathcal{P} \left( F, A \right)$, $\mathcal{R} \left(
\mathcal{N} \left( x, B \right) \right) \subseteq \mathbb{V} \left( d, w_B,
R_B \right)$ holds, then $\mathcal{P} \left( \mathcal{F} \left( \mathcal{N}
\left( x, A \right), \sigma \right), A \right) =\mathcal{P} \left( F, A
\right)$ holds.

\vspace{0.18cm}

For each member, $A$, of $\mathbb{B} \left( F \right)$, I define $N_A \equiv
\# \left( \mathcal{P} \left( F, A \right) \right)$.
\enlargethispage{0.4cm}

Choose a real number, $\tau$, such that $0 < \tau < \min \left( \left( \frac{1
- \sigma}{2 \sigma^2} \right), 1 \right)$.  (For example, $\tau = \left(
\frac{1 - \sigma}{2} \right)$, would be adequate.)

Choose a map, $M$, such that $\mathcal{D} \left( M \right) = \left( F \vdash
\left\{ \mathcal{U} \left( F \right) \right\} \right)$, and for each member,
$B$, of $\left( F \vdash \left\{ \mathcal{U} \left( F \right) \right\}
\right)$, $M_B$ is an integer, $\geq 1$, and $\leq N_{\mathcal{C} \left( F, B
\right)}$, (hence $M_B \in \mathcal{I} \left( N_{\mathcal{C} \left( F, B
\right)} \right)$), and such that, if $A$, and $B$, are two {\emph{distinct}}
members, of $F$, such that $\mathcal{C} \left( F, A \right) =\mathcal{C} \left(
F, B \right)$, then $M_A \neq M_B$.  This means that, for each member, $A$, of
$\mathbb{B} \left( F \right)$, $\mathcal{N} \left( M,\mathcal{P} \left( F, A
\right) \right)$ is a bijection, whose domain is $\mathcal{P} \left( F, A
\right)$, and whose range is $\mathcal{I} \left( N_A \right)$.

For each ordered pair, $\left( A, n \right)$, of a member, $A$, of
$\mathbb{B} \left( F \right)$, and a member, $n$, of $\mathcal{I} \left( N_A
\right)$, I define $\mathbb{J} \left( A, n \right)$ to be the unique member,
$B$, of $\mathcal{P} \left( F, A \right)$, such that $M_B = n$.

Choose a map, $u$, such that $\mathcal{D} \left( u \right) =\mathbb{B} \left(
F \right)$, and for each member, $A$, of $\mathbb{B} \left( F \right)$, $u_A$
is a unit vector.  ($u_A$ could be the same unit vector, for every member,
$A$, of $\mathbb{B} \left( F \right)$, but need not be.)

I first define, directly,
$w_{\begin{array}{c}\\[-25pt]\scriptstyle{
{\!\!\!\mathcal{U} \left( F \right)\!\!\!\!}
   }\end{array}} \equiv z$, and
$R_{\mathcal{U} \left( F \right)} \equiv L$, so that $\mathcal{Y}_3$ is
satisfied.

Now, for each member, $B$, of $\left( F \vdash \left\{ \mathcal{U} \left( F
\right) \right\} \right)$, I define $w_B$, and $R_B$, in terms of
$w_{\mathcal{C} \left( F, B \right)}$, and $R_{\mathcal{C} \left( F, B
\right)}$, by
\[ w_B \equiv w_{\mathcal{C} \left( F, B \right)} + \left( \frac{2 M_B -
   N_{\mathcal{C} \left( F, B \right)} - 1}{N_{\mathcal{C} \left( F, B
   \right)}} \right) R_{\mathcal{C} \left( F, B \right)} u_{\mathcal{C} \left(
   F, B \right)}, \]
and
\[ R_B \equiv \left( \frac{\sigma \tau}{1 + \sigma \tau} \right) \left(
   \frac{R_{\mathcal{C} \left( F, B \right)}}{N_{\mathcal{C} \left( F, B
   \right)}} \right) . \]

These relations define $w$, and $R$, completely, by induction, on the levels, in
$F$, of the members, of $F$.

The explicit formula, for $R_A$, for any member, $A$, of $F$, is
\[ R_A = L \left( \frac{\sigma \tau}{1 + \sigma \tau} \right)^{\mathbb{L}
   \left( F, A \right)} \prod_{B \in \mathcal{E} \left( F, A \right)} \left(
   \frac{1}{N_B} \right), \]
where, for any ordered pair, $\left( F, A \right)$, of a greenwood, $F$, and a
nonempty subset, $A$, of $\mathcal{U} \left( F \right)$, I defined
$\mathcal{E} \left( F, A \right)$, the stem, of $A$, in $F$, on page 4, as the
set, whose members are all the members, $B$, of $F$, such that $A \subset B$.

For each member, $A$, of $\left( F \vdash \left\{ \mathcal{U} \left( F \right)
\right\} \right)$, let $\mathcal{V} \left( F, A \right) \equiv \left( \left(
\mathcal{E} \left( F, A \right) \vdash \left\{ \mathcal{U} \left( F \right)
\right\} \right) \cup \left\{ A \right\} \right)$.  Then the explicit formula,
for $w_A$, for any member, $A$, of $\left( F \vdash \left\{ \mathcal{U} \left(
F \right) \right\} \right)$, is
\[ w_A = z + \sum_{B \in \mathcal{V} \left( F, A \right)} \left( \frac{2 M_B -
   N_{\mathcal{C} \left( F, B \right)} - 1}{N_{\mathcal{C} \left( F, B
   \right)}} \right) R_{\mathcal{C} \left( F, B \right)} u_{\mathcal{C} \left(
   F, B \right)} \qquad\qquad\qquad\qquad\qquad\qquad\quad\quad \]
\[ = z + \!\sum_{B \in \mathcal{V} \left( F, A \right)} \left( \frac{2 M_B -
   N_{\mathcal{C} \left( F, B \right)} - 1}{N_{\mathcal{C} \left( F, B
   \right)}} \right)\! \left( L \left( \frac{\sigma \tau}{1 + \sigma \tau}
   \right)^{\mathbb{L} \left( F,\mathcal{C} \left( F, B \right) \right)}
   \!\!\!\!\!\prod_{K \in \mathcal{E} \left( F,\mathcal{C} \left( F, B \right) \right)}
   \!\left( \frac{1}{N_K} \right)\! \right) u_{\mathcal{C} \left( F, B \right) .}
\]
These relations imply, that, if $A$ is any member, of $\mathbb{B} \left( F
\right)$, and $B$ is any member, of $\mathcal{P} \left( F, A \right)$, then
$w_B$ lies on the straight line, between $\left( w_A - R_A u_A \right)$, and
$\left( w_A + R_A u_A \right)$.

Furthermore, if $A$ is any member, of $\mathbb{B} \left( F \right)$, and $B$
is any member, of $\mathcal{P} \left( F, A \right)$, then $\left| w_A - w_B
\right| \leq \left( \frac{N_A - 1}{N_A} \right) R_A$ holds.  And $R_B = \left(
\frac{R_A}{N_A} \right) \left( \frac{\sigma \tau}{1 + \sigma \tau} \right)$,
hence, if $s$ is any point, in $\mathbb{V} \left( d, w_B, R_B \right)$, then,
by the triangle inequality,
\[ \left| w_A - s \right| \leq \left| w_A - w_B \right| + \left| w_B - s
   \right| \qquad\qquad\qquad\qquad\qquad\qquad\qquad\qquad\qquad\qquad\qquad\quad \]
\[ \qquad\qquad\quad\quad \leq \left( \frac{N_A - 1}{N_A} \right) R_A + \left( \frac{R_A}{N_A}
   \right) \left( \frac{\sigma \tau}{1 + \sigma \tau} \right) = R_A - \left(
   \frac{R_A}{N_A} \right) \left( \frac{1}{1 + \sigma \tau} \right) < R_A \]
holds, hence $\mathbb{V} \left( d, w_B, R_B \right) \subseteq \mathbb{V}
\left( d, w_A, R_A \right)$ holds.

Hence, if $A$, and $B$, are any members, of $F$, such that $A \subseteq B$,
and \\
$\left( \mathbb{L} \left( F, A \right) -\mathbb{L} \left( F, B \right)
\right) = 1$, both hold, then $\mathbb{V} \left( d, w_A, R_A \right) \subseteq
\mathbb{V} \left( d, w_B, R_B \right)$ holds.

Now let $n$ by any integer, $\geq 1$, such that, if $A$, and $B$, are any
members, of $F$, such that $A \subseteq B$, and $\left( \mathbb{L} \left( F,
A \right) -\mathbb{L} \left( F, B \right) \right) = n$, both hold, then
$\mathbb{V} \left( d, w_A, R_A \right) \subseteq \mathbb{V} \left( d, w_B,
R_B \right)$ holds, and let $H$, and $K$, be any members, of $F$, such that $H
\subseteq K$, and $\left( \mathbb{L} \left( F, H \right) -\mathbb{L} \left(
F, K \right) \right) = \left( n + 1 \right)$, both hold.  Then $\mathcal{C}
\left( F, H \right) \subseteq K$, and $\left( \mathbb{L} \left( F,\mathcal{C}\!
\left( F, H \right) \right) -\right.$ \\
$\left.\mathbb{L} \left( F, K \right) \right) = n$, both
hold, hence $\mathbb{V} \left( d, w_{\mathcal{C} \left( F, H \right)},
R_{\mathcal{C} \left( F, H \right)} \right) \subseteq \mathbb{V} \left( d,
w_K, R_K \right)$ holds.  Hence, since $\mathbb{V} \left( d, w_H, R_H \right)
\subseteq \mathbb{V} \left( d, w_{\mathcal{C} \left( F, H \right)},
R_{\mathcal{C} \left( F, H \right)} \right)$, also holds, $\mathbb{V} \left(
d, w_H, R_H \right) \subseteq \mathbb{V}\! \left( d, w_K, R_K \right)$ holds.

Hence, by induction, on $n$, if $A$, and $B$, are any members, of $F$, such
that $A \subseteq B$, and $\left( \mathbb{L} \left( F, A \right) -\mathbb{L}
\left( F, B \right) \right) \geq 1$, both hold, then $\mathbb{V} \left( d,
w_A, R_A \right) \subseteq \mathbb{V} \left( d, w_B, R_B \right)$ holds.

Now, if $A$, and $B$, are members, of $F$, such that $A \subseteq B$, and
$\left( \mathbb{L} \left( F, A \right) -\mathbb{L} \left( F, B \right)
\right) = 0$ holds, then $A = B$, hence $\mathbb{V} \left( d, w_A, R_A
\right) \subseteq \mathbb{V} \left( d, w_B, R_B \right)$ holds.

Hence, if $A$, and $B$, are any members, of $F$, such that $A \subseteq B$
holds, then \\
$\mathbb{V} \left( d, w_A, R_A \right) \subseteq \mathbb{V}
\left( d, w_B, R_B \right)$ holds, hence $\mathcal{Y}_4$ holds.

Now let $A$ be any member, of $\mathbb{B} \left( F \right)$, and let $y$ be
any map, such that $\mathcal{D} \left( y \right) = A$, $\mathcal{R} \left( y
\right) \subseteq \mathbb{E}_d$, and for each member, $B$, of $\mathcal{P}
\left( F, A \right)$, $\mathcal{R} \left( \mathcal{N} \left( y, B \right)
\right) \subseteq \mathbb{V} \left( d, w_B, R_B \right)$.

Let $G$ be any member, of $\mathcal{P} \left( F, A \right)$, and let $a$, and
$b$, be any members, of $G$.  Then the following inequality, $\mathcal{Z}_1$,
holds.
\[ \mathcal{Z}_1 . \hspace{4em} \left| y_a - y_b \right| \leq \left| y_a - w_G
   \right| + \left| w_G - y_b \right| \leq 2 R_G = \left( \frac{2 R_A}{N_A}
   \right) \left( \frac{\sigma \tau}{1 + \sigma \tau} \right) . \!\hspace{5em}
\]

Let $H$, and $K$, be any two {\emph{distinct}} members, of $\mathcal{P} \left(
F, A \right)$, and let $h$ be any member, of $H$, and let $k$ be any member,
of $K$.  Then
\[ \left( \frac{2 R_A}{N_A} \right) \mathbb{M} \left( M_H - M_K \right) =
   \left| w_H - w_K \right| \leq \left| w_H - y_h \right| + \left| y_h - y_k
   \right| + \left| y_k - w_K \right| \]
holds, hence, since $H \neq K$ implies that $\mathbb{M} \left( M_H - M_K
\right) \geq 1$ holds, the following inequality, $\mathcal{Z}_2$, holds.
\[ \mathcal{Z}_2 . \hspace{5em} \! \! \left| y_h - y_k \right| \geq \left(
   \frac{2 R_A}{N_A} \right) \mathbb{M} \left( M_H - M_K \right) - \left| w_H
   - y_h \right| - \left| y_k - w_K \right| \hspace{5em}\, \]
\[ \quad\quad\,\, \hspace{3em} \geq \left( \frac{2 R_A}{N_A} \right) \left( \mathbb{M}
   \left( M_H - M_K \right) - \left( \frac{\sigma \tau}{1 + \sigma \tau}
   \right) \right) \]
\[ \geq \left( \frac{2 R_A}{N_A} \right) \left( \frac{1}{1 + \sigma \tau}
   \right) . \]

Let $Q$, and $T$, be any members, of $\mathcal{P} \left( F, A \right)$, such
that $\mathbb{M} \left( M_Q - M_T \right) = 1$, and let $q$ be any member, of
$Q$, and let $t$ be any member, of $T$.  Then $\left| w_Q - w_T \right| =
\left( \frac{2 R_A}{N_A} \right)$, hence the following inequality,
$\mathcal{Z}_3$, holds.
\[ \mathcal{Z}_3 . \hspace{3em} \hspace{-1.2em} \left| y_q - y_t \right| \leq
   \left| y_q - w_Q | + \left| w_Q - w_T \right| + \left| w_T - y_t \right|
   \hspace{5em} \hspace{8em} \right.\quad\,\, \]
\[ \hspace{5.5em} \leq \left( \frac{R_A}{N_A} \right) \left( \left( \frac{\sigma
   \tau}{1 + \sigma \tau} \right) + 2 + \left( \frac{\sigma \tau}{1 + \sigma
   \tau} \right) \right) = \left( \frac{2 R_A}{N_A} \right) \left( \frac{1 + 2
   \sigma \tau}{1 + \sigma \tau} \right) . \]

\vspace{0.24cm}

Now let $B$ be any member, of $\mathcal{P} \left( F, A \right)$.
\enlargethispage{0.24cm}

Then, if $\# \left( B \right) = 1$, then $B$ is automatically a
$\sigma$-cluster, of $y$.  If $\# \left( B \right) \geq 2$, let $e$ be any
member, of $B$, and let $f$ be any member, of $\left( A \vdash B \right)$.
Then $\mathcal{K} \left( F, A, f \right) \neq B$, hence, by $\mathcal{Z}_2$,
for the case $H = B$, $K =\mathcal{K} \left( F, A, f \right)$, $h = e$, and $k
= f$, $\left| y_e - y_f \right| \geq \left( \frac{2 R_A}{N_A} \right) \left(
\frac{1}{1 + \sigma \tau} \right)$ holds.

Now $0 < \tau < 1$ holds, hence, by $\mathcal{Z}_1$, for the case $G = B$,
$\left| y_a - y_b \right| < \sigma \left| y_e - y_f \right|$ holds, for all $a
\in B$, $b \in B$, $e \in B$, and $f \in \left( A \vdash B \right)$.  Hence
$B$ is a $\sigma$-cluster, of $y$.

Hence every member, of $\mathcal{P} \left( F, A \right)$, is a
$\sigma$-cluster, of $y$.

Now let $U$ be any {\emph{strict}} subset, of $A$, such that $U$ is
{\emph{not}} a subset, of any member, $B$, of $\mathcal{P} \left( F, A
\right)$.  I will show that $U$ is {\emph{not}} a $\sigma$-cluster, of $y$.

The fact, that $U$ is {\emph{not}} a subset, of any member, of $\mathcal{P}
\left( F, A \right)$, implies that there are two distinct members, $H$, and
$K$, of $\mathcal{P} \left( F, A \right)$, such that $H \cap U$, and $K \cap U$,
are both nonempty.  Hence, by $\mathcal{Z}_2$, there exists a member, $h$, of
$U$, and a member, $k$, of $U$, such that $\left| y_h - y_k \right| \geq
\left( \frac{2 R_A}{N_A} \right) \left( \frac{1}{1 + \sigma \tau} \right)$
holds.

Now $0 < \sigma < 1$ holds, hence, by Lemma 8, no two $\sigma$-clusters, of
$y$, overlap, hence, if $U$ overlaps any member, of $\mathcal{P} \left( F, A
\right)$, then $U$ is {\emph{not}} a $\sigma$-cluster, of $y$.

Assume, now, that $U$ does not overlap any member, of $\mathcal{P} \left( F, A
\right)$.  Then, for each member, $B$, of $\mathcal{P} \left( F, A \right)$,
exactly one, of $B \subset U$, and $B \cap U = \emptyset$, holds, since, by
assumption, $U$ is not a subset, of any member, of $\mathcal{P} \left( F, A
\right)$.

Furthermore, the assumption, that $U$ is {\emph{not}} a subset, of any member,
of $\mathcal{P} \left( F, A \right)$, implies that there are at least two
members, $B$, of $\mathcal{P} \left( F, A \right)$, such that $B \subset U$
holds, and the assumption, that $U$ is a {\emph{strict}} subset, of $A$,
implies that there is at least one member, $B$, of $\mathcal{P} \left( F, A
\right)$, such that $B \cap U = \emptyset$ holds.

Now, if $\mathbb{J} \left( A, 1 \right) \subset U$ holds, let $n$ be the
{\emph{smallest}} member, of $\mathcal{I} \left( N_A \right)$, such that $U
\cap \mathbb{J} \left( A, n \right) = \emptyset$ holds, (hence $n \geq 2$),
and let $T \equiv \mathbb{J} \left( A, n \right)$, and let $Q \equiv
\mathbb{J} \left( A, \left( n - 1 \right) \right)$.

And if $U \cap \mathbb{J} \left( A, 1 \right) = \emptyset$ holds, let $n$ be
the {\emph{smallest}} member, of $\mathcal{I} \left( N_A \right)$, such that
$\mathbb{J} \left( A, n \right) \subset U$ holds, (hence $n \geq 2$), and let
$Q \equiv \mathbb{J} \left( A, n \right)$, and let $T \equiv \mathbb{J}
\left( A, \left( n - 1 \right) \right)$.

Then $Q \subseteq U$, and $T \subseteq \left( A \vdash U \right)$, both hold.
Furthermore, $\mathbb{M} \left( M_Q - M_T \right) = 1$ holds, hence, if $q$
is any member, of $Q$, and $t$ is any member, of $T$, then, by
$\mathcal{Z}_3$, $\left| y_q - y_t \right| \leq \left( \frac{2 R_A}{N_A}
\right) \left( \frac{1 + 2 \sigma \tau}{1 + \sigma \tau} \right)$ holds.

Hence, since $Q$, and $T$, are both nonempty, there exists a member, $q$, of
$U$, and a member, $t$, of $\left( A \vdash U \right)$, such that $\left| y_q
- y_t \right| \leq \left( \frac{2 R_A}{N_A} \right) \left( \frac{1 + 2 \sigma
\tau}{1 + \sigma \tau} \right)_{}$ holds.

Now I showed, above, that the fact, that $U$ is {\emph{not}} a subset, of any
member, of $\mathcal{P} \left( F, A \right)$, implies that there is a member,
$h$, of $U$, and a member, $k$, of $U$, such that $\left| y_h - y_k \right|
\geq \left( \frac{2 R_A}{N_A} \right) \left( \frac{1}{1 + \sigma \tau}
\right)$ holds.  Furthermore, $\tau \leq \left( \frac{1 - \sigma}{2 \sigma^2}
\right)$ holds, by assumption, hence $\sigma \left( 1 + 2 \sigma \tau \right)
\leq \sigma \left( 1 + \left( \frac{1 - \sigma}{\sigma} \right) \right) = 1$
holds, hence $\sigma \left| y_q - y_t \right| \leq \left| y_h - y_k \right|$
holds, hence $U$ is {\emph{not}} a $\sigma$-cluster, of $y$.

Hence, if $B$ is any member, of $\mathcal{P} \left( F, A \right)$, then $B$
{\emph{is}} a $\sigma$-cluster, of $y$, and there is no $\sigma$-cluster, $U$,
of $y$, such that $B \subset U$, and $U \subset A$, both hold, hence $B$ is a
member, of $\mathcal{P} \left( \mathcal{F} \left( y, \sigma \right), A
\right)$.  Hence $\mathcal{P} \left( F, A \right) \subseteq \mathcal{P} \left(
\mathcal{F} \left( y, \sigma \right), A \right)$ holds.  But $\mathcal{U}
\left( \mathcal{P} \left( \mathcal{F} \left( y, \sigma \right), A \right)
\right) = A =\mathcal{U} \left( \mathcal{P} \left( F, A \right) \right)$,
hence $\mathcal{P} \left( \mathcal{F} \left( y, \sigma \right), A \right)
=\mathcal{P} \left( F, A \right)$.

Now let $A$ be any member, of $\mathbb{B} \left( F \right)$, and let $x$ be
any map, such that $\mathcal{D} \left( x \right) =\mathcal{U} \left( F
\right)$, $\mathcal{R} \left( x \right) \subseteq \mathbb{E}_d$, and, for
each member, $B$, of $\mathcal{P} \left( F, A \right)$, $\mathcal{R} \left(
\mathcal{N} \left( x, B \right) \right) \subseteq \mathbb{V} \left( d, w_B,
R_B \right)$ holds.  Let $y \equiv \mathcal{N} \left( x, A \right)$.  Then $y$
is a map, such that $\mathcal{D} \left( y \right) = A$, $\mathcal{R} \left( y
\right) \subseteq \mathbb{E}_d$, and, for each member, $B$, of $\mathcal{P}
\left( F, A \right)$, $\mathcal{R} \left( \mathcal{N} \left( y, B \right)
\right) \subseteq \mathbb{V} \left( d, w_B, R_B \right)$ holds.  Hence
$\mathcal{P} \left( \mathcal{F} \left( \mathcal{N} \left( x, A \right), \sigma
\right), A \right) =\mathcal{P} \left( \mathcal{F} \left( y, \sigma \right), A
\right) =\mathcal{P} \left( F, A \right)$ holds, and $\mathcal{Y}_5$ is
verified.

Now let $x$ be any member, of $\left( \mathbb{E}_d \right)^{\mathcal{U}
\left( F \right)}$, such that, for every member, $i$, of $\mathcal{U} \left( F
\right)$, $x_i \in \mathbb{V} \left( d, w_{\left\{ i \right\}}, R_{\left\{ i
\right\}} \right)$ holds.

Let $A$ be any member, of $\mathbb{B} \left( F \right)$, and let $B$ be any
member, of $\mathcal{P} \left( F, A \right)$.  Then, for each member, $i$, of
$B$, $\left\{ i \right\} \subseteq B$ holds, hence $\mathcal{Y}_4$ implies,
that $\mathbb{V} \left( d, w_{\left\{ i \right\}}, R_{\left\{ i \right\}}
\right) \subseteq \mathbb{V} \left( d, w_B, R_B \right)$ holds, hence $x_i
\in \mathbb{V} \left( d, w_B, R_B \right)$ holds, hence $\mathcal{R} \left(
\mathcal{N}\! \left( x, B \right) \right)\! \subseteq\!
\mathbb{V} \left( d, w_B,
R_B \right)$ holds, hence $\mathcal{Y}_5$ implies, that $\mathcal{P} \left(
\mathcal{F} \left( \mathcal{N} \left( x, A \right), \sigma \right), A \right)
=\mathcal{P} \left( F, A \right)$ holds.

Note, now, that $\mathcal{U} \left( F \right) =\mathcal{D} \left( x \right)$
is a member, of $\mathbb{B} \left( F \right)$, and a $\sigma$-cluster, of
$x$, hence, at least one member, of $\mathbb{B} \left( F \right)$, is a
$\sigma$-cluster, of $x$.

Now let $A$ be any member, of $\mathbb{B} \left( F \right)$, such that $A$ is
a $\sigma$-cluster, of $x$.  Then, by Corollary 2, of Lemma 10, $\mathcal{P}
\left( \mathcal{F} \left( \mathcal{N} \left( x, A \right), \sigma \right), A
\right) =\mathcal{P} \left( \mathcal{F} \left( x, \sigma \right), A \right)$
holds.

And, by $\mathcal{Y}_5$, $\mathcal{P} \left( \mathcal{F} \left( \mathcal{N}
\left( x, A \right), \sigma \right), A \right) =\mathcal{P} \left( F, A
\right)$ holds.

Hence, if $A$ is any member, of $\mathbb{B} \left( F \right)$, such that $A$
is a $\sigma$-cluster, of $x$, then $\mathcal{P} \left( \mathcal{F} \left( x,
\sigma \right), A \right) =\mathcal{P} \left( F, A \right)$ holds.

Now $\mathcal{U} \left( F \right)$ is the only member, $A$, of $F$, such that
$\mathbb{L} \left( F, A \right) = 0$, and $\mathcal{U} \left( F \right)$ is a
$\sigma$-cluster, of $x$, hence every member, $A$, of $F$, such that
$\mathbb{L} \left( F, A \right) = 0$, is a $\sigma$-cluster, of $x$.

Now let $n$ be any integer, $\geq 0$, such that every member, $A$, of $F$, such
that $\mathbb{L} \left( F, A \right) = n$, is a $\sigma$-cluster, of $x$, and
let $B$ be any member, of $F$, such that $\mathbb{L} \left( F, B \right) =
\left( n + 1 \right)$.  Then $\mathbb{L} \left( F,\mathcal{C} \left( F, B
\right) \right) = n$, hence $\mathcal{C} \left( F, B \right)$ is a
$\sigma$-cluster, of $x$, hence \\
$\mathcal{P} \left( \mathcal{F} \left( x,
\sigma \right),\mathcal{C} \left( F, B \right) \right) =\mathcal{P} \left(
F,\mathcal{C} \left( F, B \right) \right)$, hence $B$ is a member of
$\mathcal{P} \left( \mathcal{F} \left( x, \sigma \right),\mathcal{C} \left( F,
B \right) \right)$, (since $B$ is a member of $\mathcal{P} \left(
F,\mathcal{C} \left( F, B \right) \right)$), hence $B$ is a $\sigma$-cluster,
of $x$.

Hence, if $n$ is any integer, $\geq 0$, such that every member, $A$, of $F$,
such that $\mathbb{L} \left( F, A \right) = n$, is a $\sigma$-cluster, of
$x$, then every member, $B$, of $F$, such that $\mathbb{L} \left( F, B
\right) = \left( n + 1 \right)$, is a $\sigma$-cluster, of $x$.

Now every member, $A$, of $F$, such that $\mathbb{L} \left( F, A \right) =
0$, is a $\sigma$-cluster, of $x$.  Hence, by induction, on $n$, every member,
of $F$, is a $\sigma$-cluster, of $x$.

Hence $F \subseteq \mathcal{F} \left( x, \sigma \right)$.

Furthermore, since $\mathcal{P} \left( \mathcal{F} \left( x, \sigma \right), A
\right) =\mathcal{P} \left( F, A \right)$ holds, for every member, $A$, of
$\mathbb{B} \left( F \right)$, such that $A$ is a $\sigma$-cluster, of $x$,
and every member, $A$, of $\mathbb{B} \left( F \right)$, {\emph{is}} a
$\sigma$-cluster, of $x$, $\mathcal{P} \left( \mathcal{F} \left( x, \sigma
\right), A \right) =\mathcal{P} \left( F, A \right)$ holds, for every member,
$A$, of $\mathbb{B} \left( F \right)$.

I shall now show, that $\left( \mathcal{F} \left( x, \sigma \right) \vdash F
\right)$, is empty.

If $\left( \mathcal{F} \left( x, \sigma \right) \vdash F \right)$ is nonempty,
let $B$ be a member, of $\left( \mathcal{F} \left( x, \sigma \right) \vdash F
\right)$.  Then $B \neq \mathcal{U} \left( F \right)$, hence $\mathcal{C}
\left( F, B \right)$ is defined.

Now $F \subseteq \mathcal{F} \left( x, \sigma \right)$, hence $B$ overlaps no
member of $F$, hence, for each member, $A$, of $\mathcal{P} \left(
F,\mathcal{C} \left( F, B \right) \right)$, exactly one, of the four
possibilities, $A = B$, $B \subset A$, $A \cap B = \emptyset$, and $A \subset
B$, holds.  Now $B \notin F$, hence $A = B$ cannot hold.  And $A \subset
\mathcal{C} \left( F, B \right)$ holds, hence, by the definition, of
$\mathcal{C} \left( F, B \right)$, $B \subset A$ cannot hold.

Now $B$ is nonempty, and is a subset, of $\mathcal{C} \left( F, B \right)$, and
$\mathcal{P} \left( F,\mathcal{C} \left( F, B \right) \right)$ is a partition,
of $\mathcal{C} \left( F, B \right)$, hence $A \cap B = \emptyset$ cannot
hold, for {\emph{every}} member, $A$, of $\mathcal{P} \left( F,\mathcal{C}
\left( F, B \right) \right)$, hence there must be some member, say $K$, of
$\mathcal{P} \left( F,\mathcal{C} \left( F, B \right) \right)$, such that $K
\subset B$ holds.  Hence, since $B \in \mathcal{F} \left( x, \sigma \right)$,
and $B \subset \mathcal{C} \left( F, B \right)$, both hold, $K$ is {\emph{not}}
a member, of $\mathcal{P} \left( \mathcal{F} \left( x, \sigma
\right),\mathcal{C} \left( F, B \right) \right)$.  This contradicts
$\mathcal{P} \left( \mathcal{F} \left( x, \sigma \right),\mathcal{C} \left( F,
B \right) \right) =\mathcal{P} \left( F,\mathcal{C} \left( F, B \right)
\right)$.

Hence $\left( \mathcal{F} \left( x, \sigma \right) \vdash F \right)$ is empty,
hence $\mathcal{F} \left( x, \sigma \right) = F$.

Now, let $\mathbb{D}$ be the map, such that $\mathcal{D} \left( \mathbb{D}
\right) =\mathcal{U} \left( F \right)$, and, for each member, $i$, of
$\mathcal{U} \left( F \right)$, $\mathbb{D}_i =\mathbb{V} \left( d,
w_{\left\{ i \right\}}, R_{\left\{ i \right\}} \right)$.

Note that $\mathcal{R} \left( \mathbb{D} \right)$ is a finite set, all of
whose members, are $d$-balls.

Let $\mathbb{Z} \left( F, \sigma, d, z, L \right)$ be the set, whose members
are all the maps, $x \in \left( \mathbb{E}_d \right)^{\mathcal{U} \left( F
\right)}$, such that $x_i \in \mathbb{D}_i$ holds, for every member, $i$, of
$\mathcal{U} \left( F \right)$.  Then the $\left( \left( \# \left( \mathcal{U}
\left( F \right) \right) \right) d \right)$-volume, of $\mathbb{Z} \left( F,
\sigma, d, z, L \right)$, is the product, of the $d$-volumes, of the members,
of $\mathcal{R} \left( \mathbb{D} \right)$, hence, is nonzero.  Furthermore,
$\mathcal{F} \left( x, \sigma \right) = F$ holds, for all $x \in \mathbb{Z}
\left( F, \sigma, d, z, L \right)$, and $\left| x_i - z \right| \leq L$ holds,
for every member, $i$, of $\mathcal{U} \left( F \right)$, for all $x \in
\mathbb{Z} \left( F, \sigma, d, z, L \right)$.

%\vspace{0.1cm}


\begin{thebibliography}{99}

\bibitem{BPHZ in EPS}
C. Austin, ``A BPHZ Convergence Proof in Euclidean Position Space,''
(1993).\\
A gif file scan is available, at
http://web.ukonline.co.uk/chrisaustin/.\\
\LaTeX $\;\!\!$ transcription to appear in hep-th.

\bibitem{Weinberg}
S. Weinberg, ``High-Energy Behavior In Quantum Field Theory,'' Phys.
Rev. {\bf 118} (1960) 838-849.

\bibitem{Zimmermann}
W. Zimmermann, ``Convergence of Bogolyubov's Method of Renormalization in Momentum
Space,'' Commun. Math. Phys. {\bf 15} (1969) 208-234.  Reprinted in
Lect. Notes Phys. {\bf 558} (2000) 217-243.

\bibitem{Bogoliubov Parasiuk}
N.N. Bogoliubov and O.S. Parasiuk, ``On The Multiplication Of The
Causal Function In The Quantum Theory Of Fields,'' Acta Math.
{\bf 97} (1957) 227-266.

\bibitem{Hepp}
K. Hepp, ``Proof Of The Bogolyubov-Parasiuk Theorem On Renormalization,''
Commun. Math. Phys. {\bf 2} (1966) 301-326.

\bibitem{'t Hooft}
G. 't Hooft, ``Rigorous Construction of Planar Diagram Field Theories in
Four Dimensional Euclidean Space,'' Commun. Math. Phys. {\bf 88} (1983) 1-25.

\end{thebibliography}
\end{document}